\documentclass[preprint2]{aastex}
\usepackage{array}
\usepackage{multicol}
\usepackage{amssymb}
\usepackage{epsfig}
\usepackage{ifthen}
\usepackage{rotating}

\newboolean{ispreprint}
\setboolean{ispreprint}{true}

\bibpunct[; ]{(}{)}{;}{a}{}{,}

\newlength{\nopmsp}
\newlength{\pmsp}

\newsavebox{\thisbox}
\newlength{\thiswid}

\newcommand{\etal}{\textit{et al.}}

\newcommand{\sneia}{SNe~Ia}
\newcommand{\snia}{SN~Ia}

\newcommand{\om}{\ensuremath{\Omega_\mathrm{M}}}
\newcommand{\ol}{\ensuremath{\Omega_\Lambda}}
\newcommand{\ow}{\ensuremath{\Omega_X}}
\newcommand{\scriptm}{\ensuremath{\mathcal{M}}}
\newcommand{\ebv}{\mbox{\ensuremath{E(B}-\ensuremath{V)}}}
\newcommand{\wfpc}{WFPC2}
\newcommand{\hst}{HST}

\slugcomment{to be submitted to \textit{The Astrophysical Journal}}
\shorttitle{$\Omega_\mathrm{M}$, $\Omega_\Lambda$, and $w$ from
  HST-Observed \sneia}
\shortauthors{Knop \etal}

\received{2003 May 24}
\begin{document}

\title{New Constraints on $\Omega_\mathrm{M}$, $\Omega_\Lambda$, and $w$ from
 an Independent Set of Eleven High-Redshift Supernovae Observed with
  HST$^{1}$}



\author{
R.~A.~Knop\altaffilmark{2,3,4},
G.~Aldering\altaffilmark{5,4},
R.~Amanullah\altaffilmark{6},
P.~Astier\altaffilmark{7},
G.~Blanc\altaffilmark{5,7},
M.~S.~Burns\altaffilmark{8},
A.~Conley\altaffilmark{5,9},
S.~E.~Deustua\altaffilmark{5,10},
M.~Doi\altaffilmark{11},
R.~Ellis\altaffilmark{12},
S.~Fabbro\altaffilmark{13,4},
G.~Folatelli\altaffilmark{6},
A.~S.~Fruchter\altaffilmark{14},
G.~Garavini\altaffilmark{6},
S.~Garmond\altaffilmark{5,9},
K.~Garton\altaffilmark{8},
R.~Gibbons\altaffilmark{5},
G.~Goldhaber\altaffilmark{5,9},
A.~Goobar\altaffilmark{6},
D.~E.~Groom\altaffilmark{5,4},
D.~Hardin\altaffilmark{7}, 
I.~Hook\altaffilmark{15},
D.~A.~Howell\altaffilmark{5},
A.~G.~Kim\altaffilmark{5,4},
B.~C.~Lee\altaffilmark{5},
C.~Lidman\altaffilmark{17},
J.~Mendez\altaffilmark{18,19},
S.~Nobili\altaffilmark{6},
P.~E.~Nugent\altaffilmark{5,4},
R.~Pain\altaffilmark{7},
N.~Panagia\altaffilmark{14},   
C.~R.~Pennypacker\altaffilmark{5},
S.~Perlmutter\altaffilmark{5},
R.~Quimby\altaffilmark{5},
J.~Raux\altaffilmark{7},
N.~Regnault\altaffilmark{5,23},
P.~Ruiz-Lapuente\altaffilmark{19},
G.~Sainton\altaffilmark{7},
B.~Schaefer\altaffilmark{20},
K.~Schahmaneche\altaffilmark{7},
E.~Smith\altaffilmark{2},
A.~L.~Spadafora\altaffilmark{5},
V.~Stanishev\altaffilmark{6},
M.~Sullivan\altaffilmark{21,12},
N.~A.~Walton\altaffilmark{16},
L.~Wang\altaffilmark{5},
W.~M.~Wood-Vasey\altaffilmark{5,9}, and
N.~Yasuda\altaffilmark{22} \\
(THE SUPERNOVA COSMOLOGY PROJECT)\\[12pt]
Accepted for publication in \emph{The Astrophysical Journal}
}
\altaffiltext{1}{ Based in part on observations made with the NASA/ESA
Hubble Space Telescope, obtained at the Space Telescope Science
Institute, which is operated by the Association of Universities for
Research in Astronomy, Inc., under NASA contract NAS 5-26555. These
observations are associated with programs GO-7336, GO-7590, and GO-8346.
Some of the data presented herein were obtained at the W.M. Keck
Observatory, which is operated as a scientific partnership among the
California Institute of Technology, the University of California and the
National Aeronautics and Space Administration. The Observatory was made
possible by the generous financial support of the W.M. Keck Foundation.
Based in part on observations obtained at the WIYN Observatory, which is
a joint facility of the University of Wisconsin-Madison, Indiana
University, Yale University, and the National Optical Astronomy
Observatory.  Based in part on observations made with the European
Southern Observatory telescopes (ESO programmes 60.A-0586 and
265.A-5721).  Based in part on observations made with the
Canada-France-Hawaii Telescope, operated by the National Research
Council of Canada, le Centre National de la Recherche Scientifique de
France, and the University of Hawaii.}
\altaffiltext{2}{Department of Physics and Astronomy, Vanderbilt
University, Nashville, TN 37240, USA}
\altaffiltext{3}{Visiting Astronomer, Kitt Peak National Observatory,
National Optical Astronomy Observatory, which is operated by the
Association of Universities for Research in Astronomy, Inc. (AURA) under
cooperative agreement with the National Science Foundation.}
\altaffiltext{4}{Visiting Astronomer, Cerro Tololo Interamerican
Observatory, National Optical Astronomy Observatory, which is operated
by the Association of Universities for Research in Astronomy, Inc.
(AURA) under cooperative agreement with the National Science Foundation.}
\altaffiltext{5}{E. O. Lawrence Berkeley National Laboratory, 1
Cyclotron Rd., Berkeley, CA 94720, USA }
\altaffiltext{6}{Department of Physics, Stockholm University, SCFAB,
S-106 91 Stockholm, Sweden}
\altaffiltext{7}{LPNHE, CNRS-IN2P3, University of Paris VI \& VII,
Paris, France }
\altaffiltext{8}{Colorado College ~14 East Cache La Poudre St., Colorado
Springs, CO 80903}
\altaffiltext{9}{Department of Physics, University of California
Berkeley, Berkeley, 94720-7300 CA, USA}
\altaffiltext{10}{American Astronomical Society,  2000 Florida Ave, NW,
Suite 400, Washington, DC, 20009 USA.}
\altaffiltext{11}{Department of Astronomy and Research Center for the
Early Universe, School of Science, University of Tokyo, Tokyo 113-0033, Japan}
\altaffiltext{12}{California Institute of Technology, E. California
Blvd, Pasadena,  CA 91125, USA}
\altaffiltext{13}{Centro, Multidisiplinar de Astrof\'{\i}sica, Instituo,
  Superior T\'ecnico, Lisbo}
\altaffiltext{14}{Space Telescope Science Institute, 3700 San Martin
Drive, Baltimore, MD 21218, USA}
\altaffiltext{15}{Department of Physics, University of Oxford, Nuclear
\& Astrophysics Laboratory Keble Road, Oxford, OX1 3RH, UK}
\altaffiltext{16}{Institute of Astronomy, Madingley Road, Cambridge CB3
0HA, UK }
\altaffiltext{17}{European Southern Observatory, Alonso de C ordova
3107, Vitacura, Casilla 19001, Santiago 19, Chile }
\altaffiltext{18}{Isaac Newton Group, Apartado de Correos 321, 38780
Santa Cruz de La Palma, Islas Canarias, Spain}
\altaffiltext{19}{Department of Astronomy, University of Barcelona,
Barcelona, Spain }
\altaffiltext{20}{University of Texas, Department of Astronomy, C-1400,
Austin, TX,78712, U.S.A.}
\altaffiltext{21}{Department of Physics, University of Durham, South
Road, Durham, DH1 3LE, UK}
\altaffiltext{22}{National Astronomical Observatory, Mitaka, Tokyo
181-8588, Japan}
\altaffiltext{23}{Now at LLR, CNRS-IN2P3, Ecole Polytechnique,
  Palaiseau, France}

\begin{abstract}
We report measurements of \om, \ol, and $w$ from eleven supernovae at
$z=0.36$--$0.86$ with high-quality lightcurves measured using \wfpc\ on
the HST.  This is an independent set of high-redshift supernovae that
confirms previous supernova evidence for an accelerating Universe.  The
high-quality lightcurves available from photometry on \wfpc\ make it
possible for these eleven supernovae alone to provide measurements of
the cosmological parameters comparable in statistical weight to the
previous results.  Combined with earlier Supernova Cosmology Project
data, the new supernovae yield a measurement of the mass density
\mbox{$\om=0.25^{+0.07}_{-0.06}$} (statistical) $\pm0.04$ (identified
systematics), or equivalently, a cosmological constant of
\mbox{$\ol=0.75^{+0.06}_{-0.07}$} (statistical) $\pm0.04$ (identified
systematics), under the assumptions of a flat universe and that the dark
energy equation of state parameter has a constant value $w=-1$.  When
the supernova results are combined with independent flat-universe
measurements of \om\ from CMB and galaxy redshift distortion data, they
provide a measurement of $w=-1.05^{+0.15}_{-0.20}$ (statistical)
$\pm0.09$ (identified systematic), if $w$ is assumed to be constant in
time.  In addition to high-precision lightcurve measurements, the new
data offer greatly improved color measurements of the high-redshift
supernovae, and hence improved host-galaxy extinction estimates.  These
extinction measurements show no anomalous negative \ebv\ at high
redshift.  The precision of the measurements is such that it is possible
to perform a host-galaxy extinction correction directly for individual
supernovae without any assumptions or priors on the parent \ebv\
distribution.  Our cosmological fits using full extinction corrections
confirm that dark energy is required with $P(\ol>0)>0.99$, a result
consistent with previous and current supernova analyses which rely upon
the identification of a low-extinction subset or prior assumptions
concerning the intrinsic extinction distribution.

\end{abstract}

\section{Introduction}

Five years ago, the Supernova Cosmology Project (SCP) and the High-Z
Supernova Search Team both presented studies of distant Type~Ia
supernovae (SNe~Ia) in a series of reports, which gave strong evidence
for an acceleration of the Universe's expansion, and hence for a
non-zero cosmological constant, or dark energy density \citep[for a
review, see Perlmutter \& Schmidt
2003]{per98,gar98a,schmidt98,rie98,per99}.  These results ruled out a
flat, matter-dominated (\mbox{$\om=1$}, \mbox{$\ol=0$}) universe.  For a
flat universe, motivated by inflation theory, these studies yielded a
value for the cosmological constant of $\ol\simeq0.7$.  Even in the
absence of assumptions about the geometry of the Universe, the supernova
measurements indicate the existence of dark energy with greater than 99\%
confidence.

The supernova results combined with observations of the power spectrum
of the Cosmic Microwave Background (CMB)
\citep[e.g.,][]{jaf01,ben03,spe03}, the properties of massive clusters
\cite[e.g.,][]{tur01,all02,bah03}, and dynamical redshift-space
distortions \citep{haw02} yield a consistent picture of a flat universe
with $\om\simeq0.3$ and $\ol\simeq0.7$ \citep{bah99}.  Each of these
measurements is sensitive to different combinations of the parameters,
and hence they complement each other.  Moreover, because there are three
different measurements of two parameters, the combination provides an
important consistency check.  While the current observations of galaxy
clusters and dynamics, and of high-redshift supernovae, primarily probe the
``recent'' Universe at redshifts of $z<1$, the CMB measurements probe
the early Universe at $z\sim1100$.  That consistent results are obtained
by measurements of vastly different epochs of the Universe's history
suggests a vindication of the standard model of the expanding Universe.

In the redshift range around \mbox{$z=0.4$--$0.7$}, the supernova
results are most sensitive to a linear combination of \om\ and \ol\
close to $\om-\ol$.  In contrast, galaxy clustering and dynamics are
sensitive primarily to $\om$ alone, while the CMB is most sensitive to
$\om+\ol$.  Although combinations of other measurements lead to a
separate confirmation of the Universe's acceleration
\cite[e.g.,][]{est02}, taken alone it is the supernovae that provide the
best \emph{direct} evidence for dark energy.  Therefore, it is of
importance to improve the precision of the supernova result, to confirm
the result with additional independent high-redshift supernovae, and
also to limit the possible effects of systematic errors.

\citet{per97,per99} and \citet{rie98} presented extensive accounts of,
and bounds for, possible systematic uncertainties in the supernova
measurements.  One obvious possible source of systematic uncertainty is
the effect of host-galaxy dust.  For a given mass density, the effect of
a cosmological constant on the magnitudes of high-redshift supernovae is
to make their observed brightnesses \emph{dimmer} than would have been
the case with \mbox{$\ol=0$}.  Dust extinction from within the host
galaxy of the high-redshift supernovae could have a similar effect;
however, normal dust will also redden the colors of the supernovae.
Therefore, a measurement of the color of the high-redshift supernovae,
compared to the known colors of low-redshift SNe~Ia, has been used to
provide an upper limit on the effect of host-galaxy dust extinction, or
a direct measurement of that extinction which may then be corrected.
Uncertainties on extinction corrections based on these color
measurements usually dominate the statistical error of photometric
measurements.  Previous analyses have either selected a low-extinction
subset of both low- and high-redshift supernovae and not applied
corrections directly (``Fit C,'' the primary analysis of P99), or have
used an asymmetric Bayesian prior on the intrinsic extinction
distribution to limit the propagated uncertainties from errors in color
measurements \citep[``Fit E'' of P99]{rie98}.

In \citet{sul03}, we set stronger limits on the effects of host-galaxy
extinction by comparing the extinction, cosmological parameters, and
supernova peak magnitude dispersion for subsets of the SCP supernovae
observed in different types of host galaxies, as identified from both
HST imaging and Keck spectroscopy of the hosts.  We found that
supernovae in early-type (E and S0) galaxies show a smaller dispersion
in peak magnitude at high redshift, as had previously been seen at low
redshift \citep[e.g.][]{wan97}.  This subset of the P99 sample---in
hosts unlikely to be strongly affected by extinction---independently
provided evidence at the $5\sigma$ level that $\ol>0$ in a flat
Universe and confirmed that host-galaxy dust extinction was unlikely to be
a significant systematic in the results of P99, as had been suggested
previously \citep[e.g.,][]{row02}.  The natural next step following the
work of \citet{sul03}---presented in the current paper---is to provide
high-quality individual unbiased \ebv\ measurements that allow us
to directly measure the effect of host-galaxy extinction on each
supernova event without resorting to a prior on the color excess
distribution.

The current paper presents eleven new supernovae discovered and observed
by the SCP at redshifts $0.36<z<0.86$, a range very similar to that of
the 42 high-redshift supernovae reported in \citet[hereafter
P99]{per99}.  The supernovae of that paper, with one exception, were
observed entirely with ground-based telescopes; 11 of the 14 new
supernovae reported by \citet{rie98} were also observed from the ground.
The eleven supernovae of this work have lightcurves in both the $R$ and
$I$ bands measured with the Wide-Field/Planetary Camera (\wfpc) on the
Hubble Space Telescope (HST), and represent the largest sample to date
of HST-measured SNe~Ia at high redshift.

The HST provides two primary advantages for photometry of point sources
such as supernovae.  First the sky background is much lower, allowing a
much higher signal-to-noise ratio in a single exposure.  Second, because
the telescope is not limited by atmospheric seeing, it has very high
spatial resolution.  This helps the signal-to-noise ratio by greatly
reducing the area of background emission which contributes to the noise
of the source measurement, and moreover simplifies the task of
separating the variable supernova signal from the host galaxy.  With
these advantages, the precision of the lightcurve and color measurements
is much greater for the eleven supernovae in this paper than was
possible for previous ground-based observations.  These eleven
supernovae themselves provide a high-precision \emph{new} set of
high-redshift supernovae to test the accelerating universe results.
Moreover, the higher precision lightcurve measurements in both $R$- and
$I$-bands allow us to make high-quality, unbiased, individual
host-galaxy extinction corrections to each supernova event.

We first describe the PSF-fit photometry method used for extracting the
lightcurves from the \wfpc\ images (\S~\ref{sec:hstphotometry}).  Next,
in \S~\ref{sec:lightcurvefits}, we describe the lightcurve fitting
procedure, including the methods used for calculating accurate
$K$-corrections.  So that all supernovae may be treated consistently, in
\S~\ref{sec:colorcor} we apply the slightly updated $K$-correction
procedure to all of the supernovae used in P99.  In
\S~\ref{sec:cosmofitmethod}, the cosmological fit methodology we use is
described.  In \S~\ref{sec:colorsandextinction}, we discuss the evidence
for host-galaxy extinction (only significant for three of the eleven new
supernovae) from the $R$-$I$ lightcurve colors.  In
\S~\ref{sec:cosmoparam}, we present the measurements of the cosmological
parameters \om\ and \ol\ from the new dataset alone as well as combining
this set with the data of P99.  In \S~\ref{sec:combined}, we perform a
combined fit with our data and the high-redshift SNe of \citet{rie98}.
Finally, in \S~\ref{sec:w} we present measurements of $w$, the dark
energy equation of state parameter, from these data, and from these data
combined with recent CMB and galaxy redshift distortion measurements.
These discussions of our primary results are followed by updated
analyses of systematic uncertainties for these measurements in
\S~\ref{sec:systematic}.

\section{Observations, Data Reduction, and Analysis}
\label{sec:datared}

\subsection{\wfpc\ Photometry}
\label{sec:hstphotometry}

The supernovae discussed in this paper are listed in
Table~\ref{tab:snlist}.  They were discovered during three different
supernova searches, following the techniques described in \citet{per95,
per97, per99}.  Two of the searches were conducted with the 4m Blanco
telescope at the Cerro Tololo Inter-American Observatory (CTIO), in
November/December 1997 and March/April 1998.  The final search was
conducted at the Canada-France-Hawaii Telescope (CFHT) on Mauna Kea in
Hawaii in April/May 2000.  In each case, 2--3 nights of reference images
were followed 3--4 weeks later by 2--3 nights of search images.  The two
images of each search field were seeing-matched and subtracted, and were
searched for residuals indicating a supernova candidate.  Weather
conditions limited the depth and hence the redshift range of the
March/April 1998 search.  Out of the three searches, eleven of the resulting
supernova discoveries were followed with extensive HST photometry.
These supernovae are spaced approximately evenly in the redshift range
$0.3<z<0.9$.  Nine out of the eleven supernovae were discovered very
close to maximum light; two were discovered several days before maximum
light. 

Spectra were obtained with the red side of LRIS on the Keck 10m
telescope \citep{oke95}, with FORS1 on Antu (VLT-UT1) \citep{app98}, and
with EFOSC2\footnote{\url{http://www.ls.eso.org/lasilla/sciops/efosc/}}
on the ESO 3.6m telescope.  These spectra were used to confirm the
identification of the candidates as \sneia, and to measure the redshift
of each candidate.  Nine of the eleven supernovae in the set have strong
confirmation as Type~Ia through the presence of
\ion{Si}{2}~$\lambda$6150, \ion{Si}{2}~$\lambda$4190, or \ion{Fe}{2}
features that match those of a Type~Ia observed at a similar epoch.
\mbox{SNe\,1998ay} and 1998be have spectra which are consistent with
SNe~Ia spectra, although this identification is less secure for those
two.  However, we note that the colors (measured at multiple epochs with
the HST lightcurves) are inconsistent with other non-Ia types.  (We
explore the systematic effect of removing those two supernovae from the
set in \S~\ref{sec:typecontamination}.)

Where possible, the redshift, $z$, of each candidate was measured by
matching narrow features in the host galaxy of the supernovae; the
precision of these measurements in $z$ is typically 0.001.  In cases
where there were not sufficient host-galaxy features (\mbox{SNe\,1998aw}
and 1998ba), redshifts were measured from the supernova
itself; in these cases, $z$ is measured with a (conservative) precision
of 0.01 \citep{bra93}.  Even in the latter case, redshift measurements do
not contribute significantly to the uncertainties in the final
cosmological measurements since these are dominated by the photometric
uncertainties.

\ifthenelse{\boolean{ispreprint}}{
\begin{table*}[htbp]
\renewcommand{\arraystretch}{0.8}\scriptsize
\begin{lrbox}{\thisbox}
\begin{tabular}{lcll}
\tableline
\tableline
SN     &  $z$       & F675W              & F814W        \\
Name   &            & Observations       & Observations \\
\tableline   
1997ek &  0.863     & 1998-01-05 (400s,400s)  & 1998-01-05 (500s,700s) \\
       &            & 1998-01-11 (400s,400s)  & 1998-01-11 (500s,700s) \\
       &            &                         & 1998-02-02 (1100s,1200s) \\
       &            &                         & 1998-02-14 (1100s,1200s) \\
       &            &                         & 1998-02-27 (1100s,1200s) \\
       &            &                         & 1998-11-09 (1100s,1300s) \\
       &            &                         & 1998-11-16 (1100s,1300s) \\[10pt]
1997eq &  0.538     & 1998-01-06 (300s,300s)  & 1998-01-06 (300s,300s)  \\
       &            & 1998-01-21 (400s,400s)  & 1998-01-11 (300s,300s) \\
       &            &                         & 1998-02-02 (500s,700s) \\
       &            & 1998-02-11 (400s,400s)  & 1998-02-11 (500s,700s) \\
       &            & 1998-02-19 (400s,400s)  & 1998-02-19 (500s,700s) \\[10pt]
1997ez &  0.778     & 1998-01-05 (400s,400s)  & 1998-01-05 (500s,700s) \\
       &            & 1998-01-11 (400s,400s)  & 1998-01-11 (500s,700s) \\
       &            &                         & 1998-02-02 (1100s,1200s) \\
       &            &                         & 1998-02-14 (1100s,1200s) \\
       &            &                         & 1998-02-27 (100s,1200s,1100s,1200s) \\[10pt]
1998as &  0.355     & 1998-04-08 (400s,400s)  & 1998-04-08 (500s,700s) \\
       &            & 1998-04-20 (400s,400s)  & 1998-04-20 (500s,700s) \\
       &            & 1998-05-11 (400s,400s)  & 1998-05-11 (500s,700s) \\
       &            & 1998-05-15 (400s,400s)  & 1998-05-15 (500s,700s) \\
       &            & 1998-05-29 (400s,400s)  & 1998-05-29 (500s,700s) \\[10pt]
1998aw &  0.440     & 1998-04-08 (300s,300s)  & 1998-04-08 (300s,300s)  \\
       &            & 1998-04-18 (300s,300s)  & 1998-04-18 (300s,300s)  \\
       &            & 1998-04-29 (400s,400s)  & 1998-04-29 (500s,700s) \\
       &            & 1998-05-14 (400s,400s)  & 1998-05-14 (500s,700s) \\
       &            & 1998-05-28 (400s,400s)  & 1998-05-28 (500s,700s) \\[10pt]
1998ax &  0.497     & 1998-04-08 (300s,300s)  & 1998-04-08 (300s,300s)  \\
       &            & 1998-04-18 (300s,300s)  & 1998-04-18 (300s,300s)  \\
       &            & 1998-04-29 (300s,300s)  & 1998-04-29 (500s,700s) \\
       &            & 1998-05-14 (300s,300s)  & 1998-05-14 (500s,700s) \\
       &            & 1998-05-27 (300s,300s)  & 1998-05-27 (500s,700s) \\[10pt]
1998ay &  0.638     & 1998-04-08 (400s,400s)  & 1998-04-08 (500s,700s) \\
       &            & 1998-04-20 (400s,400s)  & 1998-04-20 (500s,700s) \\
       &            &                         & 1998-05-11 (1100s,1200s) \\
       &            &                         & 1998-05-15 (1100s,1200s) \\
       &            &                         & 1998-06-03 (1100s,1200s) \\[10pt]
1998ba &  0.430     & 1998-04-08 (300s,300s)  & 1998-04-08 (300s,300s)  \\
       &            & 1998-04-19 (300s,300s)  & 1998-04-19 (300s,300s)  \\
       &            & 1998-04-29 (400s,400s)  & 1998-04-29 (500s,700s) \\
       &            & 1998-05-13 (400s,400s)  & 1998-05-13 (500s,700s) \\
       &            & 1998-05-28 (400s,400s)  & 1998-05-28 (500s,700s) \\[10pt]
1998be &  0.644     & 1998-04-08 (300s,300s)  & 1998-04-08 (300s,300s)  \\
       &            & 1998-04-19 (300s,300s)  & 1998-04-19 (300s,300s)  \\
       &            & 1998-04-30 (400s,400s)  & 1998-04-30 (500s,700s) \\
       &            & 1998-05-15 (400s,400s)  & 1998-05-15 (500s,700s) \\
       &            & 1998-05-28 (400s,400s)  & 1998-05-28 (500s,700s) \\[10pt]
1998bi &  0.740     & 1998-04-06 (400s,400s)  & 1998-04-06 (500s,700s) \\
       &            & 1998-04-18 (400s,400s)  & 1998-04-18 (500s,700s) \\
       &            &                         & 1998-04-28 (1100s,1200s) \\
       &            &                         & 1998-05-12 (1100s,1200s) \\
       &            &                         & 1998-06-02 (1100s,1200s) \\[10pt]
2000fr &  0.543     &                         & 2000-05-08 (2200s) \\
       &            & 2000-05-15 (600s,600s)  & 2000-05-15 (1100s,1100s) \\
       &            & 2000-05-28 (600s,600s)  & 2000-05-28 (600s,600s) \\
       &            & 2000-06-10 (500s,500s)  & 2000-06-10 (600s,600s)  \\
       &            & 2000-06-22 (1100s,1300s) & 2000-06-22 (1100s,1200s) \\
       &            & 2000-07-08 (1100s,1300s) & 2000-07-08 (110s,1200s) \\
\tableline
\end{tabular}
\end{lrbox}
\settowidth{\thiswid}{\usebox{\thisbox}}
\begin{center}
\begin{minipage}{\thiswid}
\caption{\wfpc\ Supernova Observations}
\label{tab:snlist}
\usebox{\thisbox}
\end{minipage}
\end{center}
\end{table*}
}{\placetable{tab:snlist}}

Each of these supernovae was imaged with two broadband filters using the
Planetary Camera (PC) CCD of the \wfpc\ on the HST, which has a scale of
0.046$''$/pixel.  Table~\ref{tab:snlist} lists the dates of these
observations.  The F675W and F814W broadband filters were chosen to have
maximum sensitivity to these faint objects, while being as close a match
as practical to the rest-frame $B$ and $V$ filters at the targeted
redshifts. (Note that all of our \wfpc\ observing parameters except the
exact target coordinates were fixed prior to the supernova discoveries.)
The effective system transmission curves provided by STScI indicate
that, when used with WFPC2, F675W is most similar to ground-based $R$
band while F814W is most similar to ground-based $I$ band.  These
filters roughly correspond to redshifted $B$- and $V$-band filters for
the supernovae at $z<0.7$, and redshifted $U$- and $B$- band filters for
the supernovae at $z>0.7$.

The HST images were reduced through the standard HST ``On-The-Fly
Reprocessing'' data reduction pipeline provided by the Space Telescope
Science Institute.  Images were then background subtracted, and images
taken in the same orbit were combined to reject cosmic rays using the
``crrej'' procedure (a part of the STSDAS IRAF package).  Photometric
fluxes were extracted from the final images using a PSF-fitting
procedure.  Traditional PSF fitting procedures assume a single isolated
point source above a constant background.  In this case, the point
source was superimposed on the image of the host galaxy.  In all
cases, the supernova image was separated from the core of the host
galaxy; however, in most cases the separation was not enough that an
annular measurement of the background would be accurate.  Because the
host-galaxy flux is the same in all of the images, we used a PSF fitting
procedure that fits a PSF \emph{simultaneously} to every image of a
given supernova observed through a given photometric filter.  The model
we fit was:

\begin{eqnarray}
f_i(x,y) = f_{0i}\times\mathrm{psf}(x-x_{0i},y-y_{0i}) + \nonumber \\
\mathrm{bg}(x-x_{0i},y-y_{0i};a_j) + p_i
\end{eqnarray}

\noindent where $f_i(x,y)$ is the measured flux in pixel $(x,y)$ of the
$i$th image, ($x_{0i},y_{0i}$) is the position of the supernova on the
$i$th image, $f_{0i}$ is the total flux in the supernova in the $i$th
image, $\mathrm{psf}(u,v)$ is a normalized point spread function,
$\mathrm{bg}(u,v;a)$ is a temporally constant background parametrized by
$a_j$, and $p_i$ is a pedestal offset for the $i$th image.  There are
$4n+m-1$ parameters in this model, where $n$ is the number of images
(typically 2, 5, or 6 previously summed images) and $m$ is the number of
parameters $a_j$ that specify the background model (typically 3 or 6).
(The $-1$ is due to the fact that a zeroth-order term in the background
is degenerate with one of the $p_i$ terms.)  Parameters varied include
$f_i$, $x_{0i}$, $y_{0i}$, $p_i$, and $a_j$.

Due to the scarcity of objects in our PC images, geometric
transformations between the images at different epochs using other
objects on the four chips of \wfpc\ together allowed an \emph{a priori}
determination of $(x_{0i},y_{0i})$ good to $\sim1$~pixel.  Allowing
those parameters to vary in the fit (effectively, using the point source
signature of the supernova to determine the offset of the image)
provided position measurements a factor of $\sim10$
better.\footnote{Note that this may introduce a bias towards higher
flux, as the fit will seek out positive fluctuations on which to center
the PSF.  However, the covariance between the peak flux and position is
typically less than $\sim4$\% of the product of the positional
uncertainty and the flux uncertainty, so the effects of this bias will
be very small in comparison to our photometric errors.}  The model was
fit to $13\times13$ pixel patches extracted from all of the images of a
time sequence of a single supernova in a single filter (except for
SN\,1998ay, which is close enough to the host galaxy that a $7\times7$
pixel patch was used to avoid having to fit the core of the galaxy with
the background model).  In four out of the 99 patches used in the fits
to the 22 lightcurves, a single bad pixel was masked from the fit.  The
series of $f_{0i}$ values, corrected as described in the rest of this
section, provided the data used in the lightcurve fits described in
\S~\ref{sec:lightcurvefits}.  For one supernova (\mbox{SN\,1997ek} at
\mbox{$z=0.86$}), the F814W background was further constrained by a
supernova-free ``final reference'' image taken 11 months after the
supernova explosion.\footnote{Although obtaining final references to
subtract the galaxy background is standard procedure for ground-based
photometry of high-redshift supernovae, the higher resolution of \wfpc\
provides sufficient separation between the supernova and host galaxy
that such images are not always necessary, particularly in this redshift
range.}

A single Tiny Tim PSF was used as $\mathrm{psf}(u,v)$ for all images of
a given filter.  The Tiny Tim PSF used was subsampled to $10\times10$
subpixels; in the fit procedure, it was shifted and integrated (properly
summing fractional subpixels).  After shifting and resampling to the PC
pixel scale, it was convolved with an empirical $3\times3$ electron
diffusion kernel with 75\% of the flux in the central element
\citep{fru00}.\footnote{See also
\url{http://www.stsci.edu/software/tinytim/tinytim\_faq.html}} The PSF
was normalized in a $0.5''$-radius aperture, chosen to match the
standard zeropoint calibration \citep{hol95,dol00}.  Although the use of
a single PSF for every image is an approximation---the PSF of \wfpc\
depends on the epoch of the observation as well as the position on the
CCD---this approximation should be valid, especially given that for
all of the observations the supernova was positioned close to the center
of the PC.  To verify that this approximation is valid, we reran the PSF
fitting procedure with individually generated PSFs for most supernovae;
we also explored using a supernova spectrum instead of a standard star
spectrum in generating the PSF.  The measured fluxes were not
significantly different, showing differences in both directions
generally within 1--2\% of the supernova peak flux value---much less
than our photometric uncertainties on individual data points.

Although one of the great advantages of the Hubble Space Telescope is
its low background, CCD photometry of faint objects over a low
background suffer from an imperfect charge transfer efficiency (CTE)
effect, which can lead to a systematic underestimate of the flux of
point sources \citep{whi99,dol00,dol03}.  On the PC, these effects can
be as large as $\sim15$\%.  The measured flux values ($f_{0i}$ above) were
corrected for the CTE of \wfpc\ following the standard procedure of
\citet{dol00}.\footnote{These CTE corrections used updated coefficients
posted on Dolphin's web page
(\url{http://www.noao.edu/staff/dolphin/wfpc2\_calib/}) in September,
2002.}  Uncertainties on the CTE corrections were propagated into the
corrected supernova fluxes, although in all cases these uncertainties
were smaller than the uncertainties in the raw measured flux values.
Because the host galaxy is a smooth background underneath the point
source, it was considered as a contribution to the background in the CTE
correction.  For an image which was a combination of several separate
exposures within the same orbit or orbits, the CTE calculation was
performed assuming that each SN image had a measured SN flux whose
fraction of the total flux was equal to the fraction of that individual
image's exposure time to the summed image's total exposure time.  This
assumption is correct most of the time, with the exception of the few
instances where Earthshine affects part of an orbit.

In addition to the HST data, there exists ground-based photometry for
each of these supernovae.  This includes the images from the search
itself, as well as a limited amount of follow-up.  The details of which
supernovae were observed with which telescopes are given with the
lightcurves in Appendix~\ref{sec:dirtylaundry}.  Ground-based
photometric fluxes were extracted from images using the same aperture
photometry procedure of P99.  A complete lightcurve in a given filter
($R$ or $I$) combined the HST data with the ground-based data (using the
color-correction procedure described below in \S~\ref{sec:colorcor}),
using measured zeropoints for the ground-based data and the Vega
zeropoints of \citet{dol00} for the HST data.  The uncertainties on
those zeropoints (0.003 for F814W or 0.006 for F675W) were added as
correlated errors between all HST data points when combining with the
ground-based lightcurve.  Similarly, the measured uncertainty in the
ground-based zeropoint was added as a correlated error to all
ground-based fluxes.  Ground-based photometric calibrations were based
on observations of \citet{lan92} standard stars observed on the same
photometric night as a supernova observation; each calibration is
confirmed over two or more nights.  Ground-based zeropoint uncertainties
are generally $\lesssim0.02$--0.03; the $R$-band ground based zeropoint
for \mbox{SN\,1998ay} is only good to $\pm0.05$.  We have compared our
ground-based aperture photometry with our HST PSF-fitting photometry
using the limited number of sufficiently bright stars present in the PC
across the eleven SNe fields.  We find the difference between the HST
and ground-based photometry to be \mbox{$0.02\pm0.02$} in both the R-
and I-bands, consistent with no offset.  The correlated uncertainties
between different supernovae arising from ground-based zeropoints based
on the same calibration data, and between the HST supernovae (which all
share the same zeropoint), were included in the covariance matrix used
in all cosmological fits (see \S~\ref{sec:cosmofitmethod}).

\subsection{Lightcurve Fits}
\label{sec:lightcurvefits}

It is the magnitude of the supernova at its lightcurve peak that serves
as a ``calibrated candle'' in estimating the cosmological parameters
from the luminosity distance relationship.  To estimate this peak
magnitude, we performed template fits to the time series of photometric
data for each supernova.  In addition to the eleven supernovae described
here, lightcurve fits were also performed to the supernovae from P99,
including 18 supernovae from \citet[hereafter H96]{ham96}, and eight
from \citet[hereafter R99]{rie99} which match the same selection
criteria used for the H96 supernovae (having data within six days of
maximum light and located at \mbox{$cz>4000$~km/s}, limiting distance
modulus error due to peculiar velocities to less than 0.15 magnitudes).
Because of new templates and $K$-corrections (see below), lightcurve
fits to the supernovae from H96 and P99 used in the analyses below were
redone for consistency.  The results of these fits are slightly
different from those quoted in P99 for the same supernovae as a result
of the change in the lightcurve template, the new $K$-corrections, and
the different fit procedure, all discussed below.  For example, because
the measured \ebv\ value was considered in the $K$-corrections
(\S~\ref{sec:colorcor}), whereas it was not in P99, one should expect to
see randomly distributed differences in fit supernova lightcurve
parameters due to scatter in the color measurements.

Lightcurve fits were performed using a $\chi^2$-minimization procedure
based on MINUIT \citep{jam75}.  For both high- and low-redshift
supernovae, color corrections and $K$-corrections are applied (see
\S~\ref{sec:colorcor}) to the photometric data.  These data were then
fit to lightcurve templates.  Fits were performed to the combined $R$-
and $I$-band data for each high-redshift supernova.  For low-redshift
supernovae, fits were performed using only the $B$- and $V$-band data
(which correspond to de-redshifted $R$- and $I$-bands for most of the
high-redshift supernovae).  The lightcurve model fit to the supernova
has four parameters to modify the lightcurve templates: time of
rest-frame $B$-band maximum light, peak flux in $R$, $R$-$I$ color at
the epoch of rest-frame maximum $B$-band light, and timescale stretch
$s$.  Stretch is a parameter which linearly scales the time axis, so
that a supernova with a high stretch has a relatively slow decay from
maximum, and a supernova with a low stretch has a relatively fast decay
from maximum \citep{per97, gol01}.  For supernovae in the redshift range
\mbox{$z=0.3$--$0.7$}, a $B$ template was fit to the $R$-band lightcurve
and a $V$ template was fit to the $I$-band lightcurve.  For supernovae
at $z>0.7$, a $U$ template was fit to the $R$-band lightcurve and a $B$
template to the $I$-band lightcurve.  Two of the high-redshift
supernovae from P99 fall at $z\sim0.18$ (\mbox{SN\,1997I} and
\mbox{SN\,1997N}); for these supernovae, $V$ and $R$ templates were fit
to the $R$- and $I$-band data.  (The peak $B$-band magnitude was
extracted by adding the intrinsic \snia\ $B$-$V$ color to the fit
$V$-band magnitude at the epoch of $B$ maximum.)

The $B$ template used in the lightcurve fits was that of \citet{gol01}.
For this paper, new $V$-band and $R$-band templates were generated
following a procedure similar to that of \citet{gol01}, by fitting a
smooth parametrized curve through the low-redshift supernova data of H96
and R99.  A new $U$-band template was generated with data from
\citet{ham91}, \citet{lir98}, \citet{ric95}, \citet{sun99}, and
\citet{wel94}; comparison of our $U$-band template shows good agreement
with the new $U$-band photometry from \citet{jha02} at the relevant
epochs.  New templates were generated by fitting a smooth curve,
$f(t')$, to the low-redshift lightcurve data, where $t'=t/(1+z)/s$; $t$
is the number of observer-frame days relative to the epoch of the
$B$-band maximum of each supernova, $z$ is the redshift of each
supernova, and $s$ is the stretch of each supernova as measured from the
$B$-band lightcurves.  Lightcurve templates had an initial parabola with
a 20-day rise time \citep{ald00}, joined to a smooth spline section to
describe the main part of the lightcurve, then joined to an exponential
decay to describe the final tail at $>\sim70$~days past maximum light.
The first 100 days of each of the three templates is listed in
Table~\ref{tab:ltcvtemplates}.

Due to a secondary ``hump'' or ``shoulder'' $\sim20$ days after maximum,
the $R$-band lightcurve does not vary strictly according to the simple
time-axis scaling parametrized by stretch which is so successful in
describing the different $U$-, $B$-, and $V$-band lightcurves.  However,
for the two $z\sim0.18$ supernova to which we fit an $R$-band template,
the peak $R$- and $I$- band magnitudes are well constrained, and the
stretch is also well measured from the rest-frame $V$-band lightcurve.

\ifthenelse{\boolean{ispreprint}}{
\begin{table*}[p]
\scriptsize
\renewcommand{\arraystretch}{1.0}\scriptsize
\begin{lrbox}{\thisbox}
\begin{tabular}{rrrr c rrrr}
\tableline
\tableline
Day\tablenotemark{a} & $U$ flux\tablenotemark{b} & $V$ flux\tablenotemark{b} & $R$ flux\tablenotemark{b} & \hspace*{0.125in}
   & Day$^1$ & $U$ flux\tablenotemark{b} & $V$ flux\tablenotemark{b} & $R$ flux\tablenotemark{b} \\
\tableline
-19 & 6.712e-03 & 4.960e-03 & 5.779e-03 &&  31 & 4.790e-02 & 2.627e-01 & 3.437e-01 \\
-18 & 2.685e-02 & 1.984e-02 & 2.312e-02 &&  32 & 4.524e-02 & 2.481e-01 & 3.238e-01 \\
-17 & 6.041e-02 & 4.464e-02 & 5.201e-02 &&  33 & 4.300e-02 & 2.345e-01 & 3.054e-01 \\
-16 & 1.074e-01 & 7.935e-02 & 9.246e-02 &&  34 & 4.112e-02 & 2.218e-01 & 2.887e-01 \\
-15 & 1.678e-01 & 1.240e-01 & 1.445e-01 &&  35 & 3.956e-02 & 2.099e-01 & 2.733e-01 \\
-14 & 2.416e-01 & 1.785e-01 & 2.080e-01 &&  36 & 3.827e-02 & 1.990e-01 & 2.592e-01 \\
-13 & 3.289e-01 & 2.430e-01 & 2.832e-01 &&  37 & 3.722e-02 & 1.891e-01 & 2.463e-01 \\
-12 & 4.296e-01 & 3.174e-01 & 3.698e-01 &&  38 & 3.636e-02 & 1.802e-01 & 2.345e-01 \\
-11 & 5.437e-01 & 4.017e-01 & 4.681e-01 &&  39 & 3.565e-02 & 1.721e-01 & 2.237e-01 \\
-10 & 6.712e-01 & 4.960e-01 & 5.779e-01 &&  40 & 3.506e-02 & 1.649e-01 & 2.137e-01 \\
 -9 & 7.486e-01 & 5.889e-01 & 6.500e-01 &&  41 & 3.456e-02 & 1.583e-01 & 2.046e-01 \\
 -8 & 8.151e-01 & 6.726e-01 & 7.148e-01 &&  42 & 3.410e-02 & 1.524e-01 & 1.962e-01 \\
 -7 & 8.711e-01 & 7.469e-01 & 7.725e-01 &&  43 & 3.365e-02 & 1.471e-01 & 1.884e-01 \\
 -6 & 9.168e-01 & 8.115e-01 & 8.236e-01 &&  44 & 3.318e-02 & 1.423e-01 & 1.813e-01 \\
 -5 & 9.524e-01 & 8.660e-01 & 8.681e-01 &&  45 & 3.266e-02 & 1.378e-01 & 1.747e-01 \\
 -4 & 9.781e-01 & 9.103e-01 & 9.062e-01 &&  46 & 3.205e-02 & 1.337e-01 & 1.687e-01 \\
 -3 & 9.940e-01 & 9.449e-01 & 9.382e-01 &&  47 & 3.139e-02 & 1.299e-01 & 1.630e-01 \\
 -2 & 1.000e+00 & 9.706e-01 & 9.639e-01 &&  48 & 3.072e-02 & 1.263e-01 & 1.578e-01 \\
 -1 & 9.960e-01 & 9.880e-01 & 9.834e-01 &&  49 & 3.005e-02 & 1.229e-01 & 1.529e-01 \\
  0 & 9.817e-01 & 9.976e-01 & 9.957e-01 &&  50 & 2.945e-02 & 1.195e-01 & 1.483e-01 \\
  1 & 9.569e-01 & 1.000e+00 & 1.000e+00 &&  51 & 2.893e-02 & 1.161e-01 & 1.440e-01 \\
  2 & 9.213e-01 & 9.958e-01 & 9.952e-01 &&  52 & 2.853e-02 & 1.128e-01 & 1.398e-01 \\
  3 & 8.742e-01 & 9.856e-01 & 9.803e-01 &&  53 & 2.830e-02 & 1.096e-01 & 1.359e-01 \\
  4 & 8.172e-01 & 9.702e-01 & 9.545e-01 &&  54 & 2.827e-02 & 1.064e-01 & 1.320e-01 \\
  5 & 7.575e-01 & 9.502e-01 & 9.196e-01 &&  55 & 2.849e-02 & 1.033e-01 & 1.282e-01 \\
  6 & 6.974e-01 & 9.263e-01 & 8.778e-01 &&  56 & 2.793e-02 & 1.003e-01 & 1.244e-01 \\
  7 & 6.375e-01 & 8.991e-01 & 8.313e-01 &&  57 & 2.738e-02 & 9.743e-02 & 1.207e-01 \\
  8 & 5.783e-01 & 8.691e-01 & 7.821e-01 &&  58 & 2.684e-02 & 9.467e-02 & 1.170e-01 \\
  9 & 5.205e-01 & 8.369e-01 & 7.324e-01 &&  59 & 2.630e-02 & 9.207e-02 & 1.133e-01 \\
 10 & 4.646e-01 & 8.031e-01 & 6.842e-01 &&  60 & 2.578e-02 & 8.964e-02 & 1.097e-01 \\
 11 & 4.113e-01 & 7.683e-01 & 6.396e-01 &&  61 & 2.527e-02 & 8.741e-02 & 1.061e-01 \\
 12 & 3.610e-01 & 7.330e-01 & 6.007e-01 &&  62 & 2.477e-02 & 8.538e-02 & 1.026e-01 \\
 13 & 3.145e-01 & 6.977e-01 & 5.691e-01 &&  63 & 2.428e-02 & 8.359e-02 & 9.910e-02 \\
 14 & 2.725e-01 & 6.629e-01 & 5.444e-01 &&  64 & 2.380e-02 & 8.207e-02 & 9.568e-02 \\
 15 & 2.356e-01 & 6.293e-01 & 5.254e-01 &&  65 & 2.333e-02 & 8.083e-02 & 9.232e-02 \\
 16 & 2.044e-01 & 5.972e-01 & 5.113e-01 &&  66 & 2.287e-02 & 7.927e-02 & 8.902e-02 \\
 17 & 1.783e-01 & 5.667e-01 & 5.011e-01 &&  67 & 2.242e-02 & 7.774e-02 & 8.579e-02 \\
 18 & 1.567e-01 & 5.376e-01 & 4.938e-01 &&  68 & 2.197e-02 & 7.624e-02 & 8.264e-02 \\
 19 & 1.388e-01 & 5.099e-01 & 4.887e-01 &&  69 & 2.154e-02 & 7.476e-02 & 7.958e-02 \\
 20 & 1.239e-01 & 4.835e-01 & 4.848e-01 &&  70 & 2.111e-02 & 7.332e-02 & 7.660e-02 \\
 21 & 1.115e-01 & 4.583e-01 & 4.814e-01 &&  71 & 2.070e-02 & 7.191e-02 & 7.373e-02 \\
 22 & 1.008e-01 & 4.342e-01 & 4.776e-01 &&  72 & 2.029e-02 & 7.052e-02 & 7.096e-02 \\
 23 & 9.144e-02 & 4.113e-01 & 4.725e-01 &&  73 & 1.989e-02 & 6.916e-02 & 6.832e-02 \\
 24 & 8.314e-02 & 3.894e-01 & 4.653e-01 &&  74 & 1.949e-02 & 6.782e-02 & 6.581e-02 \\
 25 & 7.583e-02 & 3.685e-01 & 4.552e-01 &&  75 & 1.911e-02 & 6.651e-02 & 6.344e-02 \\
 26 & 6.941e-02 & 3.486e-01 & 4.414e-01 &&  76 & 1.873e-02 & 6.523e-02 & 6.199e-02 \\
 27 & 6.380e-02 & 3.296e-01 & 4.247e-01 &&  77 & 1.836e-02 & 6.397e-02 & 6.057e-02 \\
 28 & 5.891e-02 & 3.115e-01 & 4.058e-01 &&  78 & 1.799e-02 & 6.274e-02 & 5.918e-02 \\
 29 & 5.467e-02 & 2.943e-01 & 3.855e-01 &&  79 & 1.764e-02 & 6.153e-02 & 5.783e-02 \\
 30 & 5.102e-02 & 2.781e-01 & 3.645e-01 &&  80 & 1.729e-02 & 6.034e-02 & 5.650e-02 \\
\tableline
\end{tabular}
\end{lrbox}
\settowidth{\thiswid}{\usebox{\thisbox}}
\begin{center}
\begin{minipage}{\thiswid}
\caption{$U$, $V$, and $R$ Lightcurve Templates Used}
\label{tab:ltcvtemplates}
\usebox{\thisbox}
$a$: Day is relative to the epoch of the maximum of the $B$-band
lightcurve.  The $B$-band template may be found in \citet{gol01}.

$b$: Relative fluxes.

\end{minipage}
\end{center}
\end{table*}
}{\placetable{tab:ltcvtemplates}}

Some of the high-redshift supernovae from P99 lack a supernova-free
host-galaxy image.  These supernovae were fit with an additional
variable parameter: the zero-level of the I-band lightcurve.  The
supernovae treated in this manner include \mbox{SNe\,1997O},
1997Q, 1997R, and 1997am.

The late-time lightcurve behavior may bias the result of a lightcurve
fit \citep{ald00}; it is therefore important that the low- and
high-redshift supernovae are treated in as consistent a manner as
possible.  Few or none of the high-redshift supernovae have
high-precision measurements more than $\sim$40--50 rest-frame days after
maximum light, so as in \citet{per97} and P99 these late-time points
were eliminated from the low-redshift lightcurve data before the
template-fit procedure.  Additionally, to allow for systematic offset
uncertainties on the host-galaxy subtraction, an ``error floor'' of
0.007 times the maximum lightcurve flux was applied; any lightcurve
point with an uncertainty below the error floor had its uncertainty
replaced by that value \citep{gol01}.

The final results of the lightcurve fits, including the effect of color
corrections and $K$-corrections, are listed in
Table~\ref{tab:hstsnefits} for the eleven supernovae of this paper.
Table~\ref{tab:42snefits} shows the results of new lightcurve fits to
the high-redshift supernovae of P99 used in this paper (see
\S~\ref{sec:subsets}), and Table~\ref{tab:lowzsnefits} shows the results
of lightcurve fits for the low-redshift supernovae from H96 and
R99.\footnote{These three tables are available in electronic form from
\url{http://supernova.lbl.gov}.}  Appendix~\ref{sec:dirtylaundry}
tabulates all of the lightcurve data for the eleven HST supernovae in
this paper.  The lightcurves for these supernovae (and the F675W \wfpc\
image nearest maximum light) are shown in Figures~\ref{fig:lightcurves1}
and \ref{fig:lightcurves2}.  Note that there are correlated errors
between all of the ground-based points for each supernova in these
figures, as a single ground-based zeropoint was used to scale each of
them together with the HST photometry.

\ifthenelse{\boolean{ispreprint}}{
\begin{figure*}[p]
\begin{center}
\epsfig{file=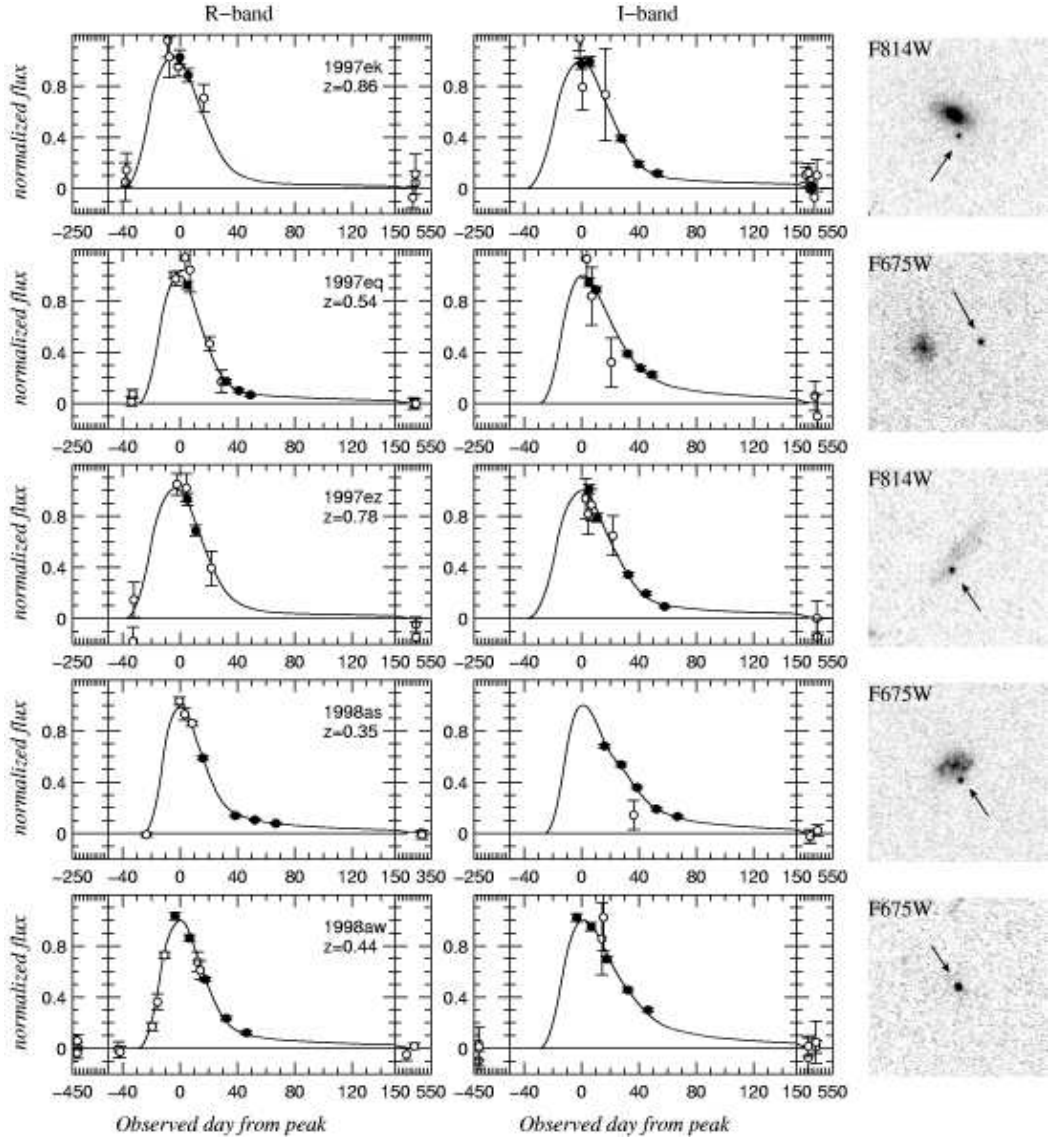 , width=5.5in}
\end{center}
\caption{Lightcurves and images from the PC CCD on \wfpc\ for the HST
  supernovae reported in this paper.  The left column shows the $R$-band
  (including F675W HST data), and the middle column shows $I$-band
  lightcurves (including F814W HST data).  Open circles represent
  ground-based data points, and filled circles represent \wfpc\ data
  points.  Note that there are correlated errors between all of the
  ground-based points for each supernova in these figures, as a single
  ground-based zeropoint was used to scale each of them together with
  the HST photometry.  The right column shows \mbox{$6''\times6''$}
  images, summed from all HST images of the supernova in the indicated
  filter.}
\label{fig:lightcurves1}
\end{figure*}
}{\placefigure{fig:lightcurves1}}

\ifthenelse{\boolean{ispreprint}}{
\begin{figure*}[p]
\begin{center}
\epsfig{file=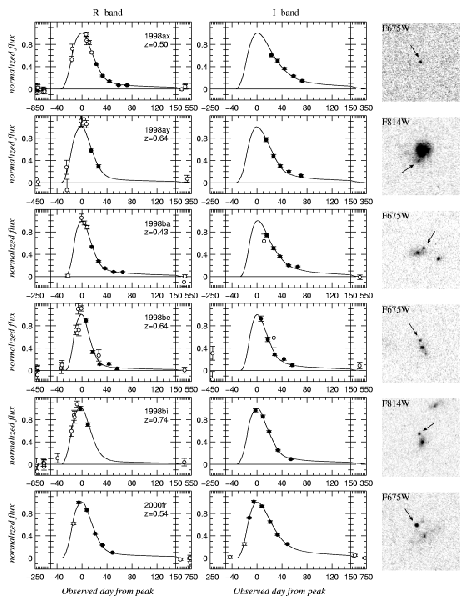 , width=5.5in}
\end{center}
\caption{Lightcurves and images from the PC CCD on \wfpc\ for the HST
  supernovae reported in this paper (continued).  The left column shows
  the $R$-band (including F675W HST data), and the middle column shows
  $I$-band lightcurves (including F814W HST data).  Open circles
  represent ground-based data points, and filled circles represent
  \wfpc\ data points.  Note that there are correlated errors between all
  of the ground-based points for each supernova in these figures, as a
  single ground-based zeropoint was used to scale each of them together
  with the HST photometry.  The right column shows \mbox{$6''\times6''$}
  images, summed from all HST images of the supernova in the indicated
  filter.}
\label{fig:lightcurves2}
\end{figure*}
}{\placefigure{fig:lightcurves2}}

\ifthenelse{\boolean{ispreprint}}{
\begin{sidewaystable*}[p]
\scriptsize
\settowidth{\nopmsp}{$-$}
\newcommand{\nopm}{\hspace*{\nopmsp}}
\caption{Supernova Lightcurve Fits: HST Supernovae from this paper}
\label{tab:hstsnefits}
\begin{tabular}{lcccccccccl}
\tableline
\tableline
SN        & z     & ${m_X}$ & ${m_B}$ & ${m_B^{eff}}$ & ${m_B^{eff}}$    & Stretch ($s$)  & $R$-$I$             & \ebv\      & $\ebv_\mathrm{host}$ & Excluded from \\
          &       & ($a$)   & ($b$)   &  ($c$)        & Ext. Corr. ($d$) &                & ($e$)               & Gal. ($f$) & ($g$)                & Subsets ($h$) \\
\tableline
1997ek   & 0.863 & 23.32 & $24.51\pm0.03$ & $24.59\pm0.19$ & $24.95\pm0.44$ & $1.056\pm0.058$ &  \nopm$0.838\pm0.054$ & $0.042$ &     $-0.091\pm0.075$ & \\
1997eq   & 0.538 & 22.63 & $23.21\pm0.02$ & $23.15\pm0.18$ & $23.02\pm0.17$ & $0.960\pm0.027$ &  \nopm$0.202\pm0.030$ & $0.044$ & \nopm$0.035\pm0.034$ & \\
1997ez   & 0.778 & 23.17 & $24.29\pm0.03$ & $24.41\pm0.18$ & $24.00\pm0.42$ & $1.078\pm0.030$ &  \nopm$0.701\pm0.048$ & $0.026$ & \nopm$0.095\pm0.068$ & \\
1998as   & 0.355 & 22.18 & $22.72\pm0.03$ & $22.66\pm0.17$ & $22.02\pm0.15$ & $0.956\pm0.012$ &  \nopm$0.226\pm0.027$ & $0.037$ & \nopm$0.158\pm0.030$ & 2,3 \\
1998aw   & 0.440 & 22.56 & $23.22\pm0.02$ & $23.26\pm0.17$ & ---            & $1.026\pm0.019$ &  \nopm$0.300\pm0.024$ & $0.026$ & \nopm$0.259\pm0.026$ & 1--3 \\
1998ax   & 0.497 & 22.63 & $23.25\pm0.05$ & $23.47\pm0.17$ & $22.96\pm0.20$ & $1.150\pm0.032$ &  \nopm$0.212\pm0.041$ & $0.035$ & \nopm$0.113\pm0.044$ & 2,3 \\
1998ay   & 0.638 & 23.26 & $23.86\pm0.08$ & $23.92\pm0.19$ & $23.85\pm0.33$ & $1.040\pm0.041$ &  \nopm$0.339\pm0.067$ & $0.035$ & \nopm$0.015\pm0.084$ & 3 \\
1998ba   & 0.430 & 22.34 & $22.97\pm0.05$ & $22.90\pm0.18$ & $22.75\pm0.18$ & $0.954\pm0.020$ &  \nopm$0.094\pm0.036$ & $0.024$ & \nopm$0.040\pm0.038$ & \\
1998be   & 0.644 & 23.33 & $23.91\pm0.04$ & $23.64\pm0.18$ & $23.26\pm0.27$ & $0.816\pm0.028$ &  \nopm$0.436\pm0.051$ & $0.029$ & \nopm$0.106\pm0.065$ & 3 \\
1998bi   & 0.740 & 22.86 & $23.92\pm0.02$ & $23.85\pm0.17$ & $23.75\pm0.37$ & $0.950\pm0.027$ &  \nopm$0.552\pm0.037$ & $0.026$ & \nopm$0.026\pm0.050$ & \\
2000fr   & 0.543 & 22.44 & $23.07\pm0.02$ & $23.16\pm0.17$ & $23.27\pm0.14$ & $1.064\pm0.011$ &  \nopm$0.135\pm0.022$ & $0.030$ &     $-0.031\pm0.025$ & \\
\tableline
\end{tabular}

$a$: Magnitude in the observed filter at the peak of the rest-frame $B$-band
lightcurve.  X=$R$ for $z<0.7$, X=$I$ for $z>0.7$.

$b$: This value has been $K$-corrected and corrected for Galactic
extinction: $m_B\equiv m_X-K_{BX}-A_X$, where $K_{BX}$ is the
cross-filter $K$-correction and $A_X$ is the Galactic extinction
correction.  These were the values used in the cosmological fits.  The
quoted error bar is the uncertainty on the peak magnitude from the
lightcurve fit.

$c$: This value includes the stretch correction: $m_B^{eff}\equiv
m_B+\alpha(s-1)$.  $\alpha$ is the best-fit value of the
stretch-luminosity slope from the fit to the primary low-extinction
subset (Fit~3 in \S~\ref{sec:cosmoresults}).  The quoted error bar
includes all uncertainties for non-extinction-corrected fits described
in \S~\ref{sec:cosmofitmethod}.  Note that these values are only
provided for convenience; they were not used directly in any
cosmological fits, since $\alpha$ is also a fit parameter.

$d$: Similar to column $c$, only with the host-galaxy extinction
correction applied.  The stretch/luminosity slope used for this value is
that from the fit to the primary subset (Fit~6 in
\S~\ref{sec:cosmoresults}).  The quoted error bar includes all
uncertainties for extinction-corrected fits described in
\S~\ref{sec:cosmofitmethod}.  A line indicates a supernova which did not
appear in the primary subset (see \S~\ref{sec:subsets}.)

$e$: This is the observed $R$-$I$ color at the epoch of the rest-frame $B$-band
lightcurve peak.

$f$: \citet{sch98}; this extinction is already included in the quoted
values of $m_B$.

$g$: Measurement uncertainty only; no intrinsic color dispersion included.

$h$: These supernovae are excluded from the indicated subsets; see
\S~\ref{sec:subsets}.
\end{sidewaystable*}
}{\placetable{tab:hstsnefits}}

\ifthenelse{\boolean{ispreprint}}{
\begin{sidewaystable*}[p]
\scriptsize\renewcommand{\arraystretch}{1.0}
\settowidth{\nopmsp}{$-$}
\newcommand{\nopm}{\hspace*{\nopmsp}}
\caption{Supernova Lightcurve Fits: New Fits to Perlmutter (1999) SNe}
\label{tab:42snefits}
\begin{tabular}{lcccccccccl}
\tableline
\tableline
SN        & z     & ${m_X}$ & ${m_B}$ & ${m_B^{eff}}$ & ${m_B^{eff}}$    & Stretch ($s$)  & $R$-$I$             & \ebv\      & $\ebv_\mathrm{host}$ & Excluded from \\
          &       & ($a$)   & ($b$)   &  ($c$)        & Ext. Corr. ($d$) &                & ($e$)               & Gal. ($f$) & ($g$)                & Subsets ($h$) \\
\tableline
1995ar   & 0.465 & 22.80 & $23.48\pm0.08$ & $23.35\pm0.22$ & $21.54\pm0.97$ & $0.909\pm0.104$ &  \nopm$0.509\pm0.222$ & $0.022$ & \nopm$0.448\pm0.242$ & \\
1995as   & 0.498 & 23.03 & $23.69\pm0.07$ & $23.74\pm0.23$ & $23.52\pm0.87$ & $1.035\pm0.090$ &  \nopm$0.155\pm0.197$ & $0.021$ & \nopm$0.051\pm0.212$ & 3 \\
1995aw   & 0.400 & 21.78 & $22.28\pm0.03$ & $22.57\pm0.18$ & $23.17\pm0.45$ & $1.194\pm0.037$ &      $-0.127\pm0.103$ & $0.040$ &     $-0.160\pm0.107$ & \\
1995ax   & 0.615 & 22.56 & $23.21\pm0.06$ & $23.38\pm0.22$ & $23.98\pm1.02$ & $1.112\pm0.073$ &  \nopm$0.152\pm0.204$ & $0.033$ &     $-0.153\pm0.249$ & \\
1995ay   & 0.480 & 22.64 & $23.07\pm0.04$ & $22.90\pm0.19$ & $22.74\pm0.70$ & $0.880\pm0.064$ &  \nopm$0.209\pm0.158$ & $0.114$ & \nopm$0.047\pm0.170$ & \\
1995az   & 0.450 & 22.46 & $22.70\pm0.07$ & $22.66\pm0.20$ & $23.04\pm0.58$ & $0.973\pm0.064$ &  \nopm$0.087\pm0.135$ & $0.181$ &     $-0.089\pm0.144$ & \\
1995ba   & 0.388 & 22.07 & $22.64\pm0.06$ & $22.60\pm0.18$ & $22.74\pm0.45$ & $0.971\pm0.047$ &  \nopm$0.006\pm0.105$ & $0.018$ &     $-0.033\pm0.110$ & \\
1996cf   & 0.570 & 22.71 & $23.31\pm0.03$ & $23.30\pm0.18$ & $23.53\pm0.45$ & $0.996\pm0.045$ &  \nopm$0.162\pm0.091$ & $0.040$ &     $-0.054\pm0.107$ & 3 \\
1996cg   & 0.490 & 22.46 & $23.09\pm0.03$ & $23.11\pm0.18$ & $22.26\pm0.45$ & $1.011\pm0.040$ &  \nopm$0.300\pm0.099$ & $0.035$ & \nopm$0.205\pm0.107$ & 3 \\
1996ci   & 0.495 & 22.19 & $22.83\pm0.02$ & $22.78\pm0.18$ & $22.92\pm0.32$ & $0.964\pm0.040$ &  \nopm$0.083\pm0.070$ & $0.028$ &     $-0.033\pm0.075$ & \\
1996cl   & 0.828 & 23.37 & $24.53\pm0.17$ & $24.49\pm0.46$ & $25.92\pm0.97$ & $0.974\pm0.239$ &  \nopm$0.549\pm0.184$ & $0.035$ &     $-0.344\pm0.251$ & \\
1996cm   & 0.450 & 22.67 & $23.26\pm0.07$ & $23.11\pm0.18$ & $22.63\pm0.77$ & $0.899\pm0.061$ &  \nopm$0.214\pm0.174$ & $0.049$ & \nopm$0.124\pm0.185$ & 3 \\
1996cn   & 0.430 & 22.58 & $23.25\pm0.03$ & $23.09\pm0.19$ & ---            & $0.890\pm0.066$ &  \nopm$0.379\pm0.090$ & $0.025$ & \nopm$0.332\pm0.097$ & 1--3 \\
1997F    & 0.580 & 22.93 & $23.51\pm0.06$ & $23.57\pm0.20$ & $23.30\pm0.95$ & $1.041\pm0.066$ &  \nopm$0.275\pm0.197$ & $0.040$ & \nopm$0.063\pm0.232$ & \\
1997H    & 0.526 & 22.70 & $23.26\pm0.04$ & $23.09\pm0.19$ & $22.51\pm0.80$ & $0.882\pm0.043$ &  \nopm$0.303\pm0.174$ & $0.051$ & \nopm$0.150\pm0.194$ & \\
1997I    & 0.172 & 20.18 & $20.34\pm0.01$ & $20.29\pm0.17$ & $20.19\pm0.28$ & $0.967\pm0.009$ &  \nopm$0.065\pm0.047$ & $0.051$ & \nopm$0.026\pm0.064$ & \\
1997N    & 0.180 & 20.39 & $20.38\pm0.02$ & $20.48\pm0.17$ & $21.28\pm0.52$ & $1.067\pm0.015$ &      $-0.141\pm0.093$ & $0.031$ &     $-0.200\pm0.123$ & \\
1997O    & 0.374 & 22.99 & $23.53\pm0.06$ & $23.60\pm0.18$ & ---            & $1.048\pm0.054$ &  \nopm$0.087\pm0.152$ & $0.029$ & \nopm$0.049\pm0.162$ & 1--3 \\
1997P    & 0.472 & 22.53 & $23.16\pm0.04$ & $22.99\pm0.18$ & $23.24\pm0.91$ & $0.888\pm0.039$ &  \nopm$0.058\pm0.207$ & $0.033$ &     $-0.052\pm0.219$ & \\
1997Q    & 0.430 & 22.01 & $22.61\pm0.02$ & $22.52\pm0.17$ & $22.55\pm0.62$ & $0.935\pm0.024$ &  \nopm$0.061\pm0.140$ & $0.030$ &     $-0.002\pm0.148$ & \\
1997R    & 0.657 & 23.29 & $23.89\pm0.05$ & $23.80\pm0.19$ & $23.68\pm0.90$ & $0.940\pm0.059$ &  \nopm$0.393\pm0.175$ & $0.030$ & \nopm$0.032\pm0.222$ & \\
1997ac   & 0.320 & 21.42 & $21.87\pm0.02$ & $21.96\pm0.17$ & $21.95\pm0.33$ & $1.061\pm0.015$ &  \nopm$0.063\pm0.065$ & $0.027$ & \nopm$0.001\pm0.072$ & \\
1997af   & 0.579 & 22.94 & $23.60\pm0.07$ & $23.38\pm0.18$ & $24.31\pm1.09$ & $0.850\pm0.045$ &  \nopm$0.045\pm0.226$ & $0.028$ &     $-0.215\pm0.265$ & \\
1997ai   & 0.450 & 22.34 & $22.94\pm0.05$ & $22.63\pm0.22$ & $22.58\pm0.59$ & $0.788\pm0.084$ &  \nopm$0.143\pm0.133$ & $0.045$ & \nopm$0.026\pm0.142$ & \\
1997aj   & 0.581 & 22.58 & $23.24\pm0.07$ & $23.16\pm0.18$ & $24.05\pm0.79$ & $0.947\pm0.045$ &  \nopm$0.045\pm0.164$ & $0.033$ &     $-0.213\pm0.193$ & \\
1997am   & 0.416 & 22.01 & $22.58\pm0.08$ & $22.63\pm0.18$ & $22.65\pm0.46$ & $1.032\pm0.060$ &  \nopm$0.037\pm0.113$ & $0.036$ &     $-0.008\pm0.119$ & \\
1997ap   & 0.830 & 23.16 & $24.35\pm0.07$ & $24.38\pm0.18$ & $23.74\pm0.50$ & $1.023\pm0.045$ &  \nopm$0.903\pm0.082$ & $0.026$ & \nopm$0.155\pm0.118$ & \\
\tableline
\end{tabular}

$a$: X=$R$ for $z<0.7$, X=$I$ for $z>0.7$

$b$: This value has been $K$-corrected and corrected for Galactic
extinction: $m_B\equiv m_X-K_{BX}-A_X$, where $K_{BX}$ is the
cross-filter $K$-correction and $A_X$ is the Galactic extinction
correction.  These were the values used in the cosmological fits.  The
quoted error bar is the uncertainty on the peak magnitude from the
lightcurve fit.

$c$: This value includes the stretch correction: $m_B^{eff}\equiv
m_B+\alpha(s-1)$.  $\alpha$ is the best-fit value of the
stretch-luminosity slope from the fit to the primary low-extinction
subset (Fit~3 in \S~\ref{sec:cosmoresults}).  The quoted error bar
includes all uncertainties for non-extinction-corrected fits described
in \S~\ref{sec:cosmofitmethod}.  Note that these values are only
provided for convenience; they were not used directly in any
cosmological fits, since $\alpha$ is also a fit parameter.

$d$: Similar to column $c$, only with the host-galaxy extinction
correction applied.  The stretch/luminosity slope used for this value is
that from the fit to the primary subset (Fit~6 in
\S~\ref{sec:cosmoresults}).  The quoted error bar includes all
uncertainties for extinction-corrected fits described in
\S~\ref{sec:cosmofitmethod}.  A line indicates a supernova which did not
appear in the primary subset (see \S~\ref{sec:subsets}.)

$e$: This is the observed $R$-$I$ color at the epoch of the rest-frame $B$-band
lightcurve peak.

$f$: \citet{sch98}; this extinction is already included in the quoted
values of $m_B$.

$g$: Measurement uncertainty only; no intrinsic color dispersion included.

$h$: These supernovae are excluded from the indicated subsets; see
\S~\ref{sec:subsets}.

\end{sidewaystable*}
}{\placetable{tab:hstsnefits}}

\ifthenelse{\boolean{ispreprint}}{
\begin{sidewaystable*}[p]
\scriptsize
\settowidth{\nopmsp}{$-$}
\newcommand{\nopm}{\hspace*{\nopmsp}}
\caption{Supernova Lightcurve Fits: Low-z SNe from Hamuy (1996) and Riess (1999)}
\label{tab:lowzsnefits}
\begin{tabular}{lcccccccccl}
\tableline
\tableline
SN        & z     & ${m_B^{meas}}$ & ${m_B}$ & ${m_B^{eff}}$ & ${m_B^{eff}}$    & Stretch ($s$)   & $R$-$I$             & \ebv\      & $\ebv_\mathrm{host}$ & Excluded from \\
($a$)     &       & ($b$)          & ($c$)   & ($d$)         & Ext. corr. ($e$) &                 & ($f$)               & Gal. ($g$) & ($h$)                & Subsets ($i$) \\
\tableline
1990O    & 0.030 & 16.58 & $16.18\pm0.03$ & $16.33\pm0.20$ & $16.30\pm0.17$ & $1.106\pm0.026$ &  \nopm$0.043\pm0.025$ & $0.098$ & \nopm$0.001\pm0.026$ & \\
1990af   & 0.050 & 17.92 & $17.76\pm0.01$ & $17.39\pm0.18$ & $17.42\pm0.13$ & $0.749\pm0.010$ &  \nopm$0.077\pm0.011$ & $0.035$ & \nopm$0.011\pm0.011$ & \\
1992P    & 0.026 & 16.12 & $16.05\pm0.02$ & $16.14\pm0.19$ & $16.16\pm0.16$ & $1.061\pm0.027$ &      $-0.045\pm0.018$ & $0.020$ &     $-0.008\pm0.019$ & \\
1992ae   & 0.075 & 18.59 & $18.42\pm0.04$ & $18.35\pm0.18$ & $18.35\pm0.15$ & $0.957\pm0.018$ &  \nopm$0.098\pm0.028$ & $0.036$ & \nopm$0.003\pm0.031$ & \\
1992ag   & 0.026 & 16.67 & $16.26\pm0.02$ & $16.34\pm0.20$ & $15.55\pm0.16$ & $1.053\pm0.015$ &  \nopm$0.220\pm0.020$ & $0.097$ & \nopm$0.189\pm0.021$ & 2,3 \\
1992al   & 0.014 & 14.61 & $14.48\pm0.01$ & $14.42\pm0.23$ & $14.53\pm0.20$ & $0.959\pm0.011$ &      $-0.054\pm0.012$ & $0.034$ &     $-0.025\pm0.013$ & \\
1992aq   & 0.101 & 19.38 & $19.30\pm0.02$ & $19.12\pm0.17$ & $19.24\pm0.15$ & $0.878\pm0.017$ &  \nopm$0.142\pm0.023$ & $0.012$ &     $-0.019\pm0.026$ & \\
1992bc   & 0.020 & 15.18 & $15.10\pm0.01$ & $15.18\pm0.20$ & $15.36\pm0.16$ & $1.053\pm0.006$ &      $-0.087\pm0.009$ & $0.022$ &     $-0.046\pm0.009$ & \\
1992bg   & 0.036 & 17.41 & $16.66\pm0.04$ & $16.66\pm0.20$ & $16.68\pm0.16$ & $1.003\pm0.014$ &  \nopm$0.128\pm0.025$ & $0.181$ &     $-0.006\pm0.026$ & \\
1992bh   & 0.045 & 17.71 & $17.60\pm0.02$ & $17.64\pm0.18$ & $17.22\pm0.14$ & $1.027\pm0.016$ &  \nopm$0.101\pm0.018$ & $0.022$ & \nopm$0.100\pm0.019$ & \\
1992bl   & 0.043 & 17.37 & $17.31\pm0.03$ & $17.03\pm0.18$ & $17.10\pm0.14$ & $0.812\pm0.012$ &  \nopm$0.017\pm0.023$ & $0.012$ &     $-0.002\pm0.024$ & \\
1992bo   & 0.018 & 15.89 & $15.78\pm0.01$ & $15.42\pm0.21$ & $15.31\pm0.17$ & $0.756\pm0.005$ &  \nopm$0.048\pm0.012$ & $0.027$ & \nopm$0.043\pm0.012$ & \\
1992bp   & 0.079 & 18.59 & $18.29\pm0.01$ & $18.16\pm0.18$ & $18.41\pm0.13$ & $0.906\pm0.014$ &  \nopm$0.088\pm0.015$ & $0.068$ &     $-0.056\pm0.017$ & \\
1992br   & 0.088 & 19.52 & $19.37\pm0.08$ & $18.93\pm0.20$ & ---            & $0.700\pm0.021$ &  \nopm$0.186\pm0.047$ & $0.027$ & \nopm$0.030\pm0.052$ & 1--3 \\
1992bs   & 0.063 & 18.26 & $18.20\pm0.04$ & $18.26\pm0.18$ & $18.37\pm0.14$ & $1.038\pm0.016$ &  \nopm$0.011\pm0.022$ & $0.013$ &     $-0.031\pm0.024$ & \\
1993B    & 0.071 & 18.74 & $18.37\pm0.04$ & $18.40\pm0.18$ & $18.10\pm0.15$ & $1.021\pm0.019$ &  \nopm$0.181\pm0.027$ & $0.080$ & \nopm$0.071\pm0.029$ & \\
1993O    & 0.052 & 17.87 & $17.64\pm0.01$ & $17.53\pm0.18$ & $17.61\pm0.13$ & $0.926\pm0.007$ &  \nopm$0.042\pm0.012$ & $0.053$ &     $-0.014\pm0.012$ & \\
1993ag   & 0.050 & 18.32 & $17.83\pm0.02$ & $17.73\pm0.18$ & $17.26\pm0.15$ & $0.936\pm0.015$ &  \nopm$0.217\pm0.020$ & $0.111$ & \nopm$0.120\pm0.021$ & 2,3 \\
1994M    & 0.024 & 16.34 & $16.24\pm0.03$ & $16.07\pm0.20$ & $15.84\pm0.16$ & $0.882\pm0.015$ &  \nopm$0.043\pm0.022$ & $0.023$ & \nopm$0.063\pm0.022$ & \\
1994S    & 0.016 & 14.85 & $14.78\pm0.02$ & $14.83\pm0.22$ & $14.86\pm0.19$ & $1.033\pm0.026$ &      $-0.061\pm0.019$ & $0.018$ &     $-0.010\pm0.019$ & \\
1995ac   & 0.049 & 17.23 & $17.05\pm0.01$ & $17.17\pm0.18$ & $17.17\pm0.13$ & $1.083\pm0.012$ &  \nopm$0.026\pm0.011$ & $0.042$ &     $-0.005\pm0.011$ & \\
1995bd   & 0.016 & 17.34 & $15.32\pm0.01$ & $15.37\pm0.30$ & ---            & $1.039\pm0.008$ &  \nopm$0.735\pm0.008$ & $0.490$ & \nopm$0.348\pm0.009$ & 1--3 \\
1996C    & 0.030 & 16.62 & $16.57\pm0.04$ & $16.74\pm0.19$ & $16.50\pm0.16$ & $1.120\pm0.020$ &  \nopm$0.012\pm0.026$ & $0.014$ & \nopm$0.051\pm0.027$ & \\
1996ab   & 0.125 & 19.72 & $19.57\pm0.04$ & $19.47\pm0.19$ & $19.82\pm0.16$ & $0.934\pm0.032$ &  \nopm$0.174\pm0.025$ & $0.032$ &     $-0.082\pm0.029$ & \\
1996bl   & 0.035 & 17.08 & $16.66\pm0.01$ & $16.71\pm0.19$ & $16.55\pm0.14$ & $1.031\pm0.015$ &  \nopm$0.093\pm0.012$ & $0.099$ & \nopm$0.036\pm0.012$ & \\
1996bo   & 0.016 & 16.18 & $15.85\pm0.01$ & $15.65\pm0.22$ & ---            & $0.862\pm0.006$ &  \nopm$0.406\pm0.008$ & $0.077$ & \nopm$0.383\pm0.008$ & 1--3 \\
\tableline
\end{tabular}

$a$: Supernovae through 1993ag are from H96, later ones from R99.

$b$: This is the measured peak magnitude of the $B$-band lightcurve.

$c$: This includes the Galactic extinction correction and a
$K$-correction: $M-B\equiv m_B^{meas}-K_B-A_B$, where $K_B$ is the
$K$-correction and $A_B$ is the Galactic extinction correction.  The
quoted error bar is the uncertainty on the peak magnitude from the
lightcurve fit.

$d$: This value includes the stretch correction: $m_B^{eff}\equiv
m_B^{meas}-K_B-A_B+\alpha(s-1)$.  $\alpha$ is the best-fit value of the
stretch/luminosity slope from the fit to the primary low-extinction
subset (Fit~3 in \S~\ref{sec:cosmoresults}).  The quoted error bar
includes all uncertainties for non-extinction corrected fits described
in \S~\ref{sec:cosmofitmethod}.  Note that these values are only
provided for convenience; they were not used directly in any
cosmological fits, since the $\alpha$ is also a fit parameter.

$e$: Similar to column $d$, only with the host-galaxy extinction
correction applied.  The stretch/luminosity slope used for this value is
that from the fit to the primary subset (Fit~6 in
\S~\ref{sec:cosmoresults}).  The quoted error bar includes all
uncertainties for extinction-corrected fits described in
\S~\ref{sec:cosmofitmethod}.  A line indicates a supernova which did not
appear in the primary subset (see \S~\ref{sec:subsets}.)

$f$: This value has been $K$-corrected and corrected for Galactic
extinction.

$g$: This is the measured $B$-$V$ color at the epoch of rest-frame $B$-band lightcurve maximum.

$h$: \citet{sch98}; this extinction is already included in the quoted
values of $m_B$ in column $c$.

$i$: These supernovae are excluded from the indicated subsets;
\S~\ref{sec:subsets}.

\end{sidewaystable*}
}{\placetable{tab:lowzsnefits}}

\subsection{Color- and $K$-Corrections}
\label{sec:colorcor}

In order to combine data from different telescopes, color corrections
were applied to remove the differences in the spectral responses of the
filters relative to the Bessell system \citep{bes90}.  For the
ground-based telescopes, the filters are close enough to the standard
Bessell filters that a single linear color term (measured at each
observatory with standard stars) suffices to put the data onto the
Bessell system, with most corrections being smaller than 0.01
magnitudes.  The \wfpc\ filters are different enough from the
ground-based filters, however, that a linear term is not
sufficient. Moreover, the differences between a \snia\ and standard star
spectral energy distribution are significant.  In this case, color
corrections were calculated by integrating template \snia\ spectra
(described below) through the system response.

In order to perform lightcurve template fitting, a cross-filter
$K$-correction must be applied to transform the data in the observed
filter into a rest-frame magnitude in the filter used for the lightcurve
template \citep{kim96}.  The color correction to the nearest standard
Bessell filter followed by a $K$-correction to a rest-frame filter is
equivalent to a direct $K$-correction from the observed filter to the
standard rest-frame filter.  In practice, we perform the two steps
separately so that all photometry may be combined to provide a
lightcurve effectively observed through a standard (e.g. $R$-band)
filter, which may then be fit with a single series of
$K$-corrections.  The data tabulated in Appendix~\ref{sec:dirtylaundry}
have all been color-corrected to the standard Bessell filters.

Color and $K$-corrections were performed following the procedure of
\citet{nug02}.  In order to perform these corrections, a template \snia\
spectrum for each epoch of the lightcurve, as described in that paper,
is necessary.  The spectral template used in this present work began
with the template of that paper.  To it was applied a smooth
multiplicative function at each day such that integration of the spectrum
through the standard filters would produce the proper intrinsic colors
for a Type~Ia supernova (including a mild dependence of those intrinsic
colors on stretch).

The proper intrinsic colors for the supernova spectral template were
determined in the $BVRI$ spectral range by smooth fits to the
low-redshift supernova data of H96 and R99.  For each color ($B$-$V$,
$V$-$R$, and $R$-$I$), every data point from those papers was
$K$-corrected and corrected for Galactic extinction.  These data were
plotted together, and then a smooth curve was fit to the plot of color
versus date relative to maximum.  This curve is given by two parameters,
each of which is a function of time and is described by a spline under
tension: an ``intercept'' $b(t)$ and a ``slope'' $m(t)$.  At any given
date the intrinsic color is
\begin{equation}
color(t') = b( t' ) + m( t' ) \times (1/s^3 - 1)
\end{equation}
where $t'=t/(s(1+z))$, $z$ is the redshift of the supernova, and $s$ is
the timescale stretch of the supernova from a simultaneous fit to the
$B$ and $V$ lightcurves (matching the procedure used for most of the
high-redshift supernovae).  This arbitrary functional form was chosen to
match the stretch vs. color distribution.

As the goal was to determine intrinsic colors without making any
assumptions about reddening, no host-galaxy extinction corrections were
applied to the literature data at this stage of the analysis.  Instead,
host-galaxy extinction was handled by performing a robust blue-side
ridge-line fit to the supernova color curves, so as to extract the
unreddened intrinsic color.  Individual color points that were outliers
were prevented from having too much weight in the fit with a small added
dispersion on each point.  The blue ridge-line was selected by allowing
any point more than $1\sigma$ to the red side of the fit model only to
contribute to the $\chi^2$ as if it were $1\sigma$ away.  Additionally,
those supernovae which were most reddened were omitted.  The resulting
fit procedure provided $B$-$V$, $V$-$R$, and $R$-$I$ as a function of
epoch and stretch; those colors were used to correct the template
spectrum as described above.

\ifthenelse{\boolean{ispreprint}}{
\begin{table*}[htb]
\settowidth{\pmsp}{$-$}
\newcommand{\nopm}{\hspace*{\pmsp}}
\begin{lrbox}{\thisbox}
\begin{tabular}{lccl}
\tableline
\tableline
SN & Raw $U$-$B^a$ & Corrected $U$-$B^b$ & Reference \\
\tableline
1980N  &     $-0.21$ & $-0.29$ & \protect\citet{ham91} \\
1989B  &   \nopm0.08 & $-0.33$ & \protect\citet{wel94} \\
1990N  &     $-0.35$ & $-0.45$ & \protect\citet{lir98} \\
1994D  &     $-0.50$ & $-0.52$ & \protect\citet{wu95} \\
1998bu &     $-0.23$ & $-0.51$ & \protect\citet{sun99} \\
\tableline
\end{tabular}
\end{lrbox}
\settowidth{\thiswid}{\usebox{\thisbox}}
\begin{center}
\begin{minipage}{\thiswid}
\caption{$U$-$B$ SN~Ia Colors at Epoch of B-band Maximum}
\label{tab:ub}
\usebox{\thisbox}

a: This is the measured $U$-$B$ value from the cited paper. \\

b: This $U$-$B$ value is $K$-corrected, and corrected for host-galaxy
and Galactic extinction.
\end{minipage}
\end{center}
\end{table*}
}{\placetable{tab:ub}}

Some of our data extend into the rest-frame $U$-band range of the
spectrum.  This is obvious for supernovae at $z>0.7$ where a $U$-band
template is fit to the $R$-band data.  However, even for supernovae at
$z\gtrsim0.55$, the de-redshifted $R$-band filter begins to overlap the
$U$-band range of the rest-frame spectrum.  Thus, it is also important
to know the intrinsic $U$-$B$ color so as to generate a proper spectral
template.  We used data from the literature, as given in
Table~\ref{tab:ub}.  Here, there is an insufficient number of supernova
lightcurves to reasonably use the sort of ridge-line analysis used above
to eliminate the effects of host-galaxy extinction in determining the
intrinsic $BVRI$ colors.  Instead, for $U$-$B$, we perform extinction
corrections using the \ebv\ values from \citet{phi99}.  Based on
Table~\ref{tab:ub}, we adopt a $U$-$B$ color of $-0.4$ at the epoch of
rest-$B$ maximum.  This value is also consistent with the data shown in
\citet{jha02} for supernovae with timescale stretch of \mbox{$s\sim1$},
although the data are not determinative.  In contrast to the other
colors, $U$-$B$ was not considered to be a function of stretch.  Even
though \citet{jha02} does show $U$-$B$ depending on lightcurve stretch,
the supernovae in this work that would be most affected (those at
$z>0.7$ where \ebv\ is estimated from the rest-frame $U$-$B$ color)
cover a small range in stretch; current low-redshift $U$-$B$ data do not
show a significant slope within that range.  See
\S~\ref{sec:systematiccolor} for the effect of systematic error in the
assumed intrinsic $U$-$B$ colors.

Any intrinsic uncertainty in $B$-$V$ is already subsumed within the
assumed intrinsic dispersion of extinction-corrected peak magnitudes
(see \S~\ref{sec:cosmofitmethod}); however, we might expect a larger
dispersion in intrinsic $U$-$B$ due to e.g., metallicity effects
\citep{hof98,len00}.  The low-redshift $U$-band photometry may also have
unmodeled scatter e.g., related to the lack of extensive UV supernova
spectrophotometry for $K$-corrections.  The effect on
extinction-corrected magnitudes will be further increased by the greater
effect of dust extinction on the bluer $U$-band light.  The scatter of
our extinction-corrected magnitudes about the best-fit cosmology
suggests an intrinsic uncertainty in $U$-$B$ of 0.04 magnitudes.  This
is also consistent with the $U$-$B$ data of \citet{jha02} over the range
of timescale stretch of our $z>0.7$ SNe~Ia, after two extreme color
outliers from \citet{jha02} are removed; there is no evidence of such
extreme color objects in our dataset. Note that this intrinsic $U$-$B$
dispersion is in addition to the intrinsic magnitude dispersion assumed
after extinction correction.

The template spectrum which has been constructed may be used to perform
color- and $K$-corrections on both the low- and high-redshift supernovae
to be used for cosmology.  However, it must be further modified to
account for the reddening effects of dust extinction in the supernova
host galaxy, and extinction of the redshifted spectrum due to Galactic
dust.  To calculate the reddening effects of both Galactic and
host-galaxy extinction, we used the interstellar extinction law of
\citet{odo94} with the standard value of the parameter \mbox{$R_V=3.1$}.
Color excess (\ebv) values due to Galactic extinction were obtained from
\citet{sch98}.

The \ebv\ values quoted in Tables~\ref{tab:hstsnefits},
\ref{tab:42snefits}, and \ref{tab:lowzsnefits} are the values necessary
to reproduce the observed $R$-$I$ color at the epoch of the maximum of
the rest-frame $B$ lightcurve.  This reproduction was performed by
modifying the spectral template exactly as described above, given the
intrinsic color of the supernova from the fit stretch, the Galactic
extinction, and the host-galaxy \ebv\ parameter.  The modified spectrum
was integrated through the Bessell $R$- and $I$-band filters, and \ebv\
was varied until the $R$-$I$ value matched the peak color from the
lightcurve fit.

For each supernova, this finally modified spectral template was
integrated through the Bessell and \wfpc\ filter transmission functions
to provide color and $K$-corrections.  The exact spectral template
needed for a given data point on a given supernova is dependent on
parameters of the fit: the stretch, the time of each point relative to
the epoch of rest-$B$ maximum, and the host-galaxy \ebv\ (measured as
described above).  Thus, color and $K$-corrections were performed
iteratively with lightcurve fitting in order to generate the final
corrections used in the fits described in \S~\ref{sec:lightcurvefits}.
An initial date of maximum, stretch, and host-galaxy extinction was
assumed in order to generate $K$-corrections for the first iteration of
the fit.  The parameters resulting from that fit were used to generate
new color and $K$-corrections, and the whole procedure was repeated
until the results of the fit converged.  Generally, the fit converged
within 2--3 iterations.

\subsection{Cosmological Fit Methodology}
\label{sec:cosmofitmethod}

Cosmological fits to the luminosity distance modulus equation from the
Friedmann-Robertson-Walker metric followed the procedure of P99.  The
set of supernova redshifts ($z$) and $K$-corrected peak $B$-magnitudes
($m_B$) were fit to the equation
\begin{equation}
m_B = \scriptm + 5\log\mathcal{D_L}(z;\om,\ol) - \alpha(s-1)
\end{equation}
where $s$ is the stretch value for the supernova, $\mathcal{D_L}\equiv
H_0d_L$ is the ``Hubble-constant-free'' luminosity distance
\citep{per97}, and $\scriptm\equiv M_B - 5\log H_0+25$ is the
``Hubble-constant-free'' $B$-band peak absolute magnitude of a
\mbox{$s=1$} \snia\ with true absolute peak magnitude $M_B$.  With this
procedure, neither $H_0$ nor $M_B$ need be known independently.  The
peak magnitude of a \snia\ is mildly dependent on the lightcurve decay
time scale, such that supernovae with a slow decay (high stretch) tend to
be over-luminous, while supernovae with a fast decay (low stretch) tend
to be under-luminous \citep{phi93}; $\alpha$ is a slope that
parameterizes this relationship.

There are four parameters in the fit: the mass density \om\ and
cosmological constant \ol, as well as the two nuisance parameters,
\scriptm\ and $\alpha$.  The four-dimensional (\om, \ol, $\mathcal{M}$,
$\alpha$) space is divided into a grid, and at each grid point a
$\chi^2$ value is calculated by fitting the luminosity distance
equation to the peak $B$-band magnitudes and redshifts of the
supernovae.  The range of parameter space explored included
\mbox{$\om=[0,3)$}, \mbox{$\ol=[-1,3)$} (for fits where host-galaxy
extinction corrections are not directly applied) or \mbox{$\om=[0,4]$},
\mbox{$\ol=[-1,4)$} (for fits with host-galaxy extinction corrections).
The two nuisance parameters are fit in the ranges $\alpha=[-1,4)$ and
$\mathcal{M}=[-3.9,3.2)$.  No further constraints are placed on the
parameters.  (These ranges for the four fit parameters contain
$>99.99$\% of the probability.)  At each point on the 4-dimensional
grid, a $\chi^2$ is calculated, and a probability is determined from
$P\propto e^{-\chi^2/2}$.  The probability of the whole 4-dimensional
grid is normalized, and then integrated over the two dimensions
corresponding to the ``nuisance'' parameters.

For each fit, all peak $m_B$ values were corrected for Galactic
extinction using \ebv\ values from \citet{sch98}, using the extinction
law of \citet{odo94} integrated through the \emph{observed}
filter.\footnote{This supersedes P99, where an incorrect dependence on
$z$ of the effective $R_R$ for Galactic extinction was applied.  The
corrected procedure decreases the flat-universe value of \om\ by 0.03.}
For our primary fits, the total effective statistical uncertainty on
each value of $m_B$ included the following contributions:
\begin{list}{$\bullet$}{\parsep=0pt\itemsep=0pt}
\renewcommand{\arraystretch}{0.6}
\item the uncertainty on $m_B$ from the lightcurve fits;
\item the uncertainty on $s$, multiplied by $\alpha$
\item the covariance between $m_B$ and $s$;
\item a contribution from the uncertainty in the redshift due to
  peculiar velocity (assumed to have a dispersion of 300~km~s$^{-1}$
  along the line of site);
\item 10\% of the Galactic extinction correction; and
\item 0.17 magnitudes of intrinsic dispersion (H96).
\end{list}
Fits where host-galaxy extinction corrections are explicitly
applied use the first five items above plus:
\begin{list}{$\bullet$}{\parsep=0pt\itemsep=0pt}
\item the uncertainty on \ebv\ multiplied by $R_B$;
\item the covariance between \ebv\ and $m_B$;
\item 0.11 magnitudes of intrinsic dispersion \citep{phi99}; and
\item an additional 0.04 magnitudes of intrinsic $U$-$B$ dispersion for
  $z>0.7$.
\end{list}
Host-galaxy extinction corrections used a value $R_B\equiv
A_B/\ebv=4.1$, which results from passing a SN~Ia spectrum through the
standard \citet{odo94} extinction law.  Except where explicitly noted
below, the \ebv\ uncertainties are \emph{not} reduced by any prior
assumptions on the intrinsic color excess distribution.  Although there
is almost certainly some intrinsic dispersion either in $R_B$, or in the
true $B$-$V$ color of a \snia\ \citep{nob03}, we do not explicitly
include such a term.  The effect of such a dispersion is included, in
principle, in the 0.11 magnitudes of intrinsic magnitude dispersion
which \citet{phi99} found after applying extinction corrections.

As discussed in \S~\ref{sec:colorcor}, the intrinsic $U$-$B$ dispersion
is likely to be greater than the intrinsic $B$-$V$ dispersion.  For
those supernovae most affected by this (i.e. those at $z>0.7$), we
included an additional uncertainty corresponding to 0.04
magnitudes of intrinsic $U$-$B$ dispersion, converted into a magnitude
error using the O'Donnell extinction law.

This set of statistical uncertainties is slightly different from that
used in P99.  For these fits, the test value of $\alpha$ was used to
propagate the stretch errors into the corrected $B$-band magnitude
errors; in contrast, P99 used a single value of $\alpha$ for
purposes of error propagation.

\subsection{Supernova Subsets}
\label{sec:subsets}

In P99, separate analyses were performed and compared for the supernova
sample before and after removing supernovae with less secure
identification as Type~Ia.  The results were shown to be consistent,
providing a cross-check of the cosmological conclusions.  For the
analyses of this paper, adding and comparing eleven very-well-measured
SNe~Ia, we only consider from P99 the more securely spectrally
identified SNe~Ia with reasonable color measurements
(i.e. $\sigma_{R-I}<0.25$); those supernovae are listed in
Table~\ref{tab:42snefits}.  Following P99, we omit one supernova which is
an outlier in the stretch distribution, with $s<0.7$
(\mbox{SN\,1992br}), and one SN which is a $>6\sigma$ outlier from the
best-fit cosmology (\mbox{SN\,1997O}).  We also omit those supernovae
which are most seriously reddened, with \mbox{$\ebv>0.25$} and
$>3\sigma$ above zero; host-galaxy extinction corrections have been
found in studies of low-redshift supernovae to overcorrect these reddest
objects \citep{phi99}.  This cut removes two SNe at low redshift
(\mbox{SNe\,1995bd} and 1996bo), one from P99 (\mbox{SN\,1996cn}), and
one of the eleven HST supernovae from this paper (\mbox{SN\,1998aw}).
The resulting ``full primary subset'' of SNe~Ia is identified as Subset
1 in the tables.

For the analyses of a ``low-extinction primary subset,'' Subset 2, we
further cull out four supernovae with host-galaxy \mbox{\ebv$>0.1$} and
$>2\sigma$ above zero, including two of the HST supernovae from this
paper (\mbox{SNe\,1992ag}, 1993ag, 1998as, and 1998ax).  The
low-extinction primary subset includes eight of the eleven new HST
supernovae presented in this paper.

Subset 3, the ``low-extinction strict Ia subset,'' makes an even more
stringent cut on spectral confirmation, including only those supernovae
whose confirmations as Type~Ia SNe are unquestionable.  This subset is
used in \S~\ref{sec:typecontamination} to estimate any possible
systematic bias resulting from type contamination.  An additional six
supernovae, including two of the HST supernovae from this paper, are
omitted from Subset 3 beyond those omitted from Subset 2; these are
\mbox{SNe\,1995as}, 1996cf, 1996cg, 1996cm, 1998ay, and 1998be.


\section{Colors and Extinction}
\label{sec:colorsandextinction}

In this section, we discuss the limits on host-galaxy extinction we can
set based on the measured colors of our supernovae.  For the primary fit
of our P99 analysis, extinction was estimated by comparing the mean
host-galaxy \ebv\ values from the low- and high-redshift samples.
Although the uncertainties on individual \ebv\ values for high-redshift
supernovae were large, the uncertainty on the mean of the distribution
was only 0.02 magnitudes. P99 showed that there was no significant
difference in the mean host-galaxy reddening between the low and
high-redshift samples of supernovae of the primary analysis (Fit C).
This tightly constrained the systematic uncertainty on the cosmological
results due to differences in extinction.  The models of \citet{hat98}
suggest that most SNe~Ia should be found with little or no host galaxy
extinction.  By making a cut to include only those objects which have
small \ebv\ values (and then verifying the consistency of low- and
high-redshift mean reddening), we are creating a subsample likely to
have quite low extinction.  The strength of this method is that it does
not depend on the exact shape of the intrinsic extinction distribution,
but only requires that most supernovae show low
extinction. Figure~\ref{fig:colorhist} (discussed below) demonstrates
that most supernovae indeed have low-extinction, as expected from the
\citet{hat98} models.  Monte Carlo simulations of our data using the
\citet{hat98} extinction distribution function and our low-extinction
\ebv\ cuts confirm the robustness of this approach, and further,
demonstrate that similarly low extinction subsamples are obtained for both
low- and high-redshift datasets despite the larger color uncertainties
for some of the P99 supernovae.

\citet{rie98} used the work of \citet{hat98} differently, by applying a
one-sided Bayesian prior to their measured \ebv\ values and
uncertainties.  A prior formed from the \citet{hat98} extinction
distribution function would have zero probability for negative values of
\ebv, a peak at \mbox{$\ebv\sim0$} with roughly 50\% of the probability,
and an exponential tail to higher extinctions.  As discussed in P99 (see
the ``Fit E'' discussion, where P99 apply the same method), when
uncertainties on high- and low-redshift supernova colors differ, use of
an asymmetric prior may introduce bias into the cosmological results,
depending on the details of the prior.  While a prior with a tight
enough peak at low extinction values introduces little bias (especially
when low- and high-redshift supernovae have comparable uncertainties),
it does reduce the apparent \ebv\ error bars on all but the most
reddened supernovae.  As we will show in Figure~\ref{fig:ebvcosmofits}
(\S~\ref{sec:cosmoparam}) the use of this prior almost completely
eliminates the contribution of color uncertainties to the size of the
cosmological confidence regions, meaning that an extinction correction
using a sharp enough prior is much more akin to simply selecting a
low-extinction subset than to performing an assumption-free extinction
correction using the \ebv\ measurement uncertainties.

The high precision measurements of the $R$-$I$ color afforded by the
\wfpc\ lightcurves for the new supernovae in this work allow a direct
estimation of the host-galaxy \ebv\ color excess without any need to
resort to any prior assumptions concerning the intrinsic extinction
distribution.

\ifthenelse{\boolean{ispreprint}}{
\begin{figure*}[p]
\begin{lrbox}{\thisbox}
\epsfig{file=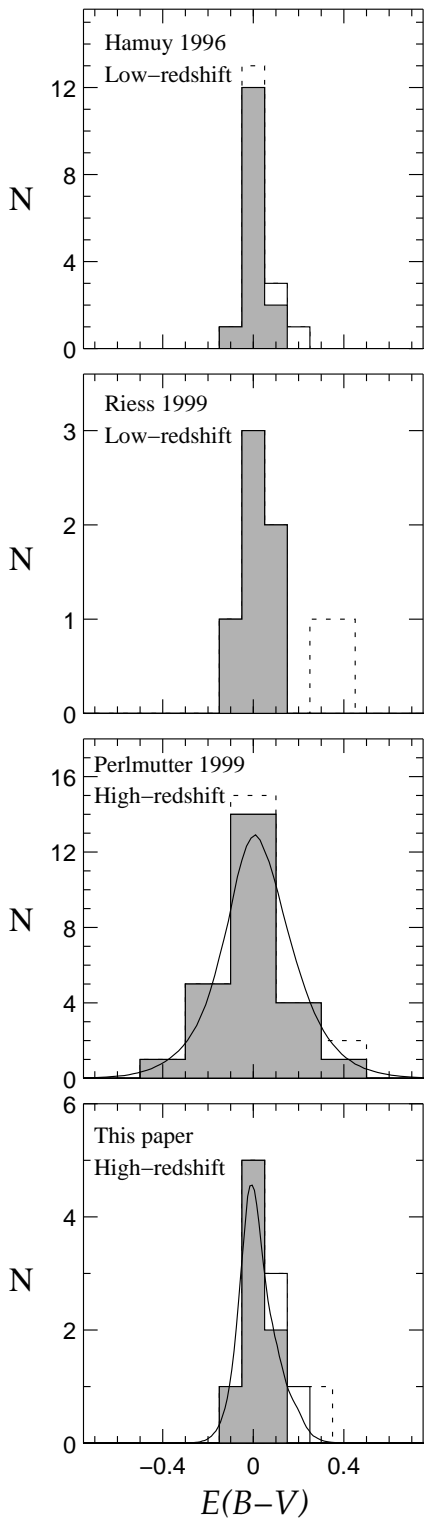,height=8in}
\end{lrbox}
\settowidth{\thiswid}{\usebox{\thisbox}}
\begin{center}
\begin{tabular}{p{\thiswid} b{2.5in}}
\usebox{\thisbox} &
\caption{Histograms of \ebv\ for the four samples of supernovae used in
  this paper.  The filled grey histogram represents just the
  low-extinction subset (Subset 2).  The open boxes on top of that
  represent supernovae which are in the primary subset (Subset 1) but
  excluded from the low-extinction subset.  Finally, the dotted
  histogram represents those supernovae which are in the full sample but
  omitted from the primary subset.  The solid lines drawn over the
  bottom two panels is a simulation of the distribution expected if the
  low-extinction subset of the H96 sample represented the true
  distribution of SN colors, given the error bars of the low-extinction
  subset of each high-redshift sample.}
  \label{fig:colorhist} \\
\end{tabular}
\end{center}
\end{figure*}
}{\placefigure{fig:colorhist}}

\ifthenelse{\boolean{ispreprint}}{
\begin{table}[tb]
\begin{lrbox}{\thisbox}
\settowidth{\nopmsp}{$-$}
\newcommand{\nopm}{\hspace*{\nopmsp}}
\begin{tabular}{lcc}
\tableline
\tableline
Sample & Complete & Low-extinction \\
       & Set      & Primary Subset \\
       &          & SNe$^a$ \\
\tableline
Low z & $+0.095\pm0.003$ &     $-0.001\pm0.003$ \\
P99   & $+0.018\pm0.024$ &     $-0.004\pm0.025$ \\
HST   & $+0.090\pm0.012$ &     $+0.012\pm0.015$ \\
\tableline
\end{tabular}
\end{lrbox}
\settowidth{\thiswid}{\usebox{\thisbox}}
\begin{center}
\begin{minipage}{\thiswid}
\caption{Mean \ebv\ Values}\label{tab:meanebv}
\usebox{\thisbox}
a: SNe omitted from our low-extinction primary subset, Subset 2,
(\S~\ref{sec:subsets}) have been omitted from these means.  This excludes
outliers, as well as supernovae with both $\ebv>0.1$ and $\ebv>2\sigma$
above zero.
\end{minipage}
\end{center}
\end{table}
}{\placetable{tab:meanebv}}

Figure~\ref{fig:colorhist} shows histograms of the host-galaxy \ebv\
values from different samples of the supernovae used in this paper.  For
the bottom two panels, a line is over-plotted that treats the H96
low-extinction subset's \ebv\ values as a parent distribution, and shows
the expected distribution for the other samples given their measurement
uncertainties.  The low-extinction subset of each sample (the grey
histogram) has a color excess distribution which is consistent with that
of the low-extinction subset of H96.  Table~\ref{tab:meanebv} lists the
variance-weighted mean \ebv\ values for the low-redshift supernovae and
for each sample of high-redshift supernovae.  Although varying amounts
of extinction are detectable in the mean colors of each full sample, the
supernovae in the low-extinction primary subset (\S~\ref{sec:subsets})
of each sample are consistent with $\ebv=0$.  This subset is consistent
with the models of \citet{hat98}, discussed above, in which most SNe~Ia
are observed in regions of very low extinction.  We will consider
cosmological fits both to this low-extinction subset and to the primary
subset with host-galaxy reddening corrections applied.

\ifthenelse{\boolean{ispreprint}}{
\begin{figure}[tb]
\begin{lrbox}{\thisbox}
\epsfig{file=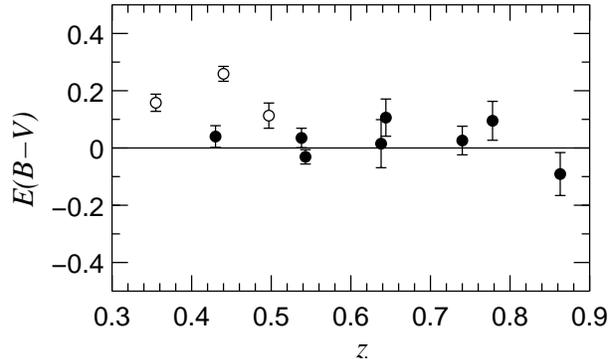}
\end{lrbox}
\settowidth{\thiswid}{\usebox{\thisbox}}
\begin{center}
\begin{minipage}{\thiswid}
\usebox{\thisbox}
\caption{A plot of \ebv\ as a function of redshift for the 11
  HST-observed supernovae of this paper shows that the blue edge of the
  distribution shows no significant evolution with redshift.  (The
  larger dispersion at lower redshifts is expected for a flux-limited
  sample.)  Error bars include only measurement errors, and no assumed
  intrinsic color dispersion.  Filled circles are those supernovae in
  the low-extinction subset (Subset 2).}
\label{fig:ebvz}
\end{minipage}
\end{center}
\end{figure}
}{\placefigure{fig:ebvz}}

Figure~\ref{fig:ebvz} shows \ebv\ vs. redshift for the eleven supernovae
of this paper.  Three of the lowest redshift SNe are likely to be
significantly reddened: \mbox{SN\,1998as} at $z=0.36$, \mbox{SN\,1998aw}
at $z=0.44$, and \mbox{SN\,1998ax} at $z=0.50$.  This higher incidence
of extincted SNe at the low-redshift end of our sample is consistent
with expectations for a flux-limited survey, where extincted supernovae
will be preferentially detected at lower redshifts.  Indeed, the
distribution of \ebv\ values versus redshift shown in
Figure~\ref{fig:ebvz} is consistent with the results of a Monte Carlo
simulation similar to that of \citet{hat98}, but including the effects
of the survey flux limit.  Several authors (including \citet{lei01} and
\citet{fal99}) have suggested that there is evidence from the \ebv\
values in \citet{rie98} that high-redshift supernovae are bluer
statistically than their low-redshift counterparts.  Our data show no
such effect (nor did our P99 SNe).

The mean host-galaxy color excess calculated for the highest redshift
supernovae is critically dependent on the assumed intrinsic $U$-$B$
color (see \S~\ref{sec:colorcor}).  An offset in this assumed
\mbox{$U$-$B$} will affect the high-redshift supernovae much more than
the low-redshift supernovae (whose measurements are primarily of the
rest frame $B$- and $V$-band lightcurves).  The $K$-corrected,
rest-frame $B$-band magnitudes are also dependent on the assumed
supernova colors that went into deriving the $K$-corrections.  If the
assumed $U$-$B$ color is too red, it will affect the cross-filter
$K$-correction applied to $R$-band data at $z\gtrsim0.5$, thereby
changing derived rest frame colors.  In \S~\ref{sec:systematic}, we
consider the effect of changing the reference $U$-$B$ color.

\ifthenelse{\boolean{ispreprint}}{
\begin{figure*}[tbp]
\begin{lrbox}{\thisbox}
\epsfig{file=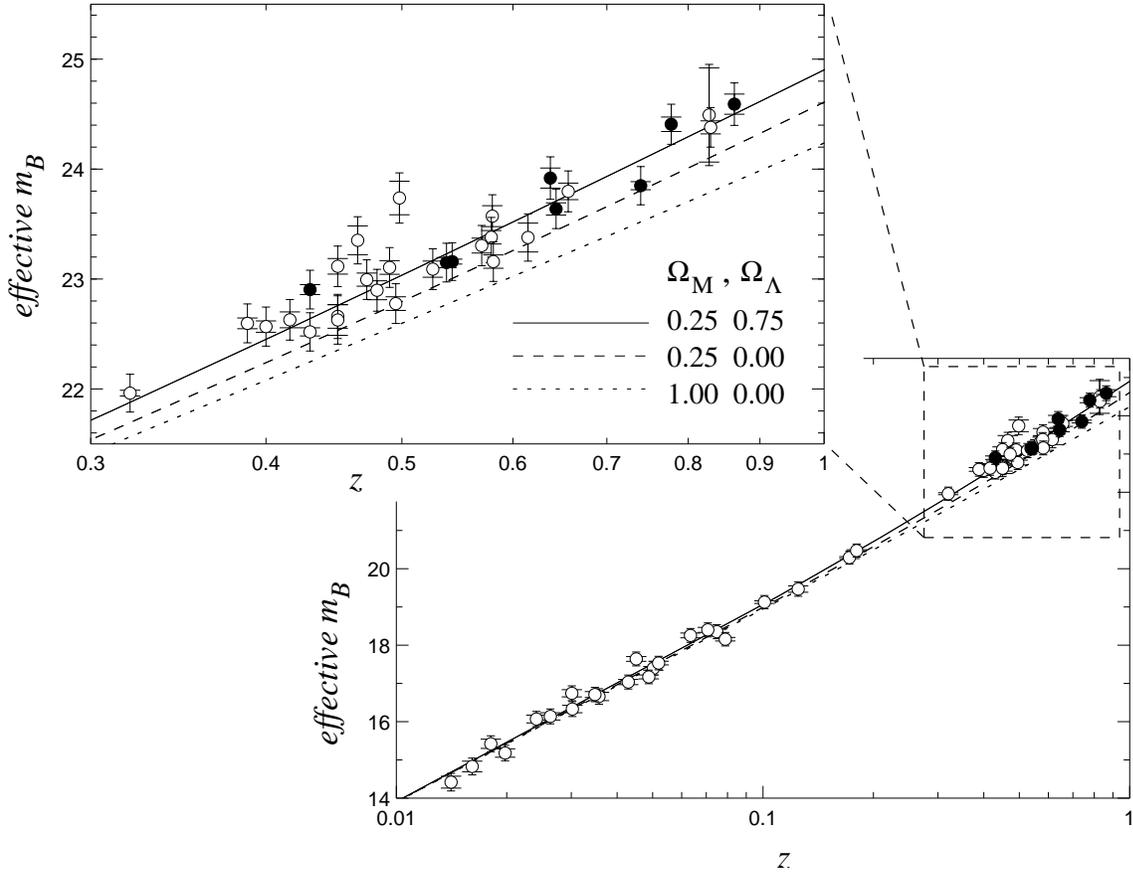 , width=6in}
\end{lrbox}
\settowidth{\thiswid}{\usebox{\thisbox}}
\begin{center}
\begin{minipage}{\thiswid}
\usebox{\thisbox}
\caption{Hubble diagram of effective $K$- and stretch-corrected $m_B$
  vs. redshift for the supernovae in the primary low-extinction subset.
  Filled circles represent the HST supernovae of this paper.  Inner
  error bars show just the measurement uncertainties; outer error bars
  include 0.17 magnitudes of intrinsic dispersion.  The solid line is
  the best-fit flat-universe cosmology from the low-extinction subset;
  the dashed and dotted lines represent the indicated cosmologies.}
\label{fig:monsterhubble}
\end{minipage}
\end{center}
\end{figure*}
}{\placefigure{fig:monsterhubble}}

\ifthenelse{\boolean{ispreprint}}{
\begin{figure*}[tbp]
\begin{lrbox}{\thisbox}
\epsfig{file=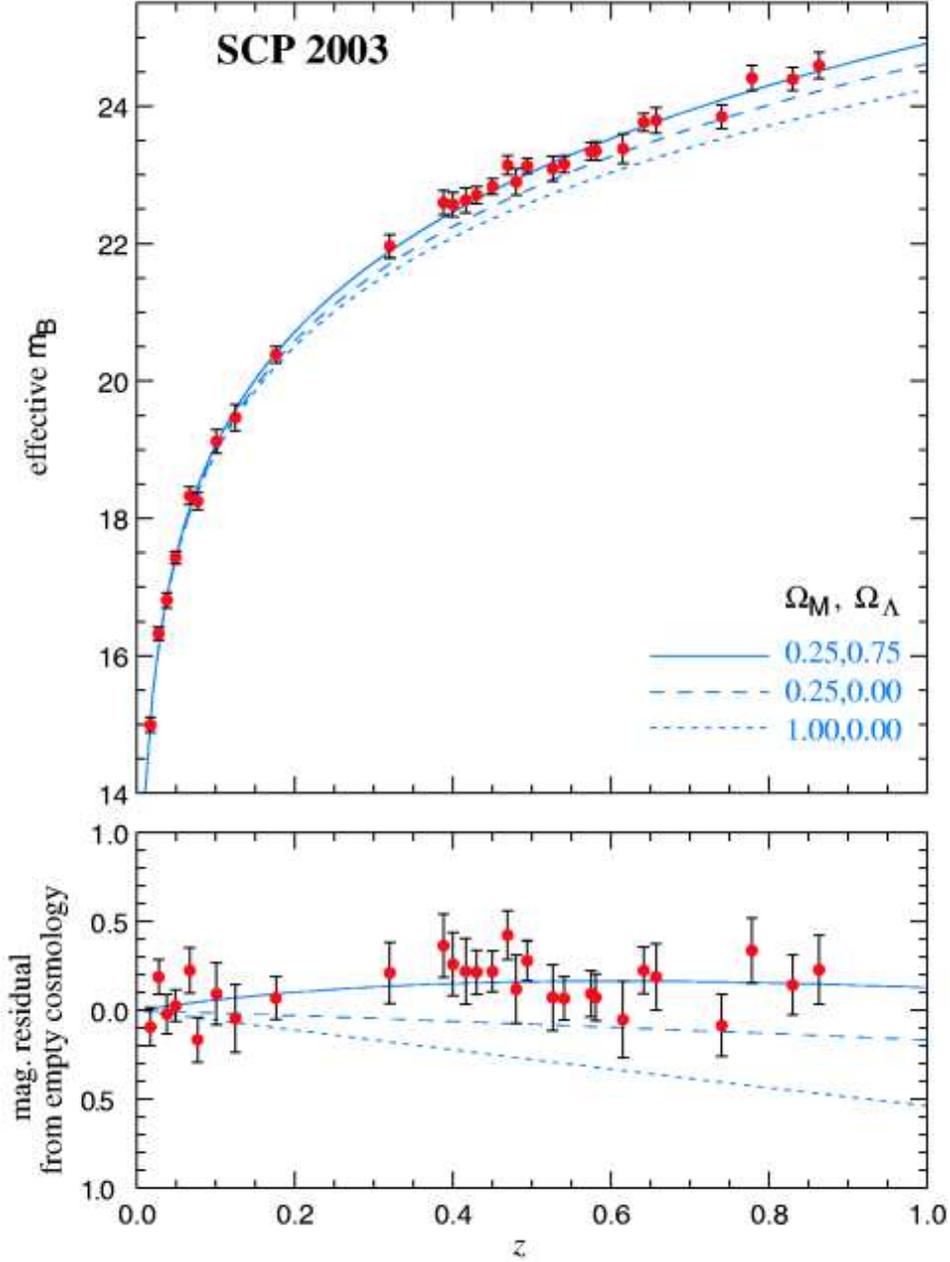 , width=5in}
\end{lrbox}
\settowidth{\thiswid}{\usebox{\thisbox}}
\begin{center}
\begin{minipage}{\thiswid}
\usebox{\thisbox}
\caption{Upper panel: Averaged Hubble diagram with a linear redshift
scale for all supernovae from our low-extinction subsample. Here
supernovae within $\Delta z < 0.01$ of each other have been combined
using a weighted average in order to more clearly show the quality and
behavior of the dataset. (Note that these averaged points are for
display only, and have not been used for any quantitative analyses.) The
solid curve overlaid on the data represents our best-fit flat-universe
model, \mbox{$(\om,\ol)=(0.25,0.75)$} (Fit~3 of
Table~\ref{tab:fits}). Two other cosmological models are shown for
comparison: \mbox{$(\om,\ol)=(0.25,0)$} and \mbox{$(\om,\ol)=(1,0)$}.
Lower panel: Residuals of the averaged data relative to an empty
universe, illustrating the strength with which dark energy has been
detected. Also shown are the suite of models from the upper panel,
including a solid curve for our best-fit flat-universe model.  }
\label{fig:binhubble}
\end{minipage}
\end{center}
\end{figure*}
}{\placefigure{fig:binhubble}}

\ifthenelse{\boolean{ispreprint}}{
\begin{figure*}[tbp]
\begin{lrbox}{\thisbox}
\epsfig{file=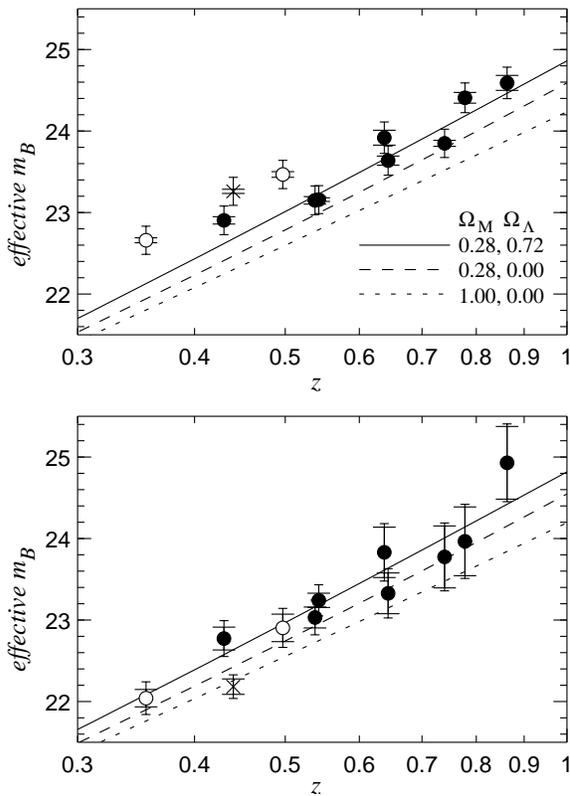 , width=3in}
\end{lrbox}
\settowidth{\thiswid}{\usebox{\thisbox}}
\begin{center}
\begin{tabular}{b{\thiswid} b{2.5in}}
\usebox{\thisbox} &
\caption{Hubble diagram of effective $K$- and stretch-corrected $m_B$
  vs. redshift for the 11 supernovae observed with \wfpc\ and reported
  in this paper.  Circles represent supernovae in the primary subset
  (Subset 1); the one point plotted as a cross (the very reddened
  supernova \mbox{SN\,1998aw}) is omitted from that subset.  Open
  circles represent reddened supernovae omitted from the low-extinction
  primary subset (Subset 2), while filled circles are in both Subsets 1
  and 2.  \textbf{Upper plot:} no host-galaxy \ebv\ extinction
  corrections have been applied.  Inner error bars only include the
  measurement error.  Outer error bars include 0.17 magnitudes of
  intrinsic dispersion.  \textbf{Lower plot:} extinction corrections
  have been applied using the standard interstellar extinction law.
  Error bars have been increased by the uncertainty in this extinction
  correction.  Again, inner error bars represent only measurement
  uncertainties, while outer error bars include 0.11 magnitudes of
  intrinsic dispersion.  Lines are for three example cosmologies with
  the indicated values of \om\ and \ol; the solid line is the best-fit
  flat-universe cosmology to our full primary subset with extinction
  corrections applied.}
\label{fig:hubs}
\end{tabular}
\end{center}
\end{figure*}
}{\placefigure{fig:hubs}}

\ifthenelse{\boolean{ispreprint}}{
\begin{figure*}[tbp]
\begin{lrbox}{\thisbox}
\epsfig{file=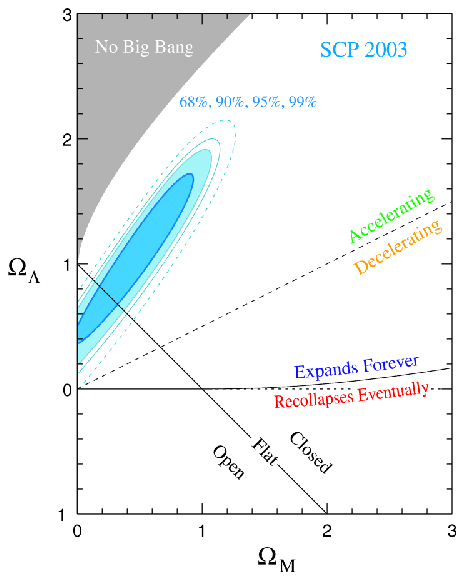, height=5.5in}
\end{lrbox}
\settowidth{\thiswid}{\usebox{\thisbox}}
\begin{center}
\begin{minipage}{\thiswid}
\usebox{\thisbox}
\caption{68\%, 90\%, 95\%, and 99\% confidence regions for \om\ and \ol\
  from this paper's primary analysis, the fit to the low-extinction
  primary subset (Fit~3).}
\label{fig:primaryconfs}
\end{minipage}
\end{center}
\end{figure*}
}{\placefigure{fig:primaryconfs}}

\section{Cosmological Results}
\label{sec:cosmoresults}

\subsection{\om\ and \ol}
\label{sec:cosmoparam}

Figures~\ref{fig:monsterhubble} through \ref{fig:hubs} show Hubble
Diagrams which effective $B$-band peak magnitudes and redshifts for the
new supernovae of this paper; these magnitudes have been $K$- and
stretch corrected, and have been corrected for Galactic extinction.
Figure~\ref{fig:monsterhubble} shows all of the data in the
low-extinction subset of supernovae.  For the sake of clarity,
Figure~\ref{fig:binhubble} shows the same subset, but for this figure
supernovae with redshifts within 0.01 of each other have been combined
in a variance-weighted average.  The lower panel of
Figure~\ref{fig:binhubble} shows the residuals from an empty universe
($\om=0$, $\ol=0$), illustrating the strength with which dark energy has
been detected.  In both Figures~\ref{fig:monsterhubble} and
\ref{fig:binhubble}, the solid line represents the flat-universe
cosmology resulting from our fits to the low-extinction subset.
Figure~\ref{fig:hubs} shows just the eleven HST supernovae from this
paper.  In the upper panel of this latter figure, the stretch- and
$K$-corrected effective $m_B$ values and uncertainties are plotted.  In
the lower panel, effective $m_B$ values have also been corrected for
host-galaxy extinction based on measured \ebv\ values.  The solid line
in this figure represents the best-fit flat-universe cosmology to the
full primary subset with extinction corrections applied.

\ifthenelse{\boolean{ispreprint}}{
\begin{table*}[htbp]
\footnotesize
\renewcommand{\arraystretch}{0.8}
\begin{lrbox}{\thisbox}
\begin{tabular}{clccccccc}
\tableline
\tableline \\[-1ex]
Fit      & High-Redshift SNe   & N$_\mathrm{SNe}$ & Min.     & \om\ for      
& \ol\ for   & $P(\ol>0)$ & $\mathcal{M}$ & $\alpha$ \\
\#       & Included in Fit$^a$ &                  & $\chi^2$ & Flat$^b$
& Flat$^b$   &            &               &          \\[0.3ex]

\tableline \\

\multicolumn{8}{l}{Fits to the Low-Extinction Primary Subset} \\[12pt]

1        & SNe from P99    & 46         & 52  & $0.25^{+0.08}_{-0.07}$ & $0.75^{+0.07}_{-0.07}$ & 0.9995  & $-3.49\pm 0.05$ & $1.58\pm 0.31$ \\[12pt]

2        & New HST SNe     & 29         & 30  & $0.25^{+0.09}_{-0.08}$ & $0.75^{+0.08}_{-0.09}$ & 0.9947  & $-3.47\pm 0.05$ & $1.06\pm 0.37$ \\
         & from this paper &            &      &                         &                         &               &               \\[12pt]

{\bfseries 3} & {\bfseries All SCP SNe} & {\bfseries 54}         & {\bfseries 60}  & $\mathbf{0.25^{+0.07}_{-0.06}}$  & $\mathbf{0.75^{+0.06}_{-0.07}}$ & {\bfseries 0.9997} & $\mathbf{-3.48\pm0.05}$ & $\mathbf{1.47\pm0.29}$ \\[12pt]

\tableline \\

\multicolumn{8}{l}{Fits to Full Primary Subset, with Extinction Correction} \\[12pt]

4        & SNe from P99    & 48         & 56  & $0.21^{+0.18}_{-0.15}$ &  $0.79^{+0.15}_{-0.18}$ & 0.9967 & $-3.55\pm 0.05$ & 1.30$\pm$0.30 \\[12pt]

5        & New HST SNe     & 33         & 39  & $0.27^{+0.12}_{-0.10}$  & $0.73^{+0.10}_{-0.12}$ & 0.9953 & $-3.54\pm 0.05$ & 1.29$\pm$0.28 \\
         & from this paper &            &      &                         &                        &                &               \\[12pt]

6        & All SCP SNe     & 58         & 65  & $0.28^{+0.11}_{-0.10}$  & $0.72^{+0.10}_{-0.11}$ & 0.9974 & $-3.53\pm 0.05$ & 1.18$\pm$0.30 \\[12pt]

%
%
%
%
%
\tableline
\end{tabular}
\end{lrbox}
\settowidth{\thiswid}{\usebox{\thisbox}}
\begin{center}
\begin{minipage}{\thiswid}
\caption{Cosmological fits}
\label{tab:fits}
\usebox{\thisbox}

a: All fits include the low-redshift SNe from H96 and R99.  See
\S~\ref{sec:subsets} for the definitions of the supernova subsets.

b: This is the intersection of the fit probability distribution with the
line \mbox{$\om+\ol=1$}.

\end{minipage}
\end{center}
\end{table*}
}{\placetable{tab:fits}}

Table~\ref{tab:fits} lists results from fits to both of our primary
subsets of supernovae.  Supernovae from both the H96 and R99
low-redshift samples were included in all fits.  The first three lines
show fits to the low-extinction primary subset.  So that the new sample
of high-redshift supernovae may be compared to those from P99, each
high-redshift sample was fit separately (Fits~1 and 2).  Fit~3 combines
all of the current SCP high-redshift supernovae from the low-extinction
subsets, and represents the primary result on \om\ and \ol\ for this
paper; Figure~\ref{fig:primaryconfs} shows the confidence regions for
\om\ vs. \ol\ from this fit.  Figure~\ref{fig:fitsonetothree} shows the
comparison of the confidence regions when each high-redshift sample is
treated separately.  Note that Fit 2 provides comparable and consistent
measurements of \om\ and \ol\ to Fit 1.  Additionally, the sizes of the
confidence regions from the 8 HST SNe in Fit 2 is similar to those in
Fit 1, which includes 25 high-redshift supernovae from P99.

\ifthenelse{\boolean{ispreprint}}{
\begin{figure}[tb]
\begin{lrbox}{\thisbox}
\epsfig{file=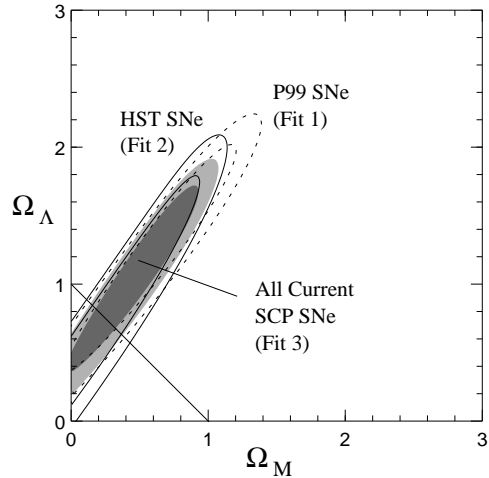 , width=2.5in}
\end{lrbox}
\settowidth{\thiswid}{\usebox{\thisbox}}
\begin{center}
\begin{minipage}{\thiswid}
\usebox{\thisbox}
\caption{Contours indicate 68\% and 90\% confidence regions for fits to
  supernovae from the low-extinction primary subset, including just the
  high-redshift SNe from P99 (dotted lines), just the new HST
  high-redshift SNe (solid lines), and all SCP high-redshift SNe (filled
  contours).  The low-redshift SNe from the primary subset are included
  in all fits.  The new, independent sample of high-redshift supernovae
  provide measurements of \om\ and \ol\ consistent with those from the
  P99 sample.}
\label{fig:fitsonetothree}
\end{minipage}
\end{center}
\end{figure}
}{\placefigure{fig:fitsonetothree}}

Fits~4--6 in Table~\ref{tab:fits} show the results for the primary
subset when host-galaxy extinction corrections have been applied.
Figure~\ref{fig:ebvcosmofits} compares these results to those of the
primary low-extinction fit.  The primary fits of
Figure~\ref{fig:fitsonetothree} are reproduced in the top row of
Figure~\ref{fig:ebvcosmofits}.  The second row has host-galaxy
extinction corrections applied using the one-sided prior used by Fit E
of P99 and \citet{rie98} discussed in \S~\ref{sec:colorsandextinction}.
The third row has full extinction corrections applied without any prior
assumptions on the intrinsic \ebv\ distribution.  Three conclusions are
apparent from this plot.  First, using a strongly peaked prior on
extinction prevents the \ebv\ error bars from being fully propagated
into the cosmological confidence regions, and hence apparently tightens
the constraints.  However, for a peaked prior, this is very similar to
assuming no extinction and not performing an extinction correction (but
without testing the assumption), while for a wider prior there is a
danger of introducing bias.  Second, the current set of supernovae
provide much smaller confidence regions on the \ol\ versus \om\ plane
than do the \sneia\ from previous high-redshift samples when unbiased
extinction corrections are applied.  Whereas
Figure~\ref{fig:fitsonetothree} shows that the current set of supernovae
give comparable measurements of \om\ and \ol\ when the low-extinction
subsample is used with no host-galaxy extinction corrections,
Figure~\ref{fig:ebvcosmofits} shows that the much higher precision color
measurements from the \wfpc\ data allow us directly to set much better
limits on the effects of host-galaxy extinction on the cosmological
results.  Finally, the cosmology which results from the
extinction-corrected fits is consistent with the fits to our
low-extinction primary subset.  Contrary to the assertion of
\citet{row02}, even when host-galaxy extinction is directly and fully
accounted for, dark energy is required with \mbox{$P(\ol>0)>0.99$}.

\ifthenelse{\boolean{ispreprint}}{
\begin{figure*}[htbp]
\begin{lrbox}{\thisbox}
\epsfig{file=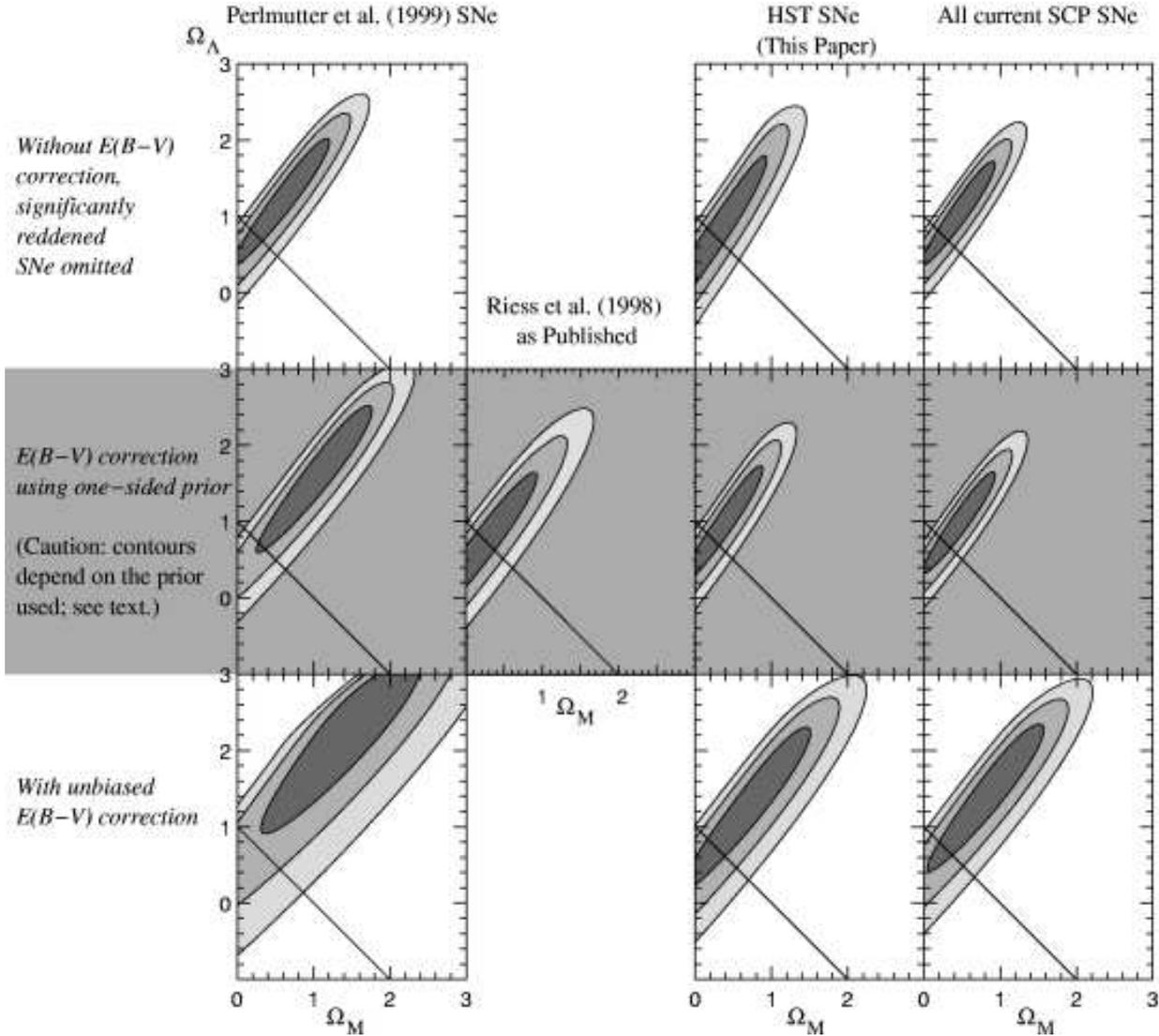 , width=6.5in}
\end{lrbox}
\settowidth{\thiswid}{\usebox{\thisbox}}
\begin{center}
\begin{minipage}{\thiswid}
\usebox{\thisbox}
\caption{\scriptsize 68.3\%, 95.4\%, and 99.7\% confidence regions for
  \om\ and \ol\ using different data subsets and methods for treating
  host-galaxy extinction corrections.  The top row represents our fits
  to the low-extinction primary subset, where significantly reddened
  supernovae have been omitted and host-galaxy extinction corrections
  are not applied.  The second row shows fits where extinction
  corrections have been applied using a one-sided extinction prior.
  These fits are sensitive to the choice of prior, and can either yield
  results equivalent to analyses assuming low extinction (but without
  testing the assumption), or yield biased results (see text).  Note
  that the published contours from \citet[][their Fig. 6, solid
  contours]{rie98} presented results from fits that included nine
  well-observed supernovae (that are comparable to the primary subsets
  used in the other panels), but also four supernovae with very sparsely
  sampled lightcurves, one supernova at $z=0.97$ without a spectral
  confirmation, as well as two supernovae from the P99 set.  The third
  row shows fits with unbiased extinction corrections applied to our
  primary subset.  The HST SNe presented in this paper show a marked
  improvement in the precision of the color measurements, and hence in
  the precision of the \om\ and \ol\ measurements when a full extinction
  correction is applied.  With full and unbiased extinction corrections,
  dark energy is still required with \mbox{$P(\ol>0)=0.99$}.}
\label{fig:ebvcosmofits}
\end{minipage}
\end{center}
\end{figure*}
}{\placefigure{fig:ebvcosmofits}}

\subsection{Combined High-Redshift Supernova Measurements}
\label{sec:combined}

Figure~\ref{fig:combinedconfs} shows measurements of \om\ and \ol\ which
combine the high-redshift supernova data of \citet{rie98} together with
the SCP data presented in this paper and in P99.  The contours show
confidence intervals from the 54 supernovae of the low-extinction
primary Subset~2 (used in Fit~3 of Table~\ref{tab:fits}), plus the nine
well-observed confirmed Type~Ia supernovae from \citet{rie98} (using the
lightcurve parameters resulting from their template-fitting analysis);
following the criteria of Subset~2, \mbox{SN\,1997ck} from that paper
has been omitted, as that supernova was not confirmed spectrally.  We
also omit from \citet{rie98} the supernovae they measured using the
``snapshot'' method (due to the very sparsely sampled lightcurve), and
two SCP supernovae that \citet{rie98} used from the P99 data set which
are redundant with our sample.  This fit has a minimum $\chi^2$ of 65
with 63 supernovae.  Under the assumption of a flat universe, it yields
a measurement of the mass density of \mbox{$\om=0.26^{+0.07}_{-0.06}$},
or equivalently a cosmological constant of $\ol=0.74^{+0.06}_{-0.07}$.
Recent ground-based data on eight new high-redshift supernovae from
\citet{ton03} (not included in this fit) are consistent with these
results.  Note that in this fit, the nine supernovae from \citet{rie98}
were not treated in exactly the same manner as the others.  The details
of the template fitting will naturally have been different, which can
introduce small differences (see \S~\ref{sec:fitmethodsystematic}).
More importantly, the $K$-corrections applied by \citet{rie98} to derive
distance moduli were almost certainly different from those used in this
paper.

\ifthenelse{\boolean{ispreprint}}{
\begin{figure}[bt]
\begin{lrbox}{\thisbox}
\epsfig{file=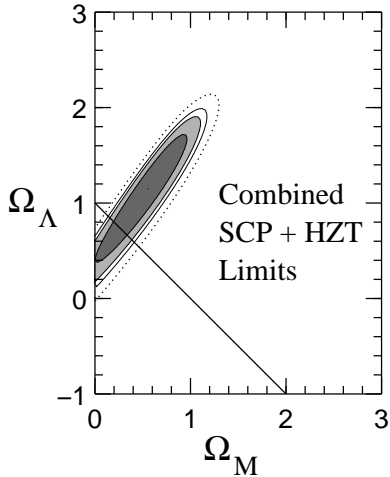}
\end{lrbox}
\settowidth{\thiswid}{\usebox{\thisbox}}
\begin{center}
\begin{minipage}{\thiswid}
\usebox{\thisbox}
\caption{68\%, 90\%, 95\%, and 99\% confidence regions for \om\ and \ol,
  combining the high-redshift data of the SCP (this paper and P99) and
  \citet{rie98}.  The fit includes Subset~2 supernovae from the SCP plus
  the nine well-observed confirmed SNe~Ia from \citet{rie98}.}
\label{fig:combinedconfs}
\end{minipage}
\end{center}
\end{figure}
}{\placefigure{fig:combinedconfs}}

\subsection{Dark Energy Equation of State}
\label{sec:w}

The fits of the previous section used a traditional constrained
cosmology where $\om$ is the energy density of non-relativistic matter
(i.e. pressure \mbox{$p=0$}), and $\ol$ is the energy density in a
cosmological constant (i.e. pressure \mbox{$p=-\rho$}, where $\rho$ is
the energy density).  In Einstein's field equations, the gravitational
effect enters in terms of $\rho+3p$.  If $w\equiv p/\rho$ is the
equation of state parameter, then for matter \mbox{$w=0$}, while for
vacuum energy (i.e. a cosmological constant) \mbox{$w=-1$}.  In fact, it
is possible to achieve an accelerating Universe so long as there is a
component with \mbox{$w<\sim-1/2$}.  (If there were no contribution from
$\om$, only $w<-1/3$ dark energy is necessary for acceleration; however,
for plausible mass densities $\om\gtrsim0.2$, the dark energy must have
a more negative value of $w$.)  The Hubble diagram for high-redshift
supernovae provides a measurement of $w$ \citep[P99,][]{gar98b}.  Panels
(a) and (b) of Figure~\ref{fig:omw} show the joint confidence regions
for $\om$ versus $w$ from the SCP supernovae, including the new HST
supernovae, under the assumptions that $w$ is constant with time, and
that the Universe is flat, i.e.  $\om+\ow=1$ (where $\ow$ is the energy
density in the component with equation of state $w$, in units of the
critical density).  The supernova alone data set a 99\% confidence limit
of \mbox{$w<-0.64$} for any positive value of $\om$, without any prior
assumptions on $w$.

A fit with extinction corrections applied to the full primary subset
(Fit~6, shown in Figure~\ref{fig:omw}b) gives a 99\% confidence limit of
\mbox{$w<-1.00$}.  However, this latter limit should be approached with
caution, because $w$ is not well bounded from below with the supernova
data alone.  Although Figure~\ref{fig:omw} only shows confidence
intervals down to $w=-2$, the 68\% confidence interval from Fit~3
extends to $w<-4$, and the 99\% confidence interval extends to $w<-10$;
these confidence intervals extend to even further negative $w$ in Fit~6.
The weight of probability at very low (and probably implausible) $w$
pulls the 68\% confidence interval in Fit~6 (Figure~\ref{fig:omw}b)
downward.  A fit which used a prior to restrict $w$ to more reasonable
values (say $w>-2$) would show similar outer confidence intervals, but a
68\% confidence interval more similar to that of the low-extinction
subset in Figure~\ref{fig:omw}a.

\ifthenelse{\boolean{ispreprint}}{
\begin{figure*}[p]
\begin{lrbox}{\thisbox}
\epsfig{file=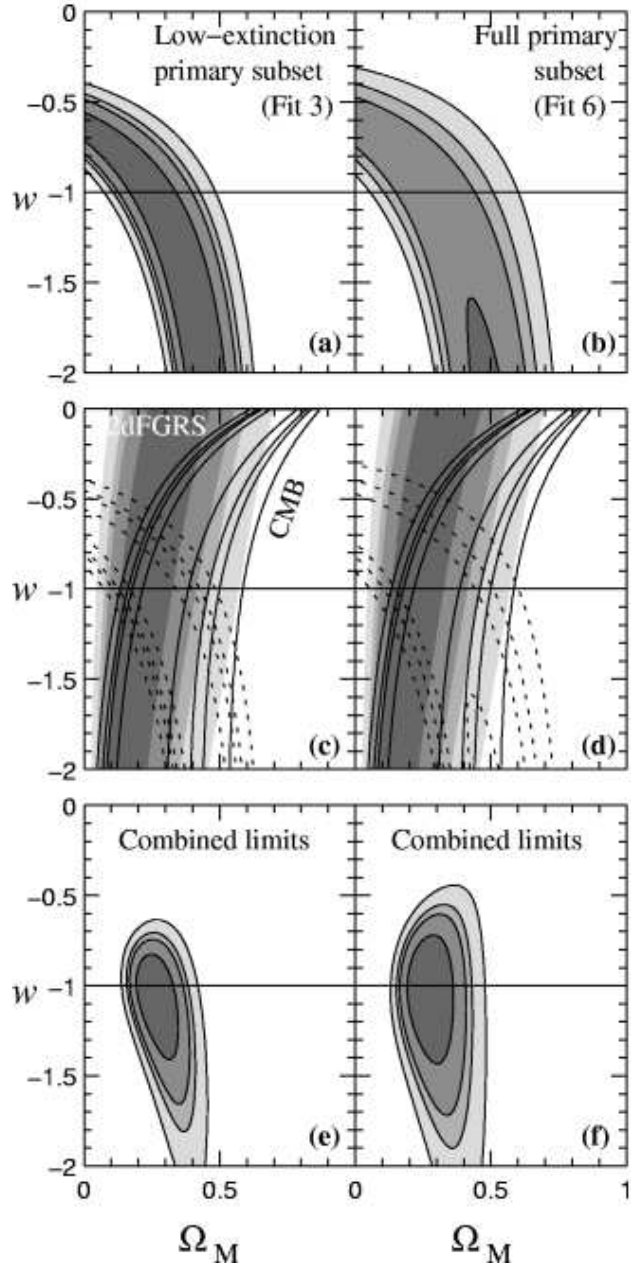, width=3.26in}
\end{lrbox}
\settowidth{\thiswid}{\usebox{\thisbox}}
\begin{center}
\usebox{\thisbox}
\caption{Joint measurements of \om\ and $w$ assuming $\om+\ow=1$ and
  that $w$ is not time-varying.  Confidence regions plotted are 68\%,
  90\%, 95\%, and 99\%.  The left column (panels a, c, and e) shows fits
  to the low-extinction primary subset; the right column (panels b, d,
  and f) shows fits to the primary subset with unbiased individual
  host-galaxy extinction corrections applied to each supernova.  The
  upper panels (a and b) show the confidence intervals from the SCP
  supernovae alone.  The middle panels (c and d) overlay this (dotted
  lines) with measurements from 2dFGRS (filled contours) \citep{haw02}
  and combined CMB measurements (solid contours) \citep{ben03,spe03}.
  The bottom panels (e and f) combine the three confidence regions to
  provide a combined measurement of \om\ and $w$.}
\label{fig:omw}
\end{center}
\end{figure*}
}{\placefigure{fig:omw}}

Other methods provide measurements of $\om$ and $w$ which are
complementary to the supernova results.  Two of these measurements are
plotted in the middle row of Figure~\ref{fig:omw}, compared with the
supernova measurements (in dotted contours).  In filled contours are
results from the redshift-distortion parameter and bias-factor
measurement of the 2dF Galaxy Redshift Survey (2dFGRS)
\citep{haw02,verde02}.  These provide a measurement of the growth
parameter, $f=0.51\pm0.11$, at the survey redshift $z=0.15$. We have
used the method of \citet{linder03} to directly solve for $f(\om,w,z)$
rather than convert $f$ to $\om$, as the conversion formula given in
\citet{haw02} is valid only for $w=-1$.  Comparison of the 2dFGRS value
of $f$ with the calculated values of $f(\om,w,z)$ yields the joint
confidence region for $\om$ and $w$.\footnote{Note that we have not used
the independent 2dFGRS power spectrum constraint on \mbox{$\om h$}
because it has not yet been generalized for different values of $w$.}

In solid lines in panels (c) and (d) of Figure~\ref{fig:omw} are
contours representing confidence regions based on the distance to the
surface of last scattering at $z=1089$ from the Wilkinson Microwave
Anisotropy Probe (WMAP) and other CMB measurements \citep{ben03,spe03}.
For a given \om\ and $w$, this reduced distance to the surface of last
scattering, $I$, is given by:
\begin{equation}
I=\int_0^{1089} [((1-\om)/\om) (1+z)^{3(1+w)} + \nonumber (1+z)^3]^{-1/2}\ dz
\end{equation}
The plotted CMB constraints come from the ``WMAPext'' sample, which
includes other CMB experiments in addition to WMAP.  They yield a
measurement of $I_0=1.76\pm0.058$, corresponding to $\om=0.29$ at
$w=-1$.  Confidence intervals are generated by calculating a probability
using $\chi^2 = \left[(I-I_0)/\sigma_{I_0}\right]^2$, where $I$ is
calculated for each $\om,w$.

As both of these measurements show mild correlations between $\om$ and
$w$ in a different sense from that of the supernova measurement, the
combined measurements provide much tighter overall constraints on both
parameters.  The confidence regions which combine these three
measurements are shown in panels (e) and (f) of Figure~\ref{fig:omw}.
When the resulting probability distribution is marginalized over \om, we
obtain a measurement of $w=-1.05^{+0.15}_{-0.20}$ (for the
low-extinction subset), or $w=-1.02^{+0.19}_{-0.24}$ (for the full
primary subset with host-galaxy extinction corrections applied).  When
the probability distribution is marginalized over $w$, we obtain a
flat-universe measurement of $\om=0.27^{+0.06}_{-0.05}$ (for the
low-extinction subset), or $\om=0.28^{+0.06}_{-0.05}$ (for the primary
subset with host-galaxy extinction corrections applied).  The 95\%
confidence limits on $w$ when our data are combined with CMB and 2dFGRS
are \mbox{$-1.61<w<-0.78$} for the low-extinction primary subset, or
\mbox{$-1.67<w<-0.62$} for the full extinction-corrected primary subset.
If we add an additional prior that $w\geq-1$, we obtain a 95\% upper
confidence limit of \mbox{$w<-0.78$} for the low-extinction primary
subset, or \mbox{$w<-0.67$} for the extinction-corrected full primary
subset.  These values may be compared with the limit in \citet{spe03}
which combines the CMB, 2dFGRS power spectrum, and HST key project $H_0$
measurements to yield a $95\%$ upper limit of \mbox{$w<-0.78$} assuming
\mbox{$w\geq-1$}.  Although both our measurement and that of
\citet{spe03} include CMB data, they are complementary in that our limit
does not include the $H_0$ prior, nor does it include any of the same
external constraints, such as those from large scale structure.

These combined measurements remain consistent with a low density
universe dominated by vacuum energy (constant $w=-1$), but are also
consistent with a wide range of other both time-varying-$w$ and
constant-$w$ dark energy models.

\section{Systematic Errors}
\label{sec:systematic}

\ifthenelse{\boolean{ispreprint}}{
\begin{figure*}[htbp]
\begin{lrbox}{\thisbox}
\epsfig{file=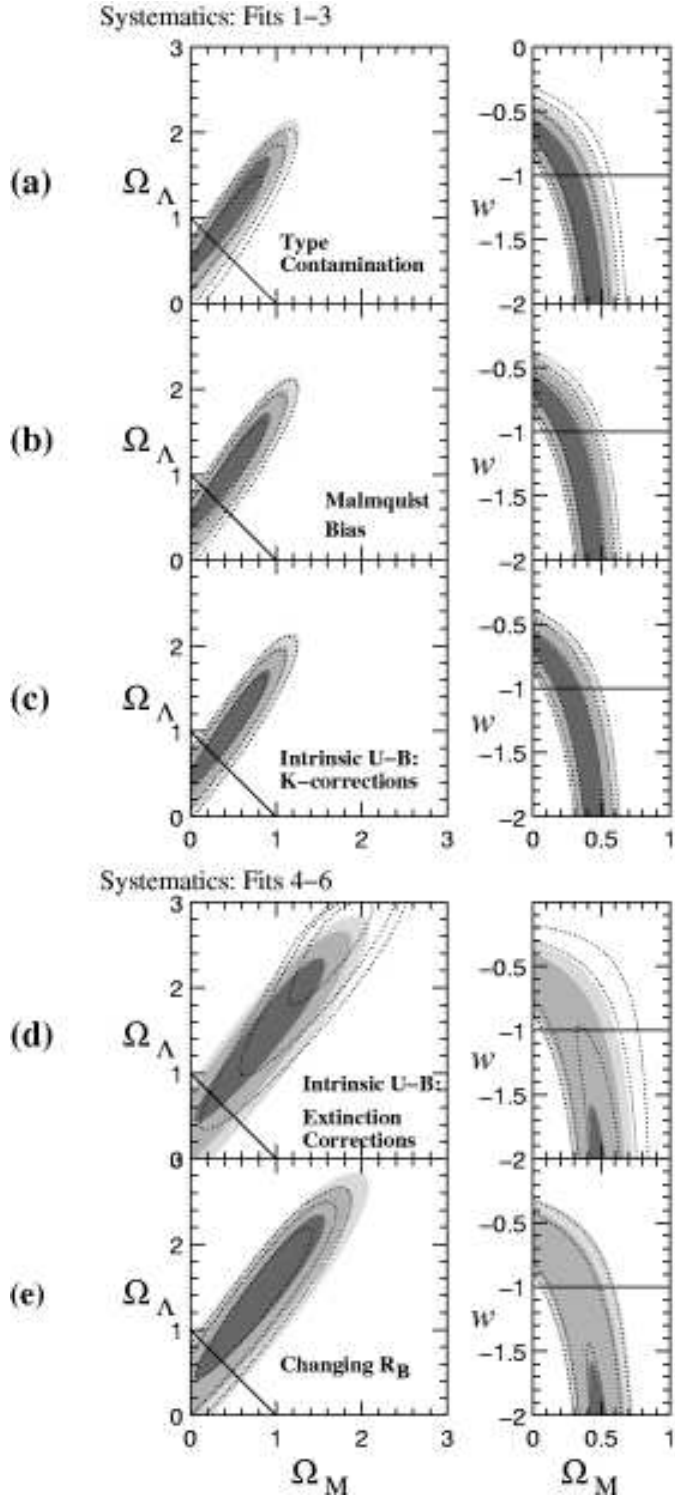, width=3.5in}
\end{lrbox}
\settowidth{\thiswid}{\usebox{\thisbox}}
\begin{center}
\begin{tabular}{b{\thiswid} b{2in}}
\usebox{\thisbox} &
\caption{\scriptsize Simulated effects of identified systematic errors
  on the cosmological parameters, estimated by applying the systematic
  effect to the supernova parameters used in the cosmological fits.  The
  left column shows fits to \om\ and \ol, and the right column to \om\
  and the dark energy equation of state parameter $w$.  Rows (a)--(c)
  show our primary fit (Fit~3) in filled contours.  (a) The dotted
  contours show the results of a fit to Subset~3, only those supernovae
  with the most secure spectral identifications as Type~Ia SNe.  (b) The
  dotted contours show a fit to Subset 1 where the supernova magnitudes
  have been dimmed to correct for Malmquist bias.  (c) The dotted
  contours show a fit to Subset 2, where $K$-corrections have been
  applied using a template spectrum with an intrinsic value of
  $U$-$B$=$-0.5$ at the epoch of B-maximum.  (d) The filled contours is
  Fit~6, the fit to the full primary subset with host-galaxy extinction
  corrections applied; the dotted contours show a fit to the same
  Subset, but using a template spectrum with an intrinsic value of
  $U$-$B$=$-0.5$ for estimating both $K$-corrections and color excesses.
  (e) The dotted contours apply extinction corrections to Subset 1 using
  a value of $R_B=3.5$ rather than the standard $R_B=4.1$ which was used
  for Fit~6 (filled contours).}
\label{fig:systematic} \\
\end{tabular}
\end{center}
\end{figure*}
}{\placefigure{fig:systematic}}

The effect of most systematic errors in the \om\ vs. \ol\ plane is
asymmetric in a manner similar to the asymmetry of our statistical
errors.  For the effects listed below, a systematic difference will tend
to move the confidence ellipses primarily along their major axis.  In
other words, these systematic effects produce a larger uncertainty in
$\om+\ol$ than in $\om-\ol$ (or, equivalently, in a measurement of $\om$
or $\ol$ alone under the assumption of a flat universe).  This means
that systematic effects do not currently hamper the cosmological
measurements from supernovae where they have the greatest weight
relative to other techniques, nor do they significantly diminish the
direct evidence from supernovae for the presence of dark energy.
However, they do limit the ability of supernovae to measure the spatial
curvature (``geometry'') of the Universe.  (Note that the semi-major
axis is not precisely in the direction of $\om+\ol$, nor is the
semi-minor axis precisely aligned with $\om-\ol$, but since these are
useful constraints we will quantify the systematic uncertainties along
these two directions.)  Figure~\ref{fig:systematic} shows the effects of
some of the systematics discussed in the following subsections.

Systematic effects on flat-universe measurements of $w$ are
smaller than the current statistical uncertainties.  The right column of
Figure~\ref{fig:systematic} shows the effect of the systematics on the
$\om$ versus $w$ confidence regions derived from our supernova data
alone.  To quantify the effect of identified systematics in the
following subsections, we determine the shift in the maximum-likelihood
value of $w$ when the supernova data is combined with the $\om$ versus
$w$ confidence regions from 2dFGRS and the CMB (See \S~\ref{sec:w}.)

\subsection{Fit Method, Subset Selection, and Choice of $\alpha$}
\label{sec:fitmethodsystematic}

There are multiple reasonable choices for lightcurve fitting methods
which yield slightly different results for the lightcurve parameters.
For the supernovae in P99, the $R$-band data on high-redshift supernovae
provided much stronger limits on the stretch (the shape of the
lightcurve) than did more sparse $I$-band lightcurves.  For consistency,
in P99 the stretch values for the low-redshift supernovae were therefore
measured using only the $B$-band lightcurves.

In this paper, there are high-quality photometric measurements from
\wfpc\ in both $R$ and $I$ bands.  Thus, data in both colors contribute
significantly to the constraints on stretch.  Additionally, photometry
is extracted from HST and ground-based images in very different
apertures, meaning that different amounts of host galaxy light will be
included; this background must be subtracted from each before the two
are combined.  As such, it is more appropriate to fit these supernovae
with fixed rather than floating lightcurve zero offsets.  As this is the
most appropriate fit method for the HST supernovae, the low-redshift
supernovae should be treated consistently.  These procedures which are
most appropriate for the HST supernovae were used for all new fits
performed in this paper, and listed in Tables~\ref{tab:hstsnefits}
through \ref{tab:lowzsnefits}.

To estimate the size of the effect due to these differences in fitting
method, cosmological confidence intervals were generated from the ``Case
C'' subset of P99 using the new fits presented in this paper and
compared to the results quoted in P99 and other variations on the
fitting method.  Differences in the fit method can change the
flat-universe value of \om\ by $\sim$0.03, and the value of $\om+\ol$ by
up to $\sim$0.8.  (This is still much less than the major-axis extent of
the statistical confidence ellipse in this direction.)  We use these
values as the ``fit-method'' systematic uncertainties.  We similarly
performed joint fits to $\om,w$ in the flat-universe, constant-$w$ case
to the supernovae from P99 with different lightcurve fit methodologies,
and from these fits we adopt a fit-method systematic uncertainty of 0.02
on constant $w$ (once combined with measurements from 2dFGRS and the
CMB).

We have also performed a fit without any stretch correction at all,
i.e. using fixed $\alpha=0$.  Although the quality of the fit is worse
($\chi^2=82$ with 54 supernovae, in comparison to $\chi^2=60$ from
Fit~3), it yields consistent cosmological results, with shifts
(\mbox{$\Delta\om^\mathrm{flat}<0.01$}) much smaller than the
already-adopted ``fit method'' systematic.  We have likewise performed a
fit to the complete set of supernovae (including all from P99 with
measured colors).  The fit cosmological values are similarly consistent
with the primary low-extinction fit.  We therefore conclude that the
effects of these choices are subsumed in the ``fit method'' systematic.

\subsection{Non-Type Ia Supernova Contamination}
\label{sec:typecontamination}

All subsets of supernovae used for cosmological fits in this paper omit
supernovae for which there is not a spectral confirmation of the
supernova type.  Nonetheless, it is possible in some cases where that
confirmation is weak that we may have contamination from non-Type~Ia
supernovae.  To estimate such an effect, we performed fits using only
those supernovae which have a firm identification as Type~Ia; this is
the ``strict-Ia subset'' from \S~\ref{sec:subsets}.  The comparison
between our primary fit (Fit 3) and this fit with a more stringent type
cut is shown in row (a) of Figure~\ref{fig:systematic}.  This fit has a
value of \om\ in a flat universe which is 0.03 higher than that of Fit
3.  The value of $\om+\ol$ is 0.48 lower than that of Fit 3.  We adopt
these values as our ``type contamination'' systematic error.

The size of this systematic for $w$ is shown in the right panel of
Figure~\ref{fig:systematic}a.  Combined with CMB and 2dFGRS
measurements, the best-fit value of $w$ is larger by $0.07$; we adopt
this as our type contamination systematic error on $w$.

\subsection{Malmquist Bias}
\label{sec:malmquist}

As most of our supernovae are from flux-limited samples, they will
suffer Malmquist bias \citep{malmquist24, malmquist36}. This effect was
discussed extensively in P99, and here we update that discussion to
include our new HST SNe~Ia.  For the measurement of the cosmological
parameters, it is the difference between the Malmquist bias of the
low-redshift and high-redshift samples which matters. In particular, the
apparent probability of $\ol>0$ is enhanced only if the low-redshift
supernovae suffer more Malmquist bias than the high-redshift supernovae,
as this makes the high-redshift SNe~Ia seem fainter.

The P99 high-redshift dataset was estimated to have little Malmquist
bias (0.01~mag) because the SN discovery magnitudes were decorrelated
with the measured peak magnitudes.  However, for the new HST sample,
nine of the eleven SNe~Ia (selected from larger samples of
supernovae found in the searches) were found almost exactly at maximum
light. This may reflect a spectroscopic flux limit superimposed on the
original search flux limit since only spectroscopically confirmed SNe~Ia
were considered, and of those, generally the higher redshift SNe~Ia from
a given search were chosen for HST for follow-up.  In particular, the
SNe~Ia selected for follow-up from the fall 1997 search were all found
at maximum light, while all but SN~1998aw from the spring 1998 search
were found at maximum light. SN~2000fr was found well before
maximum. Thus, the new high-redshift dataset is likely to suffer more
Malmquist bias than the P99 dataset.  Further complicating the
interpretation for the high-redshift supernovae is the fact that our new
HST supernovae are spread over a wide range in redshift, such that a
single brightness correction for Malmquist bias causes a more
complicated change in the fitted cosmological parameters. This is unlike
the situation in P99 in which most supernovae were at $z\sim0.5$.
Following the calculation in P99 for a high-redshift flux-limited SN
sample we estimate that the maximum Malmquist bias for the ensemble of
HST supernovae is $\sim0.03$~mag. However, we caution that it is
supernovae near the flux limit which are most strongly biased, and
therefore, that a subsample comprised of the highest-redshift members
drawn from a larger flux-limited sample will be more biased.  When
combined with the P99 high-redshift supernovae, the bias is likely to be
$\sim0.02$~mag since both samples have roughly the same statistical
weight.

As for the low-redshift SNe~Ia, in P99 we established that since most of
the SNe~Ia from the H96 flux-limited search were found near maximum,
that sample suffered about 0.04~mag of Malmquist bias.  On the other
hand, some of the R99 SNe~Ia were discovered using a galaxy-targeted
technique, which therefore is not limited by the SN flux and may be more
akin to a volume-limited sample \citep{li01}.  Thus, the addition of the
R99 SNe~Ia could slightly reduce the overall Malmquist bias of the
low-redshift sample. If we were to assume no Malmquist bias for the R99
SNe~Ia, and allowing for the fact that they contribute only $\sim1/4$
the statistical weight of the H96 supernovae, we estimate that the
Malmquist bias in the current low-redshift sample is roughly 0.03~mag.

Since Malmquist bias results in the selection of overly-bright
supernovae at the limits of a flux-limited survey, and since the
flux-limit can be strongly correlated with redshift\footnote{They are
100\% correlated for a single field, but this correlation can be diluted
by combining fields of different depths.}, this bias can result in an
apparent distortion of the shape of the Hubble diagram.  This may affect
estimates of the dark energy equation of state. The selection effects
for the current high-redshift supernovae are not sufficiently
well-defined to warrant a more detailed modeling of this effect than is
presented here. However, for future work, much better control of the
selection criteria for SNe~Ia at both low- and high-redshift will be
required in order to properly estimate the impact of this small bias.

For the current study, however, we simply note that since the
\emph{differences} in the Malmquist biases of the high- and low-redshift
subsets of SN are likely to be \emph{smaller} in this work than in P99,
the current results are less likely to be affected by Malmquist bias.
Given the above estimates of 0.03~mag of bias in the low-redshift
sample, and 0.02~mag of bias in the high-redshift sample, the
\emph{difference} in the biases is only 0.01~mag.  To perform a
quantitative estimate of the effects of Malmquist bias, we have
performed a fit by applying the mean offsets described above to each
member of a sample in our primary subset.  This fit is plotted in
Figure~\ref{fig:systematic}b.  The H96 supernovae have their magnitudes
increased (made dimmer) by 0.04, the P99 supernovae by 0.01, and six of
the eight HST supernovae in our primary subset have their magnitudes
increased by 0.04.  The two HST supernovae (\mbox{SNe\,1998bi}, and
2000fr) which were found before maximum light are assumed not to be
biased, and the other nine are offset by 0.04, yielding the above
estimated 0.03 magnitudes for the sample.  A fit with these changed
values to the supernova peak magnitudes yields a flat-universe value
which is different from our primary fit by $\om=0.01$, and a value of
\mbox{$\om+\ol$} which is different by $0.18$.  The best-fit value of
$w$, when combined with the other cosmological measurements, is $0.03$
larger.  We adopt these values---all much less than our statistical
uncertainties---as our Malmquist bias systematic error.

\subsection{$K$-corrections and Supernova Colors}
\label{sec:systematiccolor}

The generation of the spectral template used for calculating
$K$-corrections is described in \S~\ref{sec:colorcor}.  The degree to
which uncertainties in the $K$-correction introduce systematic
uncertainties into the cosmological parameters depends on whether or not
extinction corrections are being individually applied to supernovae.  In
particular, our $K$-corrections are most uncertain in the rest-frame
$U$-band range of the supernova spectrum, due to limited published
spectrophotometry.  As discussed in \S~\ref{sec:lightcurvefits}, our
primary fits use a spectral template which has a color
\mbox{$U$-$B$=$-0.4$} at the epoch of $B$-maximum.  We have investigated
the effects on our cosmology of replacing the spectral template used
both for $K$-corrections and for determining color excesses with a
template that has \mbox{$U$-$B$=$-0.5$} at the epoch of maximum $B$
light.

Figure~\ref{fig:systematic}c shows the effect on the fitted cosmology
caused by using the different template for calculating $K$-corrections
when individual host-galaxy extinction corrections are not applied.
These effects are very mild, indicating that our $K$-corrections are
robust with respect to the intrinsic $U$-$B$ color of a supernova.
Based on the comparison of these fits, we adopt a $K$-correction
systematic uncertainty of 0.13 on $\om+\ol$ and of 0.01 in $w$; the
systematic uncertainty on the flat-universe value of \om\ due to this
effect is negligible.

Although the effects of a different intrinsic $U$-$B$ color on the
$K$-corrections are mild, the effects on calculated color excesses are
much greater.  Figure~\ref{fig:systematic}d shows the difference between
Fit~6, where host-galaxy extinction corrections have been applied using
our standard color-excess values, and a fit where color-excess values
have been determined assuming the intrinsic $U$-$B$ color of a supernova
is $-0.5$ at maximum light.  As with other systematics, the primary
effect is to move the confidence intervals along their major axis.  In
this case, the large shift in $\om+\ol$ is mainly due to the fact that
with this bluer reference $U$-$B$ color, we would believe that all of
our $z>0.7$ supernovae are suffering from an amount of host-galaxy
extinction which is greater than that suffered by supernovae at lower
redshift.  Given that the more distant supernovae are dimmer and thus
closer to our detection limits than the moderate redshift supernovae,
this scenario is implausible.  If anything, one would expect the higher
redshift supernovae to be \emph{less} subject to host-galaxy extinction
due to selection effects.  Nonetheless, a value of $U$-$B$=$-0.5$ at the
epoch of $B$-band maximum is currently possible given the $U$-band
information available.  Only for those fits where extinction corrections
are applied, we have an additional intrinsic $U$-$B$ systematic error of
0.07 on the flat-universe value of \om, and a systematic error of 1.78
on $\om+\ol$.  The systematic uncertainty on $w$ is 0.10.  It is likely
that these values represent an overestimate of this systematic.

\subsection{Dust Properties}
\label{sec:dustevolution}

As discussed in \S~\ref{sec:colorsandextinction}, \citet{phi99} found
that some of the reddest supernovae at low redshift appear to be
overcorrected for extinction given the standard reddening law.  As
shown in the lower panel of Figure~\ref{fig:hubs}, our most reddened
high-redshift supernova (SN\,1998as, which is omitted from the primary
subset) is similarly overcorrected.  One possible explanation is that a
lower value of $R_B$ is appropriate for SN~Ia host galaxies.  If we use
a value of $R_B=3.5$ \citep{phi99} rather than the standard value of
$R_B=4.1$ to perform extinction corrections, it slightly changes the
best-fit cosmological values for fits where extinction correction are
applied (Fit 6); this change is shown in Figure~\ref{fig:systematic}e.
The best-fit value of $\om+\ol$ changes by 0.18, and the best-fit value
of $w$ when combined with the other cosmological measurements changes by
0.01; this systematic has a negligible effect on the flat-universe value
of \om.

A related source of systematic error is possible evolution in the
properties of the host-galaxy dust.  To examine the scale of the effect, we
consider a situation where dust in \mbox{$z<0.3$} spiral galaxies has
a \citet{car89} $R_V=3.1$ law whereas higher-redshift galaxy dust has a
ratio of selective-to-total extinction that is half as large,
i.e. $R_V=1.6$.  We use the Monte Carlo code described in \citet{kim03}
to study the bias induced when an $R_V=3.1$ extinction correction is
inappropriately applied to all supernovae.  We incorporate the redshift
and \ebv\ distributions of the supernovae considered in this paper and
an \mbox{$\ebv<0.1$} cut is applied.  For an input cosmology of
$\om=0.21$ and $\ol=0.79$, we find a modest shift in the cosmological
parameters to $\om=0.25$ and $\ol=0.77$ without assuming a flat
universe.

This bias moves almost exactly along the line
$\Omega_M+\Omega_\Lambda=1$, increasing uncertainty along the thin axis
of the error contour.  However, the extreme difference in dust
properties considered in the Monte Carlo contributes a shift in the
cosmological parameters that is less than 1 $\sigma$ of our quoted
statistical error bars.  We adopt 0.04 as the ``dust evolution''
systematic uncertainty on $\om$ in a flat universe for those fits where
host-galaxy extinction corrections are applied; this particular
systematic is insignificant along the major axis of the confidence
ellipses.

The flat-universe value of $w$, when combined with the 2dFGRS and CMB
results, increases by 0.06 under this simple model of dust evolution.
We adopt this as the dust evolution systematic on $w$ for those fits
where host-galaxy extinction corrections are applied.

\subsection{Gravitational Lensing}

Gravitational lensing decreases the modal brightness and causes
increased dispersion and positive skewness in the Hubble diagram for
high-redshift supernovae.  The size of the effect depends on the
fraction of compact objects of the total mass density of the universe,
\om.  This has been discussed in some detail in the literature
\citep{wcx97,frie97, holz98, kan98, seljak99, ms99, metcalf99, holz01,
wang02, mhh02, amm03, dhcf03, ost03}, especially in relation to the P99
and \citet{rie98} SN datasets.  A very conservative assumption of an
``empty beam'' model in a universe filled with compact objects allowed
P99 to demonstrate that gravitational lensing does not alter the case
for dark energy.

Gravitational lensing may result in a biased determination of the
cosmological parameter determination, as discussed in \citet{amm03}.
The potential bias increases with the redshift of the supernovae in the
sample.  For example, for the most distant known Type~Ia SN, SN1997ff at
z=1.7, there is evidence for significant magnification, $\Delta m \sim
-0.3$ \citep{lew01,mor01,ben02}.

As the SN sample considered in this paper does not reach as far, the
(de)magnification distortions are expected to be small, in general below
0.05 magnitudes, and less than 1\% for the cases considered in P99. To
estimate the systematic uncertainties in the cosmological parameters we
have used the SNOC package \citep{goo01} to simulate 100 realizations of
our data sets assuming a 20\% universal fraction of $\Omega_M$ in
compact objects, i.e.  of the same order as the halo fraction deduced
for the Milky Way from microlensing along the line of sight to the Large
Magellanic Cloud \citep{alc00}. The light beams are otherwise assumed to
travel through space randomly filled with galaxy halos with mass density
equally divided into SIS and NFW profiles, as described in
\citet{ber00}. According to our simulations we find that (for a flat
universe) the fitted value of $\Omega_M$ is systematically shifted by
0.01 on the average, with a statistical dispersion $\sigma_{\Delta\om} =
0.01$.  We adopt 0.01 as our gravitational lensing systematic error in
the flat-universe value of \om.  The effect on $\om+\ol$ is very small
compared to other systematics, biasing the sum by only 0.04.

The simulated offsets due to gravitational lensing, when combined with
CMB and galaxy redshift distortion measurements, increase the value of
$w$ by 0.05; we adopt this as a gravitational lensing systematic on $w$.

\subsection{Supernova Population Drift}

In P99 we discussed in detail whether the high-redshift SNe~Ia could
have systematically different properties than low-redshift SNe~Ia, and 
in particular, whether intrinsic differences might remain after
correction for stretch. One might imagine this to occur if the range of
the physical parameters controlling SN~Ia brightnesses have little
overlap between low- and high-redshift such that corrections applied to
low-redshift are inappropriate or incomplete for high-redshift SNe~Ia.
Since P99, considerable additional work has been done to address this
issue.

In addition to comparisons of stretch range (P99), as well as spectral
\citep{per98,coi00} and lightcurve \citep{gol01} features, several tests
performed directly with the P99 high-redshift SNe~Ia have shown
excellent consistency with low-redshift SNe~Ia.  Most recently, in
\citet{sul03} we have presented results on the Hubble diagram of distant
Type~Ia supernovae from P99 that have been morphologically-typed with
HST.  We found no difference in the cosmological results from their
morphologically-segregated subsamples.  In particular, E/S0
galaxies---for which one expects the tightest possible correlation
between progenitor mass and redshift---not only agree with the
cosmological fits using only spiral galaxies, but by themselves confirm
the results of P99. This is strong evidence that, while age or
metallicity could in principle affect the brightnesses of SNe~Ia,
stretch correction eliminates these differences. Likewise, the
lightcurve rise-time---a possible indicator of the energetics of the SN
explosion \citep[see][]{nug95,hof98}---while initially suggested to be
different between high- and low-redshift SNe~Ia \citep{rie99b}, has been
demonstrated to agree very well \citep[within $1.8\pm1.2$ days,
][]{ald00}.

On the theoretical side, the SN formation models of \citet{kob98} and
\citet{nom99} suggest that the progenitor binary system must have
\mbox{[Fe/H]$>-1$} in order to produce a SN~Ia. This would impose a
lower limit to the metallicities of all SNe~Ia, and thus limit the
extent of any metallicity-induced brightness differences between high-
and low-redshift SNe~Ia.  On the empirical side, the lack of a gradient
in the intrinsic luminosities of SNe~Ia with galactocentric distance,
coupled with the fact that metallicity gradients are common in spiral
galaxies \citep{hen99}, lead \citet{iva00} to suggest that metallicity
is not a key parameter in controlling SNe~Ia brightnesses at optical
wavelengths---though note that \citet{len00} show how it can affect the
ultraviolet.  In addition, \citet{ham00,ham01} find that lightcurve
width is not dependent on host-galaxy metallicity.

Alternatively, population age effects, including pre-explosion cooling
undergone by the progenitor white dwarf and other effects linked to the
mass of the primary exploding white dwarf have been suggested \citep[for
a review, see][]{rui03}.  As the local sample of SNe~Ia represents
populations of all ages and metallicities, both effects can be studied
locally.  Several low-redshift studies have presented data suggesting
that SNe~Ia intrinsic luminosities (i.e., those prior to stretch
correction) may correlate with host-galaxy environment
\citep[][R99]{ham96b,bra96,wan97,ham00,iva00,how01,wan03}.  These
findings are actually encouraging, since unlike stretch itself, there is
some hope that host-galaxy environment variations can be translated into
physical parameters such as age and metallicity.  These parameters can
help relate any drifts in the SNe~Ia population to evolution of the host
galaxies.

More importantly for cosmology, R99 used their sample of 22 local SNe~Ia
to demonstrate that any brightness variations between SNe~Ia in
different host-galaxy environments disappear after correction for
lightcurve width.  We have quantified this agreement using a larger
local sample of supernovae compiled in \citet{wan03}, 14 of which have
E/S0 hosts and 27 of which have spiral hosts.  We find that after
lightcurve-width correction there can be less than a $0.01\pm0.05$~mag
offset between SNe~Ia in local spirals and ellipticals. This indicates
that lightcurve width is able to correct for age or other differences.

Finally, \citet{wan03} demonstrate a new method, \emph{CMAGIC}, which is
able to standardize the vast majority of local SNe~Ia to within 0.08~mag
(in contrast to $\sim0.11$~mag which lightcurve-width corrections can
attain \citep{phi99}). This imposes even more severe limits on the
fraction of SNe~Ia generated by any alternate progenitor scenario, or
requires that variations in the progenitor properties have little effect
on whether the resulting SN can be standardized.

The data from the new SNe~Ia presented here do offer one new test for
consistency between low- and high-redshift SNe~Ia.  The quality of our
HST data provides measurements of the SN peak magnitudes and lightcurve
widths rivaling those for nearby SNe~Ia. This allows a direct comparison
between the stretch-luminosity relations at low- and high-redshifts.
Figure~\ref{fig:stretchplot} shows that the HST high-redshift supernovae
are found at similar stretches and luminosities as the low-redshift
supernovae.  The low- and high-redshift samples are consistent with the
same stretch-luminosity relationship, although it is primarily the
low-redshift supernovae that require a non-zero slope for this
relationship.

\ifthenelse{\boolean{ispreprint}}{
\begin{figure}[tb]
\begin{lrbox}{\thisbox}
\epsfig{file=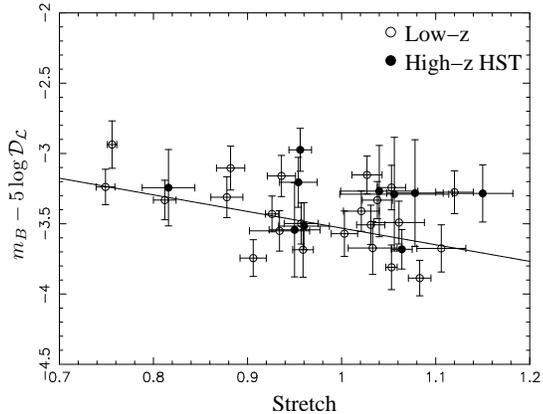 , width=2.8in}
\end{lrbox}
\settowidth{\thiswid}{\usebox{\thisbox}}
\begin{center}
\begin{minipage}{\thiswid}
\usebox{\thisbox}
\caption{Stretch-luminosity relationship for low-redshift SNe (open
  circles) and high-redshift HST SNe (filled circles).  Each point is
  the $K$-corrected and extinction-corrected $m_B$ for that supernova,
  minus $\mathcal{D_L}$, the ``Hubble-constant-free luminosity
  distance'' (see \S~\ref{sec:cosmofitmethod}), plotted against the
  stretch of that SN.  The line drawn represents the best-fit values of
  $\alpha$ and \scriptm\ from Fit~6, the fit to all Subset~1 supernovae
  with host-galaxy extinction corrections applied.  Note in particular
  that our HST SNe~Ia all have low-redshift counterparts.}
\label{fig:stretchplot}
\end{minipage}
\end{center}
\end{figure}
}{\placefigure{fig:stretchplot}}

\subsection{Possible Additional Sources of Systematic Uncertainties}

Other potential sources of systematic uncertainties have been suggested.
\citet{agu99a,agu99b} and \citet{agu00} argued that the presence of
``grey'' dust, i.e. a homogeneous intergalactic component with weak
differential extinction properties over the rest-frame optical
wavelength regime could not be ruled out by the P99 data.  Since then,
measurements of a SN~Ia at $z\simeq1.7$ \citep{rie01} were claimed to
rule out the ``grey'' dust scenario as a non-cosmological alternative
explanation to the dimming of high-redshift supernovae; however, there
remain some outstanding issues with this interpretation
\citep[e.g.,][]{goo02,bla03}.  A direct test for extinction over a wide
wavelength range, rest-frame B-I, have been performed by \citet{rie00}
on a single supernova at \mbox{$z=0.46$}, \mbox{SN\,1999Q}, which showed
no grey dust signature; however, see \citet{nob03}.  Although the
situation remains inconclusive, there is no direct evidence that
``grey'' dust is a dominant source of uncertainties.  It remains an
important issue to be addressed by future data sets including
near-infrared observations.

More recently, the possibility of axion-photon oscillations making
high-redshift supernovae appear dimmer was suggested by
\citet{csa02}.  This attenuation would be wavelength dependent, and thus
could be explored with spectroscopic studies of high-shift sources
\citep{mor02}.  Preliminary studies of QSO spectra between $z=0.15$ and
$z=5.3$ set a very conservative upper limit on the possible dimming of
z$\sim$0.8 supernovae to 0.2 magnitudes \citep{mor03}

For the current data sample, the above mentioned sources of systematic
uncertainties are difficult to quantify at present, but are believed to
be subdominant in the total error budget.

\subsection{Total Identified Systematic Uncertainty}

\ifthenelse{\boolean{ispreprint}}{
\begin{table*}[htbp]
\begin{lrbox}{\thisbox}
\begin{tabular}{lcccl}
\tableline
\tableline
\textbf{Source of}   &  \multicolumn{3}{c}{\underline{Systematic
    Uncertainty On:}} & Notes\\
\textbf{Uncertainty} &  Flat-Universe       &           &                & \\
                     &  \om or \ol$^a$      & $\om+\ol$ & constant $w^b$ & \\
\tableline  
Fit method            & 0.03  (0.5$\sigma$) & 0.80      & 0.02  & \\
Type contamination    & 0.03  (0.5$\sigma$) & 0.48      & 0.07  & \\
Malmquist Bias        & 0.01  (0.2$\sigma$) & 0.18      & 0.03  & \\
Intrinsic U-B: $K$-corrections
                      & 0.00  (0.0$\sigma$) & 0.13      & 0.01  & $c$ \\
Gravitational Lensing & 0.01  (0.2$\sigma$) & 0.04      & 0.05  & \\[12pt]

\multicolumn{5}{l}{\underline{Systematic with host-galaxy extinction corrections:}} \\[6pt]

Intrinsic U-B: color excess
                      & 0.07  (0.7$\sigma$) & 1.78      & 0.10  & $d$ \\
Extinction Slope      & 0.00  (0.0$\sigma$) & 0.18      & 0.01  & $d$ \\  
Dust Evolution        & 0.03  (0.3$\sigma$) & 0.02      & 0.06  & $d$ \\
\tableline
\end{tabular}
\end{lrbox}
\settowidth{\thiswid}{\usebox{\thisbox}}
\begin{center}
\begin{minipage}{\thiswid}
\caption{Identified Systematic Uncertainties}
\label{tab:systematic}
\usebox{\thisbox}

\footnotesize
$a$: Each systematic is given as an offset from the
flat-universe value of \om, and in terms of the smaller side of the
statistical error bar (0.06 for Fit 3 to the low-extinction subset, 0.10
for Fit 6 to the full primary subset).

$b$: This is the offset on the maximum-likelihood value of $w$ when the
the fit is combined with the 2dFGRS and CMB measurements.

$c$: Only used where host-galaxy extinction corrections are not applied.

$d$: Only used where host-galaxy extinction corrections are applied.
\end{minipage}
\end{center}
\end{table*}
}{\placetable{tab:systematic}}

The identified systematic errors are summarized in
Table~\ref{tab:systematic}.  Adding together these errors in quadrature,
we obtain a total systematic error of 0.04 on the flat-universe value of
$\om$ (along approximately the minor axis of the confidence ellipses
shown in \om\ vs. \ol\ plots); this is smaller than but approaching our
statistical uncertainty of 0.06.  The total systematic
uncertainty on \mbox{$\om+\ol$} is 0.96 (along approximately the major
axis of the confidence ellipses).  Finally, for the low-extinction
subset, we have a systematic uncertainty on constant $w$ of 0.09, less
than our high-side statistical uncertainty of 0.15.

For fits with host-galaxy extinction corrections applied, we have to
consider the additional systematic effects of an uncertainty in the
intrinsic value of $U$-$B$ on determined color excesses, and of dust
properties.  In this case, we have a total systematic error of 0.09 on
the flat-universe value of \om\ or \ol, and a total systematic error of
2.0 on $\om+\ol$; as discussed in \S~\ref{sec:systematiccolor}, this is
likely to be an overestimate of the true systematic error.  The total
systematic uncertainty on constant $w$ for the extinction-corrected full
primary sample is 0.15.


\section{Summary and Conclusions}

\begin{enumerate}

\item We present a new, independent set of eleven high-redshift supernovae
  (\mbox{$z=0.36$--$0.86$}).  These supernovae have very high-quality
  photometry measured with \wfpc\ on the HST.  The higher quality
  lightcurve measurements have small enough errors on each \ebv\
  measurement to allow an unbiased correction of host-galaxy reddening.
  We have performed improved color and $K$-corrections, necessary to
  combine \wfpc\ photometric filters with ground-based photometric
  filters.

\item The cosmological fits to $\om$ and $\ol$ are consistent with the
  SCP's previous results (P99), providing strong evidence for a
  cosmological constant.  This is a significant confirmation of the
  results of P99 and \citet{rie98}, and represents a completely new set
  of high-redshift supernovae yielding the same results as the earlier
  supernova work.  Moreover, these results are consistent with a number
  of other cosmological measurements, and together with other current
  cosmological observations is pointing towards a consensus
  \mbox{$\om\sim0.3$}, \mbox{$\ol\sim0.7$} Universe.

\item Most identified systematic errors on \om\ and \ol\ affect the
  cosmological results primarily by moving them along the direction
  where the statistical uncertainty is largest, that is, along the major
  axis of the confidence ellipses.  Systematics are much smaller along
  the minor (approximately $\om-\ol$) axis of the confidence regions,
  and may be described by giving the systematic error on \om\ or \ol\
  alone in the flat-universe case.  Our total identified systematic
  error for the low-extinction sample analysis is 0.04 on the
  flat-universe value of \om\ or \ol.  For fits with host-galaxy
  extinction corrections, a conservative estimate of the total
  identified systematic error is $0.09$.

  In the more uncertain major axis, our total identified systematic error
  is 0.96 on $\om+\ol$ for the low-extinction primary subset, and 2.0 on
  the extinction-corrected full primary subset.  Given the large size of
  these systematics in this direction, any conclusions drawn from the
  positions of supernova confidence ellipses along this direction should
  be approached with caution.

\item Under the assumption of a flat universe with vacuum energy
  (constant $w=-1$), we find a value of
  \mbox{$\om=0.25^{+0.07}_{-0.06}$} (statistical) $\pm0.04$ (identified
  systematic), or equivalently, a cosmological constant of
  \mbox{$\ol=0.75^{+0.06}_{-0.07}$} (statistical) $\pm0.04$ (identified
  systematic).  This result is robust to host-galaxy extinction, and a
  fit with full, unbiased, individual extinction corrections applied
  yields a flat-universe cosmological constant of
  \mbox{$\ol=0.72^{+0.10}_{-0.11}$} (statistical) $\pm0.09$ (identified
  systematic).  Our best confidence regions for \om\ versus \ol\ are
  shown in Figure~\ref{fig:primaryconfs}.

\item When combined with the 2dFGRS galaxy redshift distortion
  measurement and recent CMB data, we find a value for the dark energy
  equation of state parameter \mbox{$w=-1.05^{+0.15}_{-0.20}$}
  marginalizing over \om\ (or a mass
  density \mbox{$\om=0.27^{+0.06}_{-0.05}$} marginalizing over $w$), under
  the assumptions that the Universe is spatially flat and that $w$ is
  constant in time.  The identified systematic uncertainty on $w$ is
  0.09.  The current confidence regions on the flat-universe values of
  \om\ and $w$ are shown in Figure~\ref{fig:omw}.  The supernovae data
  are consistent with a low-mass Universe dominated by vacuum energy
  \mbox{($w=-1$)}, but they are also consistent with a wide range of
  constant or time-varying dark energy models.

\end{enumerate}

In summary, high-redshift supernovae continue to be the best single tool
for directly measuring the density of dark energy.  This new set of
supernovae observed with the HST confirm and strengthen previous
supernova evidence for an accelerating universe, and show that those
results are robust even when host-galaxy extinction is fully accounted
for.  High-redshift supernovae, together with other cosmological
measurements, are providing a consistent picture of a low-mass, flat
universe filled with dark energy.  The next task for cosmologists is to
better measure the properties of the dark energy, so as to further our
understanding of its nature.  Combinations of current cosmological
techniques have begun to provide measurements of its most general
property (specifically, the equation of state parameter when it is
assumed to be constant).  Future work will refine these measurements,
and in particular reduce the systematic uncertainties that will soon
limit the current series of supernova studies.  As new instruments
become available,\footnote{See, e.g., \url{http://snap.lbl.gov/}} it
will begin to be possible to relax the condition of a constant equation
of state parameter, and to question whether the properties of the dark
energy have been changing throughout the history of the Universe.

\acknowledgements

The authors wish to thank our \hst\ program coordinator, Doug Van Orsow
and the excellent HST support staff for their help in the planning,
scheduling, and execution of the observations presented herein.  Support
for this work was provided by NASA through grants HST-GO-07336.01-A and
HST-GO-08346.01-A from the Space Telescope Science Institute, which is
operated by the Association of Universities for Research in Astronomy,
Inc., under NASA contract NAS 5-26555.  The authors are indebted to
Drs. Malcolm Smith and Patrick Hall for trading several crucial hours of
observing time at the CTIO 4m, which played a key role in our SN search
in March 1998.  The authors acknowledge the tremendous help of the night
assistants and support staff at the many telescopes from which data for
this paper were obtained; we are particularly grateful to the CTIO staff
for crucial support during our key search nights, and to Di Harmer and
Paul Smith of the WIYN Queue.  We thank Gary Bernstein and Tony Tyson
for developing and supporting the Big Throughput Camera at the CTIO 4m.
This wide-field camera was important in the discovery of most of the
high-redshift supernovae presented in this paper, and enabled the high
discovery rate needed to guarantee supernovae for follow-up with \hst.
The authors are grateful to Eric Linder for the use of his
growth-parameter solver, and to Ramon Miguel for assistance with
gravitational lensing calculations.  We also wish to acknowledge NOAO
for providing and supporting the astronomical data reduction package
IRAF.  The authors wish to recognize and acknowledge the very
significant cultural role and reverence that the summit of Mauna Kea has
always had within the indigenous Hawaiian community.  We are most
fortunate to have the opportunity to conduct observations from this
mountain. This work was supported in part by the Director, Office of
Science, Office of High Energy and Nuclear Physics, of the
U.S. Department of Energy under Contract No. DE-AC03-76SF000098, by the
Center for Particle Astrophysics, an NSF Science and Technology Center
operated by the University of California, Berkeley, under Cooperative
Agreement No.  AST-91-20005.  This work was supported in part by a NASA
LTSA grant to PEN, GA, SP, and SED, and WMWV was supported in part by a
National Science Foundation Graduate Research Fellowship.  A. Goobar is
a Royal Swedish Academy Research Fellow supported by a grant from the
Knut and Alice Wallenberg Foundation.

\appendix
\section{Lightcurve Data}
\label{sec:dirtylaundry}

Tabulated below are lightcurve data for the eleven HST supernovae
presented in this paper.  For each event, there are two lightcurves, one
for $R$-band and one for $I$-band.  All photometry has been
color-corrected to the standard Bessel filters as described in
\S~\ref{sec:colorsandextinction}, using color corrections which assume
the lightcurve parameters in Table~\ref{tab:hstsnefits}.  These
lightcurves, together with a $7''\times7''$ thumbnail of the F675W
\wfpc\ image closest to maximum light, are shown in
Figures~\ref{fig:lightcurves1} and \ref{fig:lightcurves2}.  Note that
there are correlated errors between the data points.  For the
ground-based data, there is a covariance because for a given supernova
the same final reference images were subtracted from all other
ground-based points.  Similarly, the HST data include a covariance due
to a single background model having been used for all points for a given
supernova (see \S~\ref{sec:hstphotometry}).  In addition to this, the
relative photometric zeropoint magnitudes were determined separately for
the ground-based and HST photometry; in the former case, standard stars
from \citet{lan92} were used to measure magnitudes of secondary standard
stars in the supernova field of view.  In the latter case, zeropoints
from \citet{dol00} were used.  These covariance matrices will be
available from the SCP
website.\footnote{\url{http://supernova.lbl.gov/}}

Because uncertainties are flux uncertainties rather than magnitude
uncertainties, each lightcurve is presented in arbitrary flux units.
For each lightcurve, the zeropoint necessary to convert these to
magnitudes is given.  The magnitude may be calculated using the standard
formula:
\begin{equation}
m\ =\ -2.5\log{f}\ +\ m_{zp}
\end{equation}
where $m_{zp}$ is the quoted zeropoint and $f$ is the flux value from
the table.  (Because we include early-time and late-time lightcurve
points when the supernova flux is undetected given our photometry
errors, some of the measured fluxes scatter to negative values.  Note
that it is impossible to formally calculate a magnitude for these
points, and also that flux values are the proper way to quote the data as
they better reflect the units in which our photometry errors are
approximately Gaussian.)

The telescope used for each data point is indicated.  BTC = the Big
Throughput Camera on the CTIO~4m telescope.  CTIO = the prime focus
imager on the CTIO~4m telescope.  WIYN = the Nasmyth 2k$\times$2k
imager on the WIYN~3.5m telescope at Kitt Peak observatory.  INT = the
WFC (wide-field camera) on the INT 2.5m telescope at La Palma.  KECK
= the LRIS imager on the Keck 10m telescope.  NTT = the SUSI-2 imager on
the NTT 3.6m telescope at ESO.  CFHT = the CFHT12K multi-chip imager on
the 3.6m CFHT telescope on Mauna Kea in Hawaii.  Finally, HSTPC
indicates data obtained from the Planetary Camera CCD on \wfpc.

\ifthenelse{\boolean{ispreprint}}{
\twocolumn

\begin{table}[H]
\scriptsize\renewcommand{\arraystretch}{1.0}
\begin{lrbox}{\thisbox}
\begin{tabular}{rrl}
\tableline
\tableline
Julian Day & Flux$^a$ & Telescope \\
-2,400,000 \\
\tableline
50780.63 & $ 0.24\pm1.27$ & BTC \\
50780.69 & $ 0.57\pm0.93$ & BTC \\
50781.61 & $-0.28\pm1.05$ & BTC \\
50781.66 & $ 1.22\pm0.89$ & BTC \\
50781.67 & $ 0.29\pm0.89$ & BTC \\
50781.72 & $ 0.16\pm1.01$ & BTC \\
50810.58 & $ 2.71\pm1.28$ & BTC \\
50810.59 & $ 4.63\pm1.29$ & BTC \\
50810.60 & $ 5.25\pm1.24$ & BTC \\
50810.67 & $ 4.85\pm1.32$ & BTC \\
50810.68 & $ 5.04\pm1.24$ & BTC \\
50810.69 & $ 5.70\pm1.28$ & BTC \\
50811.66 & $ 4.34\pm1.10$ & BTC \\
50811.68 & $ 4.53\pm1.07$ & BTC \\
50811.69 & $ 3.55\pm1.22$ & BTC \\
50817.67 & $ 4.92\pm0.91$ & BTC \\
50817.68 & $ 5.09\pm0.84$ & BTC \\
50817.69 & $ 3.17\pm0.83$ & BTC \\
50817.70 & $ 2.65\pm0.84$ & BTC \\
50817.71 & $ 3.71\pm0.85$ & BTC \\
50817.72 & $ 3.34\pm1.02$ & BTC \\
50817.73 & $ 4.45\pm1.06$ & BTC \\
50817.73 & $ 4.77\pm1.04$ & BTC \\
50817.74 & $ 3.10\pm1.04$ & BTC \\
50818.92 & $ 4.18\pm0.23$ & HSTPC \\
50824.77 & $ 3.61\pm0.21$ & HSTPC \\
50835.67 & $ 2.49\pm0.87$ & BTC \\
50835.68 & $ 3.20\pm0.90$ & BTC \\
50835.69 & $ 2.56\pm0.99$ & BTC \\
50835.70 & $ 3.01\pm1.05$ & BTC \\
50835.70 & $ 3.26\pm1.12$ & BTC \\
51165.71 & $-0.05\pm0.60$ & BTC \\
51165.71 & $-0.67\pm0.61$ & BTC \\
51165.74 & $-0.55\pm0.71$ & BTC \\
51166.63 & $ 0.44\pm2.12$ & BTC \\
51166.65 & $ 1.20\pm1.28$ & BTC \\
51166.66 & $-0.67\pm1.49$ & BTC \\
51193.59 & $ 0.47\pm0.77$ & BTC \\
51193.60 & $-0.86\pm0.79$ & BTC \\
51193.61 & $ 0.76\pm0.70$ & BTC \\
51193.62 & $ 0.18\pm0.73$ & BTC \\
51194.65 & $ 0.46\pm0.64$ & BTC \\
\tableline
\end{tabular}
\end{lrbox}
\settowidth{\thiswid}{\usebox{\thisbox}}
\begin{center}
\begin{minipage}{\thiswid}
\caption{SN\,1997ek-R}
\label{tab:sn97201r}
\usebox{\thisbox}

$a$: Zeropoint: 25.678
\end{minipage}
\end{center}
\end{table}

\begin{table}[H]
\scriptsize\renewcommand{\arraystretch}{1.0}
\begin{lrbox}{\thisbox}
\begin{tabular}{rrl}
\tableline
\tableline
Julian Day & Flux$^a$ & Telescope \\
-2,400,000 \\
\tableline
50816.60 & $ 5.62\pm1.45$ & BTC \\
50817.56 & $ 3.22\pm1.30$ & BTC \\
50817.57 & $ 4.27\pm1.35$ & BTC \\
50817.58 & $ 4.70\pm1.40$ & BTC \\
50817.58 & $ 5.41\pm1.43$ & BTC \\
50817.59 & $ 5.82\pm1.36$ & BTC \\
50817.60 & $ 4.47\pm1.66$ & BTC \\
50817.61 & $ 5.16\pm1.52$ & BTC \\
50817.63 & $ 3.68\pm1.52$ & BTC \\
50817.64 & $ 4.48\pm1.48$ & BTC \\
50817.64 & $ 3.31\pm1.59$ & BTC \\
50817.65 & $ 5.89\pm1.23$ & BTC \\
50817.66 & $ 4.38\pm1.44$ & BTC \\
50818.93 & $ 3.83\pm0.16$ & HSTPC \\
50819.74 & $ 2.02\pm1.70$ & WIYN \\
50819.76 & $ 3.05\pm1.65$ & WIYN \\
50819.78 & $ 4.18\pm1.90$ & WIYN \\
50819.79 & $ 1.71\pm1.60$ & WIYN \\
50819.81 & $ 4.31\pm1.57$ & WIYN \\
50819.82 & $ 3.84\pm2.09$ & WIYN \\
50824.78 & $ 3.89\pm0.16$ & HSTPC \\
50835.72 & $ 2.72\pm1.96$ & BTC \\
50835.73 & $ 3.06\pm2.05$ & BTC \\
50846.74 & $ 1.54\pm0.09$ & HSTPC \\
50858.84 & $ 0.75\pm0.07$ & HSTPC \\
50871.95 & $ 0.46\pm0.06$ & HSTPC \\
51072.07 & $ 0.50\pm0.57$ & KECK \\
51072.07 & $ 0.35\pm0.58$ & KECK \\
51072.07 & $ 0.69\pm0.58$ & KECK \\
51072.11 & $ 0.31\pm0.55$ & KECK \\
51072.11 & $ 0.94\pm0.58$ & KECK \\
51072.12 & $-0.23\pm0.57$ & KECK \\
51101.99 & $-0.37\pm0.54$ & KECK \\
51102.00 & $ 0.51\pm0.58$ & KECK \\
51102.00 & $ 0.58\pm0.59$ & KECK \\
51102.05 & $ 1.20\pm0.75$ & KECK \\
51102.06 & $ 1.53\pm0.90$ & KECK \\
51126.93 & $-0.04\pm0.06$ & HSTPC \\
51134.26 & $ 0.06\pm0.05$ & HSTPC \\
51165.70 & $-0.66\pm1.15$ & BTC \\
51165.72 & $ 0.21\pm1.06$ & BTC \\
51165.73 & $-0.44\pm1.12$ & BTC \\
51193.64 & $ 0.01\pm1.12$ & BTC \\
51193.65 & $-0.28\pm1.13$ & BTC \\
51193.67 & $-0.46\pm1.50$ & BTC \\
51194.59 & $ 0.99\pm1.17$ & BTC \\
51194.60 & $ 1.34\pm1.30$ & BTC \\
51194.60 & $ 0.73\pm1.15$ & BTC \\
\tableline
\end{tabular}
\end{lrbox}
\settowidth{\thiswid}{\usebox{\thisbox}}
\begin{center}
\begin{minipage}{\thiswid}
\caption{SN\,1997ek-I}
\label{tab:sn97201i}
\usebox{\thisbox}

$a$: Zeropoint: 24.801
\end{minipage}
\end{center}
\end{table}

\clearpage
\begin{table}[H]
\scriptsize\renewcommand{\arraystretch}{1.0}
\begin{lrbox}{\thisbox}
\begin{tabular}{rrl}
\tableline
\tableline
Julian Day & Flux$^a$ & Telescope \\
-2,400,000 \\
\tableline
50780.60 & $ 0.01\pm0.12$ & BTC \\
50780.66 & $ 0.21\pm0.12$ & BTC \\
50781.60 & $-0.08\pm0.10$ & BTC \\
50781.63 & $ 0.19\pm0.10$ & BTC \\
50781.68 & $ 0.09\pm0.10$ & BTC \\
50781.72 & $ 0.14\pm0.11$ & BTC \\
50810.61 & $ 1.76\pm0.12$ & BTC \\
50810.62 & $ 1.80\pm0.12$ & BTC \\
50810.63 & $ 1.88\pm0.13$ & BTC \\
50810.64 & $ 1.87\pm0.11$ & BTC \\
50810.70 & $ 1.91\pm0.12$ & BTC \\
50810.71 & $ 1.82\pm0.11$ & BTC \\
50811.70 & $ 1.78\pm0.10$ & BTC \\
50818.34 & $ 2.23\pm0.28$ & INT \\
50818.36 & $ 1.98\pm0.24$ & INT \\
50819.85 & $ 1.69\pm0.05$ & HSTPC \\
50821.66 & $ 2.14\pm0.54$ & WIYN \\
50821.67 & $ 1.79\pm0.39$ & WIYN \\
50835.41 & $ 0.85\pm0.13$ & INT \\
50835.42 & $ 0.87\pm0.18$ & INT \\
50835.43 & $ 0.85\pm0.34$ & INT \\
50843.68 & $ 0.37\pm0.18$ & WIYN \\
50843.70 & $ 0.02\pm0.40$ & WIYN \\
50846.81 & $ 0.32\pm0.02$ & HSTPC \\
50855.82 & $ 0.18\pm0.02$ & HSTPC \\
50863.82 & $ 0.12\pm0.02$ & HSTPC \\
51165.56 & $ 0.01\pm0.12$ & BTC \\
51165.61 & $ 0.01\pm0.41$ & BTC \\
51165.62 & $-0.61\pm0.67$ & BTC \\
51165.64 & $ 0.00\pm0.12$ & BTC \\
51193.58 & $-0.03\pm0.10$ & BTC \\
51193.63 & $ 0.02\pm0.09$ & BTC \\
\tableline
\end{tabular}
\end{lrbox}
\settowidth{\thiswid}{\usebox{\thisbox}}
\begin{center}
\begin{minipage}{\thiswid}
\caption{SN\,1997eq-R}
\label{tab:sn97198r}
\usebox{\thisbox}

$a$: Zeropoint: 23.284
\end{minipage}
\end{center}
\end{table}

\begin{table}[H]
\scriptsize\renewcommand{\arraystretch}{1.0}
\begin{lrbox}{\thisbox}
\begin{tabular}{rrl}
\tableline
\tableline
Julian Day & Flux$^a$ & Telescope \\
-2,400,000 \\
\tableline
50818.37 & $ 1.15\pm0.50$ & INT \\
50818.38 & $ 1.05\pm0.32$ & INT \\
50818.39 & $ 1.20\pm0.32$ & INT \\
50818.41 & $ 0.94\pm0.49$ & INT \\
50818.43 & $ 1.20\pm0.48$ & INT \\
50818.46 & $ 1.05\pm0.25$ & INT \\
50819.87 & $ 0.91\pm0.03$ & HSTPC \\
50821.68 & $ 0.93\pm0.35$ & WIYN \\
50821.69 & $ 0.83\pm0.41$ & WIYN \\
50821.70 & $ 0.65\pm0.38$ & WIYN \\
50824.90 & $ 0.86\pm0.02$ & HSTPC \\
50835.54 & $ 0.59\pm0.27$ & INT \\
50835.56 & $ 0.13\pm0.29$ & INT \\
50835.58 & $-0.11\pm0.50$ & INT \\
50846.82 & $ 0.38\pm0.02$ & HSTPC \\
50855.83 & $ 0.27\pm0.02$ & HSTPC \\
50863.83 & $ 0.22\pm0.01$ & HSTPC \\
51165.57 & $ 0.03\pm0.29$ & BTC \\
51165.60 & $ 0.06\pm0.34$ & BTC \\
51165.63 & $ 0.07\pm0.20$ & BTC \\
51165.65 & $ 0.06\pm0.17$ & BTC \\
51193.58 & $-0.10\pm0.17$ & BTC \\
\tableline
\end{tabular}
\end{lrbox}
\settowidth{\thiswid}{\usebox{\thisbox}}
\begin{center}
\begin{minipage}{\thiswid}
\caption{SN\,1997eq-I}
\label{tab:sn97198i}
\usebox{\thisbox}

$a$: Zeropoint: 22.388
\end{minipage}
\end{center}
\end{table}

\begin{table}[H]
\scriptsize\renewcommand{\arraystretch}{1.0}
\begin{lrbox}{\thisbox}
\begin{tabular}{rrl}
\tableline
\tableline
Julian Day & Flux$^a$ & Telescope \\
-2,400,000 \\
\tableline
50780.75 & $-0.41\pm1.15$ & BTC \\
50780.82 & $-0.88\pm0.96$ & BTC \\
50781.74 & $-1.46\pm1.01$ & BTC \\
50781.79 & $ 0.29\pm1.18$ & BTC \\
50781.79 & $ 1.09\pm0.96$ & BTC \\
50811.77 & $ 6.05\pm1.04$ & BTC \\
50811.77 & $ 3.90\pm1.89$ & WIYN \\
50811.77 & $ 5.82\pm1.03$ & BTC \\
50811.78 & $ 5.62\pm1.02$ & BTC \\
50811.78 & $ 5.82\pm2.22$ & WIYN \\
50811.79 & $ 3.97\pm4.73$ & WIYN \\
50811.81 & $ 5.97\pm1.04$ & BTC \\
50811.81 & $ 4.83\pm1.16$ & BTC \\
50817.84 & $ 5.51\pm1.22$ & BTC \\
50817.85 & $ 7.72\pm1.63$ & BTC \\
50817.86 & $ 4.58\pm2.15$ & BTC \\
50818.70 & $ 4.93\pm1.13$ & INT \\
50818.72 & $ 5.04\pm1.09$ & INT \\
50819.06 & $ 4.96\pm0.25$ & HSTPC \\
50824.97 & $ 3.65\pm0.22$ & HSTPC \\
50835.66 & $ 4.69\pm1.49$ & INT \\
50835.67 & $ 2.88\pm1.68$ & INT \\
50835.81 & $ 1.81\pm1.49$ & BTC \\
50835.82 & $-0.07\pm1.66$ & BTC \\
50835.83 & $ 0.52\pm1.70$ & BTC \\
51193.75 & $-0.14\pm0.74$ & BTC \\
51193.76 & $ 0.37\pm0.69$ & BTC \\
51193.76 & $ 0.00\pm1.08$ & BTC \\
51193.77 & $-1.23\pm0.85$ & BTC \\
51193.78 & $-0.20\pm0.83$ & BTC \\
51193.79 & $-0.21\pm0.78$ & BTC \\
51193.80 & $-1.80\pm1.63$ & WIYN \\
51195.73 & $-1.37\pm1.26$ & WIYN \\
51195.75 & $-0.21\pm1.40$ & WIYN \\
51195.77 & $-0.58\pm1.18$ & WIYN \\
51195.78 & $-0.92\pm1.36$ & WIYN \\
\tableline
\end{tabular}
\end{lrbox}
\settowidth{\thiswid}{\usebox{\thisbox}}
\begin{center}
\begin{minipage}{\thiswid}
\caption{SN\,1997ez-R}
\label{tab:sn97226r}
\usebox{\thisbox}

$a$: Zeropoint: 25.688
\end{minipage}
\end{center}
\end{table}

\begin{table}[H]
\scriptsize\renewcommand{\arraystretch}{1.0}
\begin{lrbox}{\thisbox}
\begin{tabular}{rrl}
\tableline
\tableline
Julian Day & Flux$^a$ & Telescope \\
-2,400,000 \\
\tableline
50816.74 & $ 2.05\pm1.90$ & BTC \\
50816.76 & $ 4.83\pm2.03$ & BTC \\
50816.77 & $ 4.64\pm1.89$ & BTC \\
50816.78 & $ 6.11\pm1.90$ & BTC \\
50816.78 & $ 5.02\pm2.02$ & BTC \\
50816.85 & $ 6.84\pm2.14$ & BTC \\
50818.63 & $ 4.19\pm2.23$ & INT \\
50818.65 & $ 4.24\pm1.55$ & INT \\
50818.66 & $ 4.12\pm1.54$ & INT \\
50818.68 & $ 4.30\pm1.54$ & INT \\
50819.07 & $ 5.23\pm0.18$ & HSTPC \\
50820.79 & $ 4.42\pm1.56$ & WIYN \\
50820.81 & $ 5.69\pm1.50$ & WIYN \\
50820.83 & $ 3.92\pm1.46$ & WIYN \\
50820.84 & $ 4.22\pm1.42$ & WIYN \\
50820.86 & $ 6.08\pm1.67$ & WIYN \\
50820.87 & $ 3.26\pm1.70$ & WIYN \\
50824.99 & $ 4.07\pm0.17$ & HSTPC \\
50835.60 & $ 5.27\pm1.77$ & INT \\
50835.61 & $ 0.53\pm2.03$ & INT \\
50835.63 & $ 5.55\pm1.94$ & INT \\
50835.64 & $ 5.62\pm2.52$ & INT \\
50835.84 & $ 3.39\pm2.13$ & BTC \\
50835.85 & $ 1.78\pm2.23$ & BTC \\
50835.86 & $-0.47\pm2.56$ & BTC \\
50846.55 & $ 1.77\pm0.09$ & HSTPC \\
50858.98 & $ 1.00\pm0.08$ & HSTPC \\
50871.89 & $ 0.48\pm0.04$ & HSTPC \\
51189.97 & $ 0.80\pm1.13$ & WIYN \\
51189.98 & $-0.74\pm1.22$ & WIYN \\
51190.00 & $-0.20\pm1.35$ & WIYN \\
51191.90 & $-0.54\pm1.34$ & WIYN \\
51191.92 & $-1.64\pm1.16$ & WIYN \\
51191.93 & $ 0.15\pm1.28$ & WIYN \\
51194.70 & $-3.19\pm2.44$ & BTC \\
51194.71 & $-1.06\pm2.73$ & BTC \\
51194.72 & $-0.60\pm2.43$ & BTC \\
51194.73 & $-0.52\pm2.81$ & BTC \\
51194.74 & $-1.26\pm2.28$ & BTC \\
51194.75 & $-0.84\pm2.49$ & BTC \\
51194.76 & $-0.27\pm1.90$ & BTC \\
51194.77 & $-2.00\pm2.19$ & BTC \\
51194.78 & $-1.89\pm2.02$ & BTC \\
51194.78 & $-1.58\pm2.61$ & BTC \\
51194.79 & $-0.68\pm2.38$ & BTC \\
\tableline
\end{tabular}
\end{lrbox}
\settowidth{\thiswid}{\usebox{\thisbox}}
\begin{center}
\begin{minipage}{\thiswid}
\caption{SN\,1997ez-I}
\label{tab:sn97226i}
\usebox{\thisbox}

$a$: Zeropoint: 24.954
\end{minipage}
\end{center}
\end{table}

\begin{table}[H]
\scriptsize\renewcommand{\arraystretch}{1.0}
\begin{lrbox}{\thisbox}
\begin{tabular}{rrl}
\tableline
\tableline
Julian Day & Flux$^a$ & Telescope \\
-2,400,000 \\
\tableline
50872.63 & $-0.10\pm0.10$ & BTC \\
50872.66 & $-0.07\pm0.09$ & BTC \\
50872.67 & $ 0.06\pm0.09$ & BTC \\
50872.72 & $-0.07\pm0.10$ & BTC \\
50872.73 & $-0.06\pm0.11$ & BTC \\
50873.57 & $ 0.06\pm0.11$ & BTC \\
50873.58 & $ 0.03\pm0.10$ & BTC \\
50895.58 & $ 2.33\pm0.12$ & BTC \\
50895.62 & $ 2.47\pm0.15$ & BTC \\
50896.58 & $ 2.64\pm0.12$ & BTC \\
50899.70 & $ 2.24\pm0.12$ & BTC \\
50904.68 & $ 2.15\pm0.11$ & BTC \\
50904.69 & $ 2.05\pm0.10$ & BTC \\
50904.70 & $ 2.20\pm0.10$ & BTC \\
50904.71 & $ 1.95\pm0.11$ & BTC \\
50904.72 & $ 2.00\pm0.10$ & BTC \\
50912.29 & $ 1.42\pm0.04$ & HSTPC \\
50935.01 & $ 0.33\pm0.02$ & HSTPC \\
50948.52 & $ 0.25\pm0.02$ & HSTPC \\
50963.17 & $ 0.19\pm0.02$ & HSTPC \\
51193.83 & $ 0.06\pm0.08$ & BTC \\
51193.84 & $-0.07\pm0.08$ & BTC \\
51193.86 & $ 0.04\pm0.08$ & BTC \\
51196.03 & $ 0.21\pm0.13$ & WIYN \\
51196.04 & $-0.19\pm0.12$ & WIYN \\
51196.05 & $-0.11\pm0.16$ & WIYN \\
\tableline
\end{tabular}
\end{lrbox}
\settowidth{\thiswid}{\usebox{\thisbox}}
\begin{center}
\begin{minipage}{\thiswid}
\caption{SN\,1998as-R}
\label{tab:sn98122r}
\usebox{\thisbox}

$a$: Zeropoint: 23.139
\end{minipage}
\end{center}
\end{table}

\begin{table}[H]
\scriptsize\renewcommand{\arraystretch}{1.0}
\begin{lrbox}{\thisbox}
\begin{tabular}{rrl}
\tableline
\tableline
Julian Day & Flux$^a$ & Telescope \\
-2,400,000 \\
\tableline
50912.31 & $ 9.24\pm0.21$ & HSTPC \\
50924.07 & $ 7.27\pm0.19$ & HSTPC \\
50932.65 & $ 1.95\pm1.56$ & WIYN \\
50935.02 & $ 4.86\pm0.17$ & HSTPC \\
50948.53 & $ 2.57\pm0.14$ & HSTPC \\
50963.19 & $ 1.79\pm0.12$ & HSTPC \\
51194.86 & $-1.02\pm0.98$ & BTC \\
51194.87 & $ 0.60\pm1.12$ & BTC \\
51196.93 & $-0.55\pm1.23$ & WIYN \\
51196.94 & $ 0.73\pm1.12$ & WIYN \\
51196.96 & $-1.44\pm1.28$ & WIYN \\
51280.50 & $ 0.53\pm1.60$ & BTC \\
51280.51 & $-2.08\pm1.50$ & BTC \\
51280.51 & $ 0.67\pm1.50$ & BTC \\
51280.52 & $ 0.60\pm1.33$ & BTC \\
51280.53 & $ 1.32\pm1.45$ & BTC \\
51280.54 & $ 0.72\pm1.46$ & BTC \\
\tableline
\end{tabular}
\end{lrbox}
\settowidth{\thiswid}{\usebox{\thisbox}}
\begin{center}
\begin{minipage}{\thiswid}
\caption{SN\,1998as-I}
\label{tab:sn98122i}
\usebox{\thisbox}

$a$: Zeropoint: 24.788
\end{minipage}
\end{center}
\end{table}

\clearpage
\begin{table}[H]
\scriptsize\renewcommand{\arraystretch}{1.0}
\begin{lrbox}{\thisbox}
\begin{tabular}{rrl}
\tableline
\tableline
Julian Day & Flux$^a$ & Telescope \\
-2,400,000 \\
\tableline
50513.71 & $ 0.08\pm0.14$ & BTC \\
50513.73 & $-0.08\pm0.16$ & BTC \\
50513.75 & $ 0.06\pm0.13$ & BTC \\
50514.71 & $ 0.08\pm0.14$ & BTC \\
50517.74 & $-0.19\pm0.14$ & BTC \\
50517.76 & $ 0.04\pm0.16$ & BTC \\
50518.79 & $ 0.31\pm0.17$ & BTC \\
50518.81 & $-0.02\pm0.17$ & BTC \\
50872.56 & $-0.03\pm0.21$ & BTC \\
50872.59 & $-0.03\pm0.22$ & BTC \\
50873.73 & $-0.03\pm0.18$ & BTC \\
50873.74 & $-0.09\pm0.15$ & BTC \\
50895.60 & $ 0.02\pm0.16$ & BTC \\
50895.64 & $ 0.55\pm0.16$ & BTC \\
50896.58 & $ 0.67\pm0.15$ & BTC \\
50896.60 & $ 0.39\pm0.16$ & BTC \\
50899.69 & $ 0.89\pm0.15$ & BTC \\
50904.63 & $ 1.87\pm0.14$ & BTC \\
50904.64 & $ 1.66\pm0.14$ & BTC \\
50904.65 & $ 1.75\pm0.13$ & BTC \\
50904.66 & $ 1.82\pm0.14$ & BTC \\
50904.67 & $ 1.82\pm0.14$ & BTC \\
50912.03 & $ 2.53\pm0.07$ & HSTPC \\
50922.11 & $ 2.11\pm0.06$ & HSTPC \\
50927.56 & $ 2.05\pm0.38$ & BTC \\
50927.57 & $ 1.80\pm0.34$ & BTC \\
50927.60 & $ 1.69\pm0.36$ & BTC \\
50927.61 & $ 0.96\pm0.41$ & BTC \\
50929.64 & $ 1.48\pm0.28$ & WIYN \\
50929.65 & $ 1.06\pm0.33$ & WIYN \\
50929.67 & $ 1.90\pm0.31$ & WIYN \\
50933.07 & $ 1.32\pm0.04$ & HSTPC \\
50947.71 & $ 0.58\pm0.03$ & HSTPC \\
50961.83 & $ 0.30\pm0.03$ & HSTPC \\
51192.96 & $-0.19\pm0.26$ & WIYN \\
51192.98 & $-0.14\pm0.39$ & WIYN \\
51193.00 & $ 0.18\pm0.28$ & WIYN \\
51193.02 & $-0.14\pm0.24$ & WIYN \\
51193.03 & $-0.29\pm0.28$ & WIYN \\
51279.60 & $ 0.01\pm0.13$ & BTC \\
51279.61 & $ 0.04\pm0.14$ & BTC \\
51279.63 & $-0.04\pm0.12$ & BTC \\
51279.66 & $ 0.01\pm0.13$ & BTC \\
51280.56 & $ 0.14\pm0.16$ & BTC \\
51280.57 & $ 0.17\pm0.15$ & BTC \\
\tableline
\end{tabular}
\end{lrbox}
\settowidth{\thiswid}{\usebox{\thisbox}}
\begin{center}
\begin{minipage}{\thiswid}
\caption{SN\,1998aw-R}
\label{tab:sn9855r}
\usebox{\thisbox}

$a$: Zeropoint: 23.536
\end{minipage}
\end{center}
\end{table}

\begin{table}[H]
\scriptsize\renewcommand{\arraystretch}{1.0}
\begin{lrbox}{\thisbox}
\begin{tabular}{rrl}
\tableline
\tableline
Julian Day & Flux$^a$ & Telescope \\
-2,400,000 \\
\tableline
50513.76 & $-0.33\pm0.25$ & BTC \\
50514.74 & $-0.10\pm0.22$ & BTC \\
50514.76 & $-0.12\pm0.21$ & BTC \\
50514.78 & $ 0.06\pm0.23$ & BTC \\
50518.73 & $ 0.18\pm0.42$ & BTC \\
50518.75 & $-0.08\pm0.34$ & BTC \\
50912.04 & $ 1.79\pm0.05$ & HSTPC \\
50922.12 & $ 1.67\pm0.05$ & HSTPC \\
50929.70 & $ 1.50\pm0.49$ & WIYN \\
50930.71 & $ 1.80\pm0.46$ & WIYN \\
50933.08 & $ 1.23\pm0.03$ & HSTPC \\
50947.73 & $ 0.80\pm0.03$ & HSTPC \\
50961.84 & $ 0.53\pm0.03$ & HSTPC \\
51194.03 & $-0.07\pm0.32$ & WIYN \\
51194.05 & $-0.26\pm0.51$ & WIYN \\
51195.97 & $-0.21\pm0.32$ & WIYN \\
51195.98 & $ 0.13\pm0.27$ & WIYN \\
51196.00 & $ 0.10\pm0.29$ & WIYN \\
51196.02 & $ 0.05\pm0.27$ & WIYN \\
51279.59 & $-0.03\pm0.21$ & BTC \\
51279.62 & $-0.06\pm0.25$ & BTC \\
51279.64 & $ 0.15\pm0.21$ & BTC \\
51279.65 & $ 0.01\pm0.23$ & BTC \\
51279.66 & $ 0.19\pm0.25$ & BTC \\
51280.55 & $ 0.14\pm0.31$ & BTC \\
51280.57 & $-0.02\pm0.28$ & BTC \\
51280.59 & $-0.30\pm0.29$ & BTC \\
51280.60 & $ 0.09\pm0.29$ & BTC \\
\tableline
\end{tabular}
\end{lrbox}
\settowidth{\thiswid}{\usebox{\thisbox}}
\begin{center}
\begin{minipage}{\thiswid}
\caption{SN\,1998aw-I}
\label{tab:sn9855i}
\usebox{\thisbox}

$a$: Zeropoint: 22.874
\end{minipage}
\end{center}
\end{table}

\begin{table}[H]
\scriptsize\renewcommand{\arraystretch}{1.0}
\begin{lrbox}{\thisbox}
\begin{tabular}{rrl}
\tableline
\tableline
Julian Day & Flux$^a$ & Telescope \\
-2,400,000 \\
\tableline
50138.65 & $-0.03\pm0.09$ & CTIO \\
50138.67 & $-0.09\pm0.10$ & CTIO \\
50159.64 & $-0.09\pm0.08$ & CTIO \\
50159.66 & $ 0.03\pm0.07$ & CTIO \\
50160.67 & $ 0.01\pm0.07$ & CTIO \\
50160.68 & $ 0.02\pm0.06$ & CTIO \\
50168.59 & $-0.03\pm0.07$ & CTIO \\
50168.65 & $ 0.14\pm0.06$ & CTIO \\
50169.64 & $ 0.13\pm0.15$ & CTIO \\
50169.67 & $-0.01\pm0.08$ & CTIO \\
50432.83 & $-0.06\pm0.06$ & CTIO \\
50453.84 & $-0.01\pm0.08$ & CTIO \\
50454.77 & $ 0.01\pm0.06$ & CTIO \\
50459.82 & $-0.02\pm0.04$ & CTIO \\
50459.83 & $-0.02\pm0.05$ & CTIO \\
50459.84 & $ 0.02\pm0.05$ & CTIO \\
50490.79 & $ 0.01\pm0.06$ & BTC \\
50490.79 & $ 0.07\pm0.06$ & BTC \\
50490.80 & $-0.04\pm0.06$ & BTC \\
50490.80 & $-0.04\pm0.06$ & BTC \\
50513.71 & $-0.03\pm0.06$ & BTC \\
50514.72 & $-0.06\pm0.06$ & BTC \\
50872.54 & $ 0.72\pm0.12$ & BTC \\
50872.57 & $ 0.58\pm0.12$ & BTC \\
50873.53 & $ 0.84\pm0.17$ & BTC \\
50873.55 & $ 0.95\pm0.10$ & BTC \\
50895.52 & $ 1.42\pm0.09$ & BTC \\
50895.55 & $ 1.06\pm0.19$ & BTC \\
50895.71 & $ 1.24\pm0.07$ & BTC \\
50896.53 & $ 1.14\pm0.10$ & BTC \\
50900.70 & $ 1.14\pm0.07$ & BTC \\
50900.71 & $ 1.04\pm0.07$ & BTC \\
50904.59 & $ 0.91\pm0.06$ & BTC \\
50904.60 & $ 0.84\pm0.06$ & BTC \\
50904.61 & $ 0.81\pm0.06$ & BTC \\
50904.62 & $ 0.84\pm0.06$ & BTC \\
50904.63 & $ 0.89\pm0.06$ & BTC \\
50911.96 & $ 0.59\pm0.03$ & HSTPC \\
50922.04 & $ 0.31\pm0.02$ & HSTPC \\
50933.00 & $ 0.18\pm0.02$ & HSTPC \\
50947.65 & $ 0.09\pm0.01$ & HSTPC \\
50961.23 & $ 0.09\pm0.01$ & HSTPC \\
51193.80 & $-0.00\pm0.05$ & BTC \\
51193.81 & $-0.00\pm0.05$ & BTC \\
51193.82 & $-0.01\pm0.06$ & BTC \\
51279.52 & $-0.01\pm0.08$ & BTC \\
51279.57 & $ 0.11\pm0.08$ & BTC \\
51280.61 & $ 0.06\pm0.06$ & BTC \\
\tableline
\end{tabular}
\end{lrbox}
\settowidth{\thiswid}{\usebox{\thisbox}}
\begin{center}
\begin{minipage}{\thiswid}
\caption{SN\,1998ax-R}
\label{tab:sn98109r}
\usebox{\thisbox}

$a$: Zeropoint: 22.922
\end{minipage}
\end{center}
\end{table}

\begin{table}[H]
\scriptsize\renewcommand{\arraystretch}{1.0}
\begin{lrbox}{\thisbox}
\begin{tabular}{rrl}
\tableline
\tableline
Julian Day & Flux$^a$ & Telescope \\
-2,400,000 \\
\tableline
50911.97 & $ 1.95\pm0.10$ & HSTPC \\
50922.05 & $ 1.62\pm0.10$ & HSTPC \\
50933.01 & $ 1.18\pm0.06$ & HSTPC \\
50947.66 & $ 0.75\pm0.05$ & HSTPC \\
50961.24 & $ 0.47\pm0.04$ & HSTPC \\
\tableline
\end{tabular}
\end{lrbox}
\settowidth{\thiswid}{\usebox{\thisbox}}
\begin{center}
\begin{minipage}{\thiswid}
\caption{SN\,1998ax-I}
\label{tab:sn98109i}
\usebox{\thisbox}

$a$: Zeropoint: 23.685
\end{minipage}
\end{center}
\end{table}

\begin{table}[H]
\scriptsize\renewcommand{\arraystretch}{1.0}
\begin{lrbox}{\thisbox}
\begin{tabular}{rrl}
\tableline
\tableline
Julian Day & Flux$^a$ & Telescope \\
-2,400,000 \\
\tableline
50521.85 & $ 0.02\pm0.50$ & WIYN \\
50521.86 & $ 0.17\pm0.56$ & WIYN \\
50872.54 & $ 2.11\pm1.08$ & BTC \\
50872.57 & $ 1.27\pm0.97$ & BTC \\
50873.53 & $ 0.57\pm1.81$ & BTC \\
50873.55 & $-0.70\pm1.04$ & BTC \\
50895.52 & $ 5.69\pm0.90$ & BTC \\
50895.55 & $ 6.69\pm1.91$ & BTC \\
50895.71 & $ 6.10\pm0.78$ & BTC \\
50896.53 & $ 6.70\pm1.24$ & BTC \\
50900.70 & $ 5.74\pm0.76$ & BTC \\
50900.71 & $ 6.74\pm0.91$ & BTC \\
50904.59 & $ 5.48\pm0.78$ & BTC \\
50904.60 & $ 5.64\pm0.75$ & BTC \\
50904.61 & $ 5.61\pm0.78$ & BTC \\
50904.62 & $ 5.76\pm0.82$ & BTC \\
50904.63 & $ 5.91\pm0.79$ & BTC \\
50912.16 & $ 3.11\pm0.20$ & HSTPC \\
50923.99 & $ 1.58\pm0.17$ & HSTPC \\
51193.80 & $-0.09\pm0.60$ & BTC \\
51193.81 & $ 0.61\pm0.48$ & BTC \\
51193.82 & $ 0.53\pm0.64$ & BTC \\
\tableline
\end{tabular}
\end{lrbox}
\settowidth{\thiswid}{\usebox{\thisbox}}
\begin{center}
\begin{minipage}{\thiswid}
\caption{SN\,1998ay-R}
\label{tab:sn98104r}
\usebox{\thisbox}

$a$: Zeropoint: 25.093
\end{minipage}
\end{center}
\end{table}

\begin{table}[H]
\scriptsize\renewcommand{\arraystretch}{1.0}
\begin{lrbox}{\thisbox}
\begin{tabular}{rrl}
\tableline
\tableline
Julian Day & Flux$^a$ & Telescope \\
-2,400,000 \\
\tableline
50912.17 & $ 1.56\pm0.08$ & HSTPC \\
50924.00 & $ 0.96\pm0.07$ & HSTPC \\
50934.68 & $ 0.61\pm0.04$ & HSTPC \\
50948.59 & $ 0.40\pm0.04$ & HSTPC \\
50967.81 & $ 0.26\pm0.04$ & HSTPC \\
\tableline
\end{tabular}
\end{lrbox}
\settowidth{\thiswid}{\usebox{\thisbox}}
\begin{center}
\begin{minipage}{\thiswid}
\caption{SN\,1998ay-I}
\label{tab:sn98104i}
\usebox{\thisbox}

$a$: Zeropoint: 23.685
\end{minipage}
\end{center}
\end{table}

\clearpage

\begin{table}[H]
\scriptsize\renewcommand{\arraystretch}{1.0}
\begin{lrbox}{\thisbox}
\begin{tabular}{rrl}
\tableline
\tableline
Julian Day & Flux$^a$ & Telescope \\
-2,400,000 \\
\tableline
50873.79 & $ 0.03\pm0.09$ & BTC \\
50873.80 & $ 0.09\pm0.09$ & BTC \\
50873.81 & $ 0.01\pm0.09$ & BTC \\
50873.82 & $ 0.03\pm0.09$ & BTC \\
50873.83 & $ 0.01\pm0.08$ & BTC \\
50873.84 & $-0.03\pm0.09$ & BTC \\
50895.78 & $ 1.50\pm0.14$ & BTC \\
50895.85 & $ 1.64\pm0.15$ & BTC \\
50899.75 & $ 1.52\pm0.11$ & BTC \\
50899.84 & $ 1.43\pm0.14$ & BTC \\
50899.90 & $ 1.20\pm0.21$ & BTC \\
50900.74 & $ 1.54\pm0.10$ & BTC \\
50900.75 & $ 1.32\pm0.10$ & BTC \\
50904.77 & $ 1.36\pm0.11$ & BTC \\
50904.78 & $ 1.20\pm0.11$ & BTC \\
50904.79 & $ 1.42\pm0.13$ & BTC \\
50904.80 & $ 1.30\pm0.09$ & BTC \\
50904.81 & $ 1.34\pm0.11$ & BTC \\
50912.10 & $ 0.79\pm0.03$ & HSTPC \\
50923.12 & $ 0.41\pm0.02$ & HSTPC \\
50933.21 & $ 0.22\pm0.02$ & HSTPC \\
50947.12 & $ 0.12\pm0.01$ & HSTPC \\
50961.90 & $ 0.12\pm0.01$ & HSTPC \\
51258.01 & $-0.15\pm0.11$ & WIYN \\
51279.82 & $ 0.07\pm0.08$ & BTC \\
51279.85 & $-0.05\pm0.10$ & BTC \\
51280.69 & $-0.02\pm0.07$ & BTC \\
51280.70 & $ 0.03\pm0.06$ & BTC \\
\tableline
\end{tabular}
\end{lrbox}
\settowidth{\thiswid}{\usebox{\thisbox}}
\begin{center}
\begin{minipage}{\thiswid}
\caption{SN\,1998ba-R}
\label{tab:sn9819r}
\usebox{\thisbox}

$a$: Zeropoint: 22.779
\end{minipage}
\end{center}
\end{table}

\begin{table}[H]
\scriptsize\renewcommand{\arraystretch}{1.0}
\begin{lrbox}{\thisbox}
\begin{tabular}{rrl}
\tableline
\tableline
Julian Day & Flux$^a$ & Telescope \\
-2,400,000 \\
\tableline
50907.82 & $ 3.18\pm1.99$ & WIYN \\
50907.83 & $ 3.96\pm1.75$ & WIYN \\
50907.84 & $ 6.80\pm1.81$ & WIYN \\
50907.85 & $ 6.04\pm2.36$ & WIYN \\
50912.11 & $ 5.74\pm0.22$ & HSTPC \\
50923.13 & $ 3.95\pm0.21$ & HSTPC \\
50933.22 & $ 2.81\pm0.12$ & HSTPC \\
50947.13 & $ 1.57\pm0.10$ & HSTPC \\
50961.92 & $ 1.37\pm0.10$ & HSTPC \\
51279.83 & $-1.51\pm1.00$ & BTC \\
51279.84 & $ 0.88\pm1.09$ & BTC \\
51280.69 & $-1.04\pm0.83$ & BTC \\
51280.71 & $ 0.66\pm0.72$ & BTC \\
51280.72 & $-0.06\pm0.68$ & BTC \\
51280.73 & $ 0.13\pm0.68$ & BTC \\
\tableline
\end{tabular}
\end{lrbox}
\settowidth{\thiswid}{\usebox{\thisbox}}
\begin{center}
\begin{minipage}{\thiswid}
\caption{SN\,1998ba-I}
\label{tab:sn9819i}
\usebox{\thisbox}

$a$: Zeropoint: 24.477
\end{minipage}
\end{center}
\end{table}

\begin{table}[H]
\scriptsize\renewcommand{\arraystretch}{1.0}
\begin{lrbox}{\thisbox}
\begin{tabular}{rrl}
\tableline
\tableline
Julian Day & Flux$^a$ & Telescope \\
-2,400,000 \\
\tableline
50490.86 & $ 0.49\pm0.55$ & BTC \\
50490.87 & $-0.39\pm0.54$ & BTC \\
50513.83 & $-0.02\pm0.52$ & BTC \\
50513.84 & $ 0.15\pm0.54$ & BTC \\
50514.83 & $ 0.53\pm0.60$ & BTC \\
50514.86 & $-0.51\pm0.53$ & BTC \\
50517.88 & $ 0.33\pm0.70$ & BTC \\
50517.90 & $-0.26\pm0.71$ & BTC \\
50517.90 & $ 0.69\pm0.81$ & BTC \\
50518.86 & $ 0.22\pm0.62$ & BTC \\
50518.87 & $ 0.57\pm0.66$ & BTC \\
50872.74 & $-0.75\pm0.91$ & BTC \\
50872.89 & $ 1.36\pm0.93$ & BTC \\
50873.87 & $ 0.63\pm0.53$ & BTC \\
50895.78 & $ 4.22\pm0.70$ & BTC \\
50895.84 & $ 5.34\pm0.88$ & BTC \\
50899.75 & $ 7.13\pm0.79$ & BTC \\
50899.82 & $ 6.98\pm0.91$ & BTC \\
50900.76 & $ 4.64\pm0.65$ & BTC \\
50904.73 & $ 6.58\pm0.65$ & BTC \\
50904.74 & $ 6.90\pm0.67$ & BTC \\
50904.75 & $ 6.31\pm0.72$ & BTC \\
50904.75 & $ 7.32\pm0.73$ & BTC \\
50904.76 & $ 8.29\pm0.76$ & BTC \\
50904.86 & $ 7.95\pm0.89$ & BTC \\
50912.23 & $ 5.73\pm0.25$ & HSTPC \\
50923.19 & $ 2.11\pm0.18$ & HSTPC \\
50932.74 & $ 2.04\pm0.89$ & WIYN \\
50932.77 & $ 1.38\pm0.93$ & WIYN \\
50934.08 & $ 0.73\pm0.12$ & HSTPC \\
50949.00 & $ 0.76\pm0.13$ & HSTPC \\
50962.17 & $ 0.21\pm0.13$ & HSTPC \\
51279.68 & $-0.16\pm0.67$ & BTC \\
51279.71 & $ 0.31\pm0.68$ & BTC \\
51279.75 & $ 0.21\pm0.73$ & BTC \\
51279.77 & $-0.30\pm0.79$ & BTC \\
\tableline
\end{tabular}
\end{lrbox}
\settowidth{\thiswid}{\usebox{\thisbox}}
\begin{center}
\begin{minipage}{\thiswid}
\caption{SN\,1998be-R}
\label{tab:sn9878r}
\usebox{\thisbox}

$a$: Zeropoint: 25.350
\end{minipage}
\end{center}
\end{table}

\begin{table}[H]
\scriptsize\renewcommand{\arraystretch}{1.0}
\begin{lrbox}{\thisbox}
\begin{tabular}{rrl}
\tableline
\tableline
Julian Day & Flux$^a$ & Telescope \\
-2,400,000 \\
\tableline
50514.85 & $-0.21\pm0.83$ & BTC \\
50514.87 & $-1.02\pm0.78$ & BTC \\
50518.84 & $ 2.00\pm0.90$ & BTC \\
50518.85 & $ 1.47\pm0.86$ & BTC \\
50518.85 & $ 0.31\pm0.82$ & BTC \\
50912.25 & $ 3.66\pm0.18$ & HSTPC \\
50923.20 & $ 2.19\pm0.17$ & HSTPC \\
50932.80 & $ 2.35\pm1.10$ & WIYN \\
50932.85 & $ 2.26\pm0.92$ & WIYN \\
50934.09 & $ 1.13\pm0.09$ & HSTPC \\
50949.01 & $ 0.80\pm0.08$ & HSTPC \\
50962.19 & $ 0.37\pm0.08$ & HSTPC \\
51279.69 & $ 0.81\pm0.89$ & BTC \\
51279.70 & $ 0.49\pm0.87$ & BTC \\
51279.72 & $ 1.51\pm0.73$ & BTC \\
51279.73 & $-0.02\pm0.71$ & BTC \\
51279.76 & $ 0.62\pm0.83$ & BTC \\
51279.77 & $ 0.58\pm0.85$ & BTC \\
51280.64 & $-0.87\pm0.82$ & BTC \\
51280.64 & $ 0.36\pm0.84$ & BTC \\
51280.65 & $ 0.12\pm0.73$ & BTC \\
51280.66 & $-0.13\pm0.78$ & BTC \\
51280.67 & $ 1.24\pm0.76$ & BTC \\
51280.68 & $-0.62\pm0.76$ & BTC \\
\tableline
\end{tabular}
\end{lrbox}
\settowidth{\thiswid}{\usebox{\thisbox}}
\begin{center}
\begin{minipage}{\thiswid}
\caption{SN\,1998be-I}
\label{tab:sn9878i}
\usebox{\thisbox}

$a$: Zeropoint: 24.384
\end{minipage}
\end{center}
\end{table}

\begin{table}[H]
\scriptsize\renewcommand{\arraystretch}{1.0}
\begin{lrbox}{\thisbox}
\begin{tabular}{rrl}
\tableline
\tableline
Julian Day & Flux$^a$ & Telescope \\
-2,400,000 \\
\tableline
50138.79 & $-1.04\pm0.91$ & CTIO \\
50138.82 & $ 0.85\pm0.86$ & CTIO \\
50168.80 & $-0.68\pm0.66$ & CTIO \\
50490.86 & $ 0.40\pm0.49$ & BTC \\
50490.87 & $-0.09\pm0.48$ & BTC \\
50513.83 & $ 0.26\pm0.51$ & BTC \\
50513.84 & $-0.10\pm0.53$ & BTC \\
50514.83 & $-1.06\pm0.58$ & BTC \\
50514.86 & $-0.05\pm0.50$ & BTC \\
50517.88 & $ 0.13\pm0.65$ & BTC \\
50517.89 & $-0.11\pm0.60$ & BTC \\
50517.89 & $ 0.93\pm0.60$ & BTC \\
50517.90 & $-0.29\pm0.68$ & BTC \\
50517.90 & $-0.35\pm0.74$ & BTC \\
50872.89 & $ 0.52\pm0.81$ & BTC \\
50873.87 & $ 0.60\pm0.51$ & BTC \\
50895.78 & $ 3.15\pm0.63$ & BTC \\
50895.84 & $ 3.11\pm0.79$ & BTC \\
50899.75 & $ 4.93\pm0.65$ & BTC \\
50899.82 & $ 4.28\pm0.70$ & BTC \\
50900.76 & $ 4.44\pm0.55$ & BTC \\
50904.73 & $ 6.10\pm0.61$ & BTC \\
50904.75 & $ 5.30\pm0.61$ & BTC \\
50904.75 & $ 5.38\pm0.64$ & BTC \\
50904.76 & $ 6.21\pm0.66$ & BTC \\
50904.86 & $ 5.27\pm0.77$ & BTC \\
50910.15 & $ 5.27\pm0.20$ & HSTPC \\
50922.18 & $ 3.75\pm0.18$ & HSTPC \\
51279.71 & $ 0.94\pm0.73$ & BTC \\
51279.74 & $ 0.63\pm0.67$ & BTC \\
51279.75 & $-1.14\pm0.68$ & BTC \\
51279.77 & $ 0.47\pm0.76$ & BTC \\
\tableline
\end{tabular}
\end{lrbox}
\settowidth{\thiswid}{\usebox{\thisbox}}
\begin{center}
\begin{minipage}{\thiswid}
\caption{SN\,1998bi-R}
\label{tab:sn98142r}
\usebox{\thisbox}

$a$: Zeropoint: 25.213
\end{minipage}
\end{center}
\end{table}

\begin{table}[H]
\scriptsize\renewcommand{\arraystretch}{1.0}
\begin{lrbox}{\thisbox}
\begin{tabular}{rrl}
\tableline
\tableline
Julian Day & Flux$^a$ & Telescope \\
-2,400,000 \\
\tableline
50910.16 & $ 2.07\pm0.06$ & HSTPC \\
50922.20 & $ 1.83\pm0.06$ & HSTPC \\
50931.99 & $ 1.25\pm0.04$ & HSTPC \\
50946.38 & $ 0.54\pm0.03$ & HSTPC \\
50966.88 & $ 0.20\pm0.02$ & HSTPC \\
\tableline
\end{tabular}
\end{lrbox}
\settowidth{\thiswid}{\usebox{\thisbox}}
\begin{center}
\begin{minipage}{\thiswid}
\caption{SN\,1998bi-I}
\label{tab:sn98142i}
\usebox{\thisbox}

$a$: Zeropoint: 23.685
\end{minipage}
\end{center}
\end{table}

\begin{table}[H]
\scriptsize\renewcommand{\arraystretch}{1.0}
\begin{lrbox}{\thisbox}
\begin{tabular}{rrl}
\tableline
\tableline
Julian Day & Flux$^a$ & Telescope \\
-2,400,000 \\
\tableline
51671.77 & $ 1.02\pm0.07$ & KECK \\
51671.77 & $ 1.05\pm0.07$ & KECK \\
51671.78 & $ 1.06\pm0.07$ & KECK \\
51671.78 & $ 0.99\pm0.07$ & KECK \\
51679.98 & $ 1.66\pm0.04$ & HSTPC \\
51692.91 & $ 1.43\pm0.03$ & HSTPC \\
51706.26 & $ 0.73\pm0.02$ & HSTPC \\
51718.04 & $ 0.39\pm0.01$ & HSTPC \\
51733.86 & $ 0.16\pm0.01$ & HSTPC \\
52014.72 & $-0.01\pm0.07$ & NTT \\
52014.73 & $-0.08\pm0.07$ & NTT \\
52014.74 & $ 0.04\pm0.08$ & NTT \\
52014.75 & $-0.04\pm0.06$ & NTT \\
52014.76 & $-0.04\pm0.07$ & NTT \\
52014.77 & $-0.08\pm0.10$ & NTT \\
52014.78 & $-0.07\pm0.09$ & NTT \\
52014.79 & $-0.04\pm0.10$ & NTT \\
52014.80 & $-0.16\pm0.14$ & NTT \\
52376.98 & $ 0.01\pm0.04$ & CFHT \\
52376.99 & $-0.00\pm0.03$ & CFHT \\
52377.04 & $ 0.01\pm0.04$ & CFHT \\
52377.05 & $-0.02\pm0.04$ & CFHT \\
52382.01 & $ 0.03\pm0.05$ & CFHT \\
52384.98 & $-0.00\pm0.09$ & CFHT \\
52386.85 & $-0.14\pm0.10$ & CFHT \\
\tableline
\end{tabular}
\end{lrbox}
\settowidth{\thiswid}{\usebox{\thisbox}}
\begin{center}
\begin{minipage}{\thiswid}
\caption{SN\,2000fr-R}
\label{tab:C00-008r}
\usebox{\thisbox}

$a$: Zeropoint: 22.998
\end{minipage}
\end{center}
\end{table}

\begin{table}[H]
\scriptsize\renewcommand{\arraystretch}{1.0}
\begin{lrbox}{\thisbox}
\begin{tabular}{rrl}
\tableline
\tableline
Julian Day & Flux$^a$ & Telescope \\
-2,400,000 \\
\tableline
51641.99 & $ 0.03\pm0.04$ & CFHT \\
51664.95 & $ 0.40\pm0.05$ & CFHT \\
51664.99 & $ 0.40\pm0.06$ & CFHT \\
51672.86 & $ 1.14\pm0.02$ & HSTPC \\
51679.97 & $ 1.59\pm0.03$ & HSTPC \\
51692.91 & $ 1.46\pm0.03$ & HSTPC \\
51706.20 & $ 1.02\pm0.03$ & HSTPC \\
51717.98 & $ 0.66\pm0.02$ & HSTPC \\
51733.79 & $ 0.40\pm0.02$ & HSTPC \\
51997.93 & $ 0.05\pm0.06$ & CFHT \\
51997.94 & $ 0.01\pm0.06$ & CFHT \\
51997.99 & $ 0.19\pm0.05$ & CFHT \\
51998.00 & $ 0.03\pm0.06$ & CFHT \\
51998.01 & $ 0.08\pm0.06$ & CFHT \\
52376.96 & $ 0.04\pm0.06$ & CFHT \\
52376.97 & $-0.06\pm0.06$ & CFHT \\
52377.00 & $ 0.13\pm0.06$ & CFHT \\
52377.00 & $-0.09\pm0.06$ & CFHT \\
52377.01 & $-0.01\pm0.06$ & CFHT \\
52377.03 & $ 0.01\pm0.07$ & CFHT \\
\tableline
\end{tabular}
\end{lrbox}
\settowidth{\thiswid}{\usebox{\thisbox}}
\begin{center}
\begin{minipage}{\thiswid}
\caption{SN\,2000fr-I}
\label{tab:C00-008i}
\usebox{\thisbox}

$a$: Zeropoint: 22.805
\end{minipage}
\end{center}
\end{table}

\clearpage
}{}


\end{document}